\documentclass{aa}
\usepackage[varg]{txfonts}
\usepackage{graphicx}
\usepackage{natbib}
\usepackage[colorlinks=true,     linkcolor=blue, citecolor=blue, filecolor=blue, urlcolor=blue]{hyperref}
\usepackage{pifont}

\newcommand{\HII}{H\,{\footnotesize II}{}}
\newcommand{\hi}{H\,{\footnotesize I}{}}

\newcommand{\HI}{H\,{\footnotesize I}{}}
\newcommand{\htwo}{\mathrm{H_2}{}}

\newcommand{\mum}{\mathrm{\mu m}}

\newcommand{\kkms}{\mathrm{K\,km\,s^{-1}}}

\newcommand{\cotwoone}{\mathrm{CO(2-1)}}

\newcommand{\Xco}{$X_\mathrm{CO}$}
\newcommand{\mathCO}{\mathrm{CO}}

\newcommand{\Nhtwo}{N_\htwo}

\newcommand*{\command}[1]{\texttt{#1}}

\newcommand{\cmKkms}{\mathrm{cm^{-2}/(K\,km\,s^{-1})}}

\newcommand{\cothreesigma}{1.046}

\newcommand{\numstructdendrodust}{$326${} }
\newcommand{\numstructdendroCO}{$199${} }

\defcitealias{Keilmann2024}{Paper I}
\defcitealias{Nguyen2016}{NL16}

\begin{document}

\title{Molecular cloud matching in CO and dust in M33}
\subtitle{II. Physical properties of giant molecular clouds}

\author{
  Eduard Keilmann,  \inst{1}
  Slawa Kabanovic, \inst{1}
  Nicola Schneider, \inst{1}
  Volker Ossenkopf-Okada, \inst{1}
  J\"urgen Stutzki, \inst{1}
  Masato I. N. Kobayashi, \inst{1}  
  Robert Simon, \inst{1}
  Christof Buchbender, \inst{1}
  Dominik Riechers, \inst{1}
  Frank Bigiel, \inst{2}
  Fatemeh Tabatabaei \inst{3,4} 
}
\institute{
I. Physikalisches Institut, Universität zu K\"{o}ln,  Z\"ulpicher Stra\ss{}e 77, 50937 K\"oln, Germany \\
\email{keilmann@ph1.uni-koeln.de}
\and Argelander-Institut für Astronomie, Universität Bonn, Auf dem H\"ugel 71, 53121 Bonn, Germany 
\and School of Astronomy, Institute for Research in Fundamental Sciences (IPM), PO Box 19395-5531, Tehran, Iran 
\and Max-Planck-Institut für Astronomie, K\"{o}nigstuhl 17, 69117, Heidelberg, Germany 
}
\date{}

\abstract{
Understanding the physical properties such as mass, size, and surface mass density of giant molecular clouds or associations (GMCs/GMAs) in galaxies is crucial for gaining deeper insights into the molecular cloud and star formation (SF) processes. We determine these quantities for the Local Group flocculent spiral galaxy M33 using {\sl Herschel} dust and archival $^{12}\mathrm{CO(2-1)}$ data from the IRAM 30m telescope, and compare them to GMC/GMA properties of the Milky Way derived from CO literature data. For M33, we apply the Dendrogram algorithm on a novel 2D dust-derived $\Nhtwo$ map at an angular resolution of $18.2''$ and on the $^{12}\mathrm{CO(2-1)}$ data and employ an \Xco\ factor map instead of a constant value.
Dust and CO-derived values are similar, with mean radii of $\sim\,$$58\,$pc for the dust and $\sim\,$$68\,$pc for CO, respectively.
However, the largest GMAs have a radius of around $150\,$pc, similar to what was found in the Milky Way and other galaxies, suggesting a physical process that limits the size of GMAs. 
The less massive and smaller M33 galaxy also hosts less massive and lower-density GMCs compared to the Milky Way by an order of magnitude. Notably, the most massive ($>$ a few $10^6\,\mathrm{M_\odot}$) GMC population observed in the Milky Way is mainly missing in M33. The mean surface mass density of M33 is significantly smaller than that of the Milky Way and this is attributed to higher column densities of the largest GMCs in the Milky Way, despite similar GMC areas. 
We find no systematic gradients in physical properties with the galactocentric radius in M33. However, surface mass densities and masses are higher near the center, implying increased SF activity.
In both galaxies, the central region contains $\sim\,$30\% of the total molecular mass. 
The index of the power-law spectrum of the GMC masses across the entire disk of M33 is $\alpha=2.3\pm0.1$ and $\alpha=1.9\pm0.1$ for dust- and CO-derived data, respectively. 
We conclude that GMC properties in M33 and the Milky Way are largely similar, though M33 lacks high-mass GMCs, for which there is no straightforward explanation. Additionally, GMC properties are only weakly dependent on the galactic environment, with stellar feedback playing a role that needs further investigation. }

\titlerunning{Cloud matching in M33 II}
\authorrunning{Keilmann et al.}
\maketitle

\section{Introduction} \label{sec:introduction}

Molecular clouds (MCs) are the birthplaces of stars in galaxies and their formation is a complex process influenced by various physical mechanisms. One key process is the gravitational collapse of dense regions within the interstellar medium (ISM) of galaxies. These regions often arise from instabilities within the ISM, which are triggered by processes such as spiral density waves and stellar feedback~\citep{McKee2007,Dobbs2014,Renaud2015}. 
Spiral density waves in galaxies like the Milky Way are mediated by gravitational interactions between stars, gas, and dark matter in the galactic disk. As these waves propagate through the disk, they compress and shock the gas along the trailing edge of a spiral arm~\citep{Fujimoto1968,Roberts1969}, leading to the formation of dense MCs. These clouds serve as sites for star formation (SF) due to their high density and low temperatures. Consequently, spiral arms are expected to exhibit a higher star formation efficiency (SFE) than less dense galaxy regions~\citep{Lord1990,SilvaVilla2012,Yu2021}. 
Indeed, a greater number of young stars are found in the spiral arms, suggesting a higher star formation rate (SFR) in these areas~\citep{Bigiel2008,Schinnerer2013}. 
However, the rise in the SFR in spiral arms may simply result from higher surface densities, with the spiral's gravitational potential restructuring and concentrating the gas rather than influencing SF directly.
\citet{Elmegreen1986} and~\citet{Querejeta2024} found no increase in the SFR in galaxies with strong spiral patterns. 
If this is true in general, the SFE should remain constant regardless of the galaxy, as various studies have noted~\citep{Moore2012,Ragan2016,Urquhart2021,Querejeta2021}.

Stellar feedback ---in the form of stellar winds, supernova explosions, and radiation pressure from massive stars--- also plays a significant role in the formation, evolution, and lifetimes of MCs. These processes inject energy and momentum into the ISM, creating turbulence and disrupting the equilibrium of the gas. The compression and turbulence induced by stellar feedback, as well as the shear induced by galactic dynamics~\citep{Chevance2020}, can trigger the collapse of MCs, initiating the formation of new stars. However, stellar feedback can also lead to the destruction of MCs~\citep[e.g.,][]{Bonne2023}. \citet{Chevance2022} suggested that the main causes of cloud destruction in galaxies are early stellar feedback mechanisms, which take place prior to supernova explosions. 
It is still a matter of debate as to whether SF is more influenced by the environment ---for example, central regions versus spiral arms--- or by stellar feedback mechanisms~\citep{Corbelli2017,ReyRaposo2017,Kruijssen2019,Chevance2022,Liu2022,Choi2023}.
Additionally, other factors such as magnetic fields and turbulence within the ISM can influence the formation and evolution of MCs; magnetic fields provide support against gravitational collapse and can regulate the dynamics of the gas, while turbulence contributes to the fragmentation and structure of MCs. 

Linking the physical properties of giant molecular clouds (GMCs) in different large-scale galactic environments, such as spiral arms versus central regions, allows the systematic exploration of how the morphology of a galaxy impacts initial SF conditions. Observations on ``cloud scales''~\citep{Schinnerer2024} of $50-100\,$pc  match the sizes of GMCs or giant molecular associations (GMAs), which are up to a few hundred parsecs in size~\citep{Nguyen2016} (\citetalias{Nguyen2016} hereafter). The present study focuses mainly on what these latter authors refer to as molecular cloud associations (MCCs) and ``mini-starburst'' GMCs with an elevated SFR.

Molecular clouds consist mostly of molecular hydrogen $\htwo$; 
however, it is difficult to detect cool $\htwo$ directly due to its symmetry and small moment of inertia. 
One approach to determining the H$_2$ distribution in a galaxy is to employ observations of dust in the far-infrared (FIR) ---for example using the {\sl Herschel} satellite--- and to derive a total hydrogen column density map from a spectral energy distribution (SED) fit to the fluxes. The H$_2$ distribution is then obtained by subtracting the \HI\ component. 
H$_2$ maps can also be obtained using carbon monoxide (CO), the second-most abundant molecule, and applying the $\mathCO$-to-H$_2$ conversion factor, \Xco~\citep{BolattoWolfire2013}.
The low-$J$ rotational transitions of CO are established as a good tracer of the cold regions of MCs because these lines have low excitation temperatures (up to a few $10\,\mathrm{K}$) and low critical densities 
(a few $100\,\mathrm{cm^{-3}}$ up to a few $10^3\,\mathrm{cm^{-3}}$) for collisional excitation. 

\begin{figure*}[htbp]
  \centering
  \includegraphics[width=0.45\linewidth]{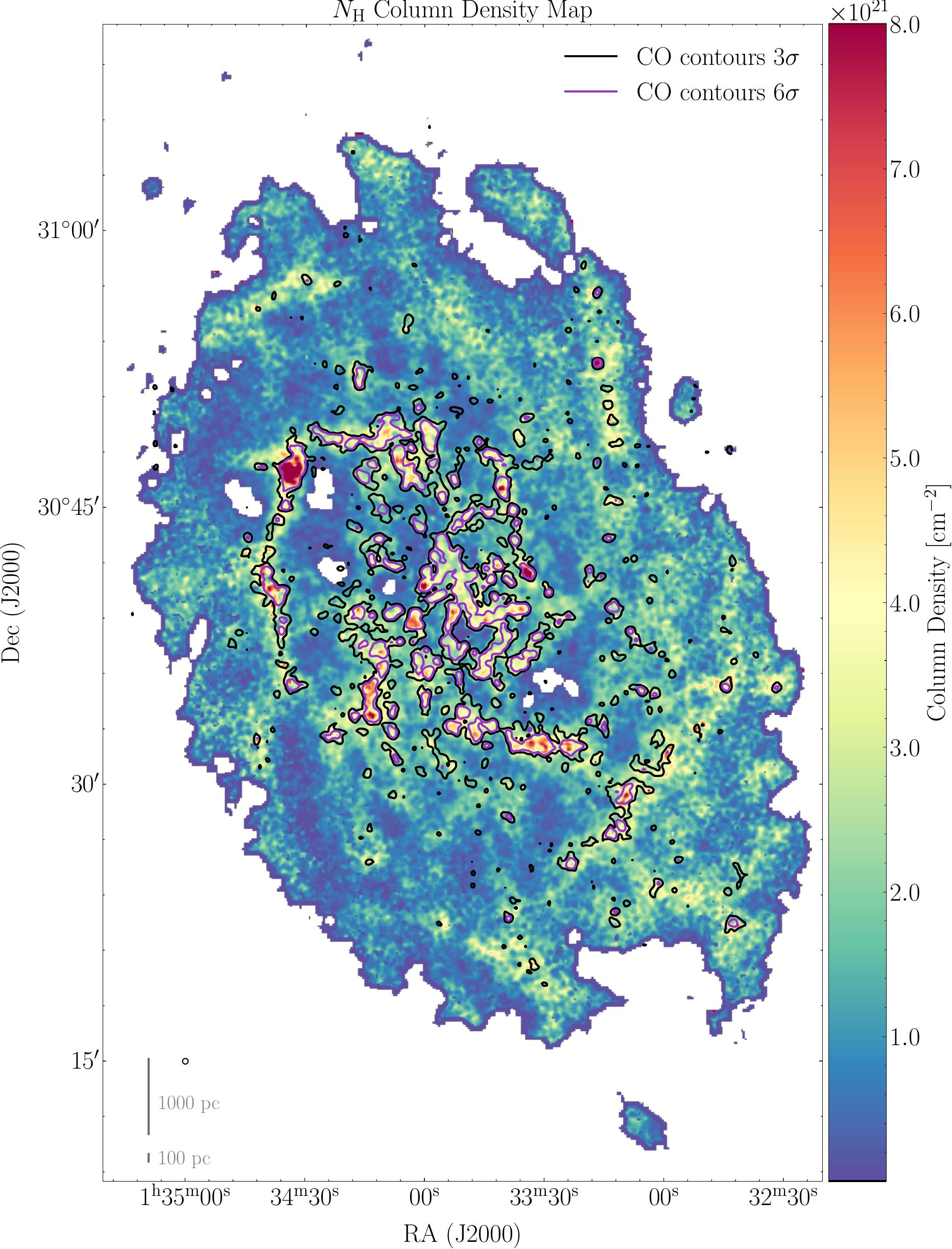}
  \includegraphics[width=0.45\linewidth]{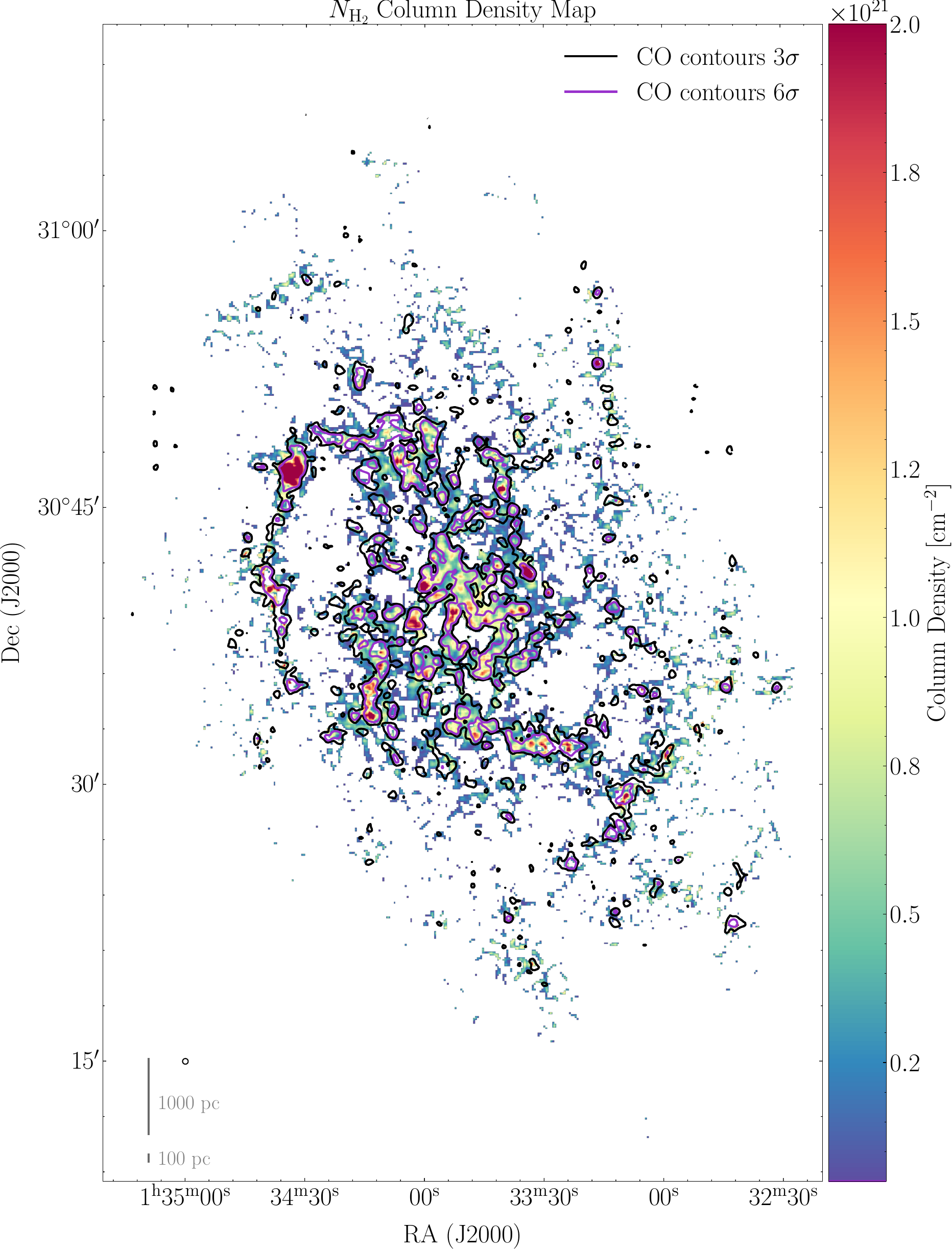} 
  \caption
      {Total and molecular hydrogen column density maps.
      Left: High-resolution $N(\mathrm{H})$ total gas column density map obtained from the {\sl Herschel} flux maps of M33 with $18.2''$ angular resolution (indicated by the circle in the lower left corner) using the $\beta$ map from~\citet{Tabatabaei2014}. Values below and above a minimum and maximum threshold ($10^{18}\, \mathrm{cm}^{-2}$ and $10^{22}\, \mathrm{cm}^{-2}$, respectively) are blanked. 
      Right: High-resolution H$_2$ column density map derived from the total $N(\mathrm{H})$ map by subtracting the \HI\ component. In both maps, the CO contour levels $3$ and $6\,\sigma$ of the CO map (Fig.~\ref{fig:co_intint_map} in Appendix~\ref{app_a:COmap}) are shown.}
    \label{fig:colden_co_map}
\end{figure*}
%

However, metallicity significantly affects MCs. Lower metallicity leads to less dust and therefore less dust shielding and self-shielding of ultraviolet (UV) radiation. Thus, the far-UV field can penetrate deeper into MCs, photo-dissociating $\mathCO$, and leaving a larger envelope of emitting ionized carbon (C$^+$) around a smaller CO-rich core~\citep{Poglitsch1995,Stark1997,Wilson1997,BolattoWolfire2013}. This is intensified by reduced CO self-shielding due to lower $\mathCO$ abundance. H$_2$ also photo-dissociates via absorption of Lyman-Werner band photons but achieves sufficient column densities to become self-shielded at moderate extinctions ($A_{\rm V}$) within C$^+$-emitting gas and thus can become a significant mass reservoir. Hence, there exists a substantial molecular hydrogen reservoir outside the $\mathCO$-emitting area that is referred to as $\mathCO$-dark $\htwo$ gas~\citep{Roellig2006,Wolfire2010}. This must be taken into account when comparing the mass estimates of dust and CO, as done in this study.

M33, classified as a flocculent Sc-type spiral galaxy, lies at a distance of $847\,$kpc~\citep{Karachentsev2004}, has an inclination of $\sim\,$$56^{\circ}$ ~\citep{ReganVogel1994}, and contains numerous massive SF regions. Our $18.2''$ angular resolution corresponds to $\sim\,$$75\,\mathrm{pc}$, which is the size of GMCs or GMAs in the Milky Way~\citepalias{Nguyen2016}, allowing us to resolve individual GMCs on a 2D image. 
Recent interferometric observations~\citep{Peltonen2023,Muraoka2023} have even higher resolution, resolving $<\,$$50\,$pc scales, but are not discussed here because we focus on large GMCs. The metallicity of M33 was measured using neon and oxygen abundances in \HII\ regions~\citep{Willner2002,Crockett2006,Rosolowsky2008a,Magrini2010} and varies widely, ranging from values comparable to the Milky Way to lower ones (see discussion and references in~\citealt{Magrini2010}). However, the average metallicity is approximately half solar, which is frequently cited in the literature~\citep{Gratier2012,Druard2014,Corbelli2019,Kramer2020}, and the total mass (gas and stars) of M33 is $\sim\,$$10^{11} \mathrm{M_{\odot}}$, roughly $10\%$ of the Milky Way mass~\citep{Marel2012,Patel2018}.

The objective of the present paper is to analyze and compare GMC properties based on dust- and CO-derived $\Nhtwo$ maps, considering the galactocentric radius and environment within M33. We also aim to establish a GMC mass spectrum and compare our findings with Milky Way studies from~\citetalias{Nguyen2016},~\citet{Rice2016} and PHANGS~\citep{Leroy2021}, a survey studying galaxy formation and evolution.

The paper is organized as follows. Section~\ref{sec:data} summarizes the data and methods for producing H$_2$ maps at 18.2$''$ resolution presented by~\citet{Keilmann2024} (\citetalias{Keilmann2024} hereafter). Section~\ref{subsec:sources} introduces the Dendrograms algorithm~\citep{Rosolowsky2008b} and presents the extraction results. The equations for determining MC properties (mass, size, density, pressure, etc.) are given in Sect.~\ref{subsubsec:prop}. These results are presented in Sect.~\ref{sec:analysis}, where we also compare with Milky Way studies. Section~\ref{sec:comparison_Rgal} discusses cloud mass distributions and properties with respect to the galactic radius and environment. Section~\ref{sec:summary} summarizes the paper.

\section{Data and methods} \label{sec:data}

In~\citetalias{Keilmann2024}, we presented two techniques using FIR {\sl Herschel} data to produce a total hydrogen column density map of M33. The first procedure (Method I) consists of a multiwavelength SED fit and is briefly summarized in the following. A more detailed description of this method can be found in~\citetalias{Keilmann2024}.
The second approach (Method II) mainly served as a consistency check because it uses only the SPIRE $250\,\mum$ data to obtain the $\Nhtwo$ map and does not account for the variable emissivity index $\beta$ of the dust. 
Therefore, we only use the $\Nhtwo$ map derived from Method I for the current study. 

We make use of \HI\ data acquired by~\citet{Koch2018}. The primary benefit of these data lies in the short-spacing corrections, which are absent in the \HI\ data from~\citet{Gratier2010} and which were utilized to generate the $\kappa_0$ map and final column density maps in~\citetalias{Keilmann2024}. The incorporation of the \HI\ data from~\citet{Koch2018} for this present study has not resulted in any significant changes to the generated maps, especially in the molecular phase of the maps. In these areas, both \hi\ maps are equal to within $\sim\,$$10\%$. The deviation increases in the diffuse region of M33 beyond the molecular regions, where GMCs are not detected anyway. All findings and conclusions of the initial~\citetalias{Keilmann2024} remain unchanged. The updated data products are available at the CDS.

\subsection{{\sl Herschel} dust data} \label{subsec:herschel-data}

For Method I, we use the level 2.5 archive data at 160, 250, 350 and 500$\,\mum$ from the Herschel Key Project HerM33es\footnote{http://archives.esac.esa.int/hsa/whsa/ (PACS observation ID: 1638302627, SPIRE observation ID: 1638304642). The SPIRE data reduction was optimized for extended emission.}~\citep{Kramer2010}.  
All maps are in units of MJy$\,$sr$^{-1}$ and reprojected to a grid of $6''$.

Method I involved several steps. First, we performed spatial decomposition using a modified Planck function to fit the dust temperature and surface density, $\Sigma$, to the SED derived from flux densities within the $160$ to $500\,\mum$ range. We then applied Eq.~17 from~\citetalias{Keilmann2024},
\begin{equation}
    \label{eq:highresMethodfinal}
    \Sigma_\mathrm{high} = {{\Sigma}}_\mathrm{500} + \left({{\Sigma}}_\mathrm{350} - {{\Sigma}}_\mathrm{350}\ast G_\mathrm{500\_350}\right) + \left({{\Sigma}}_\mathrm{250} - {{\Sigma}}_\mathrm{250}\ast G_\mathrm{350\_250}\right)~,
\end{equation}
to generate a high-resolution map of gas surface density $\Sigma_\mathrm{high}$ at a resolution of $18.2''$. This equation combines surface density distributions at $250\,\mum$, $350\,\mum$ and $500\,\mum$, convolved with a Gaussian kernel to the respective resolutions\footnote{The angular resolutions are $18.2''$, $24.9''$ and $36.3''$ for 250$\,\mum$, 350$\,\mum$ and 500$\,\mum$, respectively.} in Eq.~\ref{eq:highresMethodfinal}. 
$G_{\lambda_c\_\lambda_o}$ are the Gaussian kernels of width 
$\sqrt{{\theta_c}^2 - {\theta_o}^2}$ 
for the convolution, commonly denoted as $\ast$. The beam at the required resolution is specified by $\theta_c$ and the beam at the original resolution by $\theta_o$, while the index $\lambda_c\_\lambda_o$ denotes the corresponding wavelengths. SED fitting was conducted for each map using a pixel-by-pixel modified blackbody function with the specific intensity given by Eq.~10 in~\citetalias{Keilmann2024} with 
 \begin{equation}
    I_\nu = {\kappa_{0,\mathrm{DGR}}}(\lambda/250\,\mu\mathrm{m})^{-\beta} \, \mu_m\, m_\mathrm{H}\, N_\mathrm{H} \, B_\nu(T_d)~,
    \label{eq:specificIntensitySurface}
\end{equation}
assuming optically thin emission. The mean molecular weight is indicated as $\mu_m$, $m_\mathrm{H}$ is the mass of the hydrogen atom, $N_\mathrm{H}$ the total hydrogen column density and $B_\nu(T_d)$ the Planck function. A dust opacity law similar to~\citet{Kruegel1994} has been used with 
\begin{equation}
\kappa_g(\nu) = {\kappa_{0,\mathrm{DGR}}} \,\times\, (\lambda/250\,\mu\mathrm{m})^{-\beta}~.
\label{eq:kappa}
\end{equation}
The dust-to-gas ratio (DGR) is inherently included in our definition of the dust absorption coefficient in units of ($\mathrm{cm^2/g}$), which we denote hereafter simply as $\kappa_{0}$, and $\beta$ is the emissivity index determined by~\citet{Tabatabaei2014}. We use this $\beta$ map alongside the dust temperature of the cold component that are given 
in Fig.~8 and~9, respectively, in~\citet{Tabatabaei2014}. We derived the dust absorption coefficient $\kappa_{0}$ pixel-by-pixel using Eq.~15 in~\citetalias{Keilmann2024}. Further details on the determination of $\kappa_{0}$ and $\beta$, including interpolation techniques, are described in Section~3.2 of~\citetalias{Keilmann2024}. Our approach avoids assumptions regarding the \Xco\ factor or DGR, allowing for a more accurate evaluation of these factors.

Following our application of this technique to M33, we obtained a total column density map of the galaxy with a spatial resolution of $\sim\,$$75\,$pc or an angular resolution of $18.2''$ (Fig.~\ref{fig:colden_co_map}, left). From this map, we derived a molecular gas column density map (Fig.~\ref{fig:colden_co_map}, right) by subtracting \HI\ data from the VLA~\citep{Koch2018}. The \HI\ data at $12''$ angular resolution have been smoothed to $18.2''$ and then transformed to column density using Eq.~1 in~\citetalias{Keilmann2024}, based on~\citet{Rohlfs1996}.
The total H$_2$ gas mass (Fig~\ref{fig:colden_co_map}, right) is $1.6\times10^8\,\mathrm{M_\odot}$.

\subsection{IRAM $30$m telescope $^{12}\cotwoone$ data} \label{subsec:co_data}

M33 was observed using the HERA multibeam dual-polarization receiver in the On-The-Fly mapping mode, targeting the $^{12}\cotwoone$ line. The observations were conducted as part of the IRAM 30m Large Program titled ``The complete $\mathCO\mathrm{(2-1)}$ map of M33''~\citep{Gratier2010,Druard2014}. The $\mathCO$ line-integrated map has been acquired from the \href{https://www.iram.fr/ILPA/LP006/}{IRAM repository}. 
The archive data have been converted from antenna temperature scale to main beam brightness temperature using a forward efficiency of $F_\mathrm{eff} = 0.92$ and a beam efficiency of $B_\mathrm{eff} = 0.56$~\citep{Druard2014}.  We calculated an rms noise level of $0.35\,\kkms$, equivalent to $3\sigma=\cothreesigma\,\kkms$, for the smoothed map with a resolution of $18.2''$ (Fig.~\ref{fig:co_intint_map} in Appendix~\ref{app_a:COmap}). 
We only utilize the IRAM 30m line-integrated intensity map of $\mathCO$, since we do not have velocity information from the dust-derived data. Moreover, due to the inclination of
$56^\circ$, M33 appears mostly face-on, resulting in small line of sight effects. Hence, employing only the line-integrated $\mathCO$ data is justified and serves as a meaningful comparison with the dust-derived data.

\subsection{The \Xco\ conversion factor map}

Using dust-derived $\Nhtwo$ data and IRAM $\mathCO$ data, we computed the \Xco\ conversion factor map by dividing the $\Nhtwo$ map by the $\mathCO\mathrm{(2-1)}$ map, scaled to $\mathCO\mathrm{(1-0)}$ with a line ratio of $\mathrm{CO(2-1)/CO(1-0)} = 0.8$~\citep{Druard2014}. This generates a pixel-by-pixel \Xco\ map, avoiding a uniform value across the galaxy, as often used in the literature~\citep{Gratier2012,Druard2014,Clark2023,Muraoka2023}. The \Xco\ map, based on the dust-derived $\Nhtwo$ map, is thus influenced by assumptions in the dust-derived results. Fig.~\ref{fig:ratioMap} shows the \Xco\ map outlining the main spiral arms of M33. Minimum values in the outer area are about $10^{19}\,\cmKkms$, with maxima in the northern and southern spiral arms exceeding $10^{21}\,\cmKkms$.

To compare the \Xco\ values to those reported in the literature for M33, we calculated a single \Xco\ value in~\citetalias{Keilmann2024} using a simple mean and a binned histogram with a log-normal fit. We also conducted scatter plots and a radial line profile to investigate radial dependency. All methods show considerable variability in \Xco\ with no significant correlation with galactocentric radius. However, the simple mean, log-normal fit and scatter plot indicate values approximately at or below the Galactic \Xco, between $1.6\times10^{20}\,\cmKkms$ and $2.1\times10^{20}\,\cmKkms$. Despite the diversity, this spread is expected due to natural fluctuations in the $\mathrm{CO}$-to-H$_{2}$ ratio across the interstellar medium, as noted by~\citet{BolattoWolfire2013} and recent studies~\citep{Ramambason2024,Chiang2024}.

\section{Identification of coherent structures} \label{subsec:sources}

To derive physical quantities such as sizes, densities, and masses and to compare structures in H$_2$ column density maps derived from {\sl Herschel} with CO data from the IRAM 30m, we need to identify coherent structures in the 2D maps.

The ISM of a galaxy is a complex multi-phase environment, from hot, tenuous atomic gas to cold, dense molecular gas. The simplest approach defines MCs as entities with well-defined borders ~\citep{Elmegreen1985} but complex substructure consisting of clumps and filaments~\citep[e.g.,][]{Stutzki1990,Schneider2011,Pineda2023}. Extraction algorithms separate dense gas into distinct clouds/clumps, often using velocity information from spectral line observations for statistical analysis. Various methods identify point sources, clumps, clouds and filaments in the Milky Way~\citep{Stutzki1990,Williams1995,Rosolowsky2006,Rosolowsky2008b,Henshaw2019,Li2020,Sasha2021}. \citet{Li2020} concluded that GAUSSCLUMPS~\citep{Stutzki1990} and Dendrograms~\citep{Rosolowsky2008b} perform best in extracting clumps in synthetic data cubes, including noise.  
\begin{figure*}[htbp]
  \centering
\includegraphics[width=0.45\linewidth]{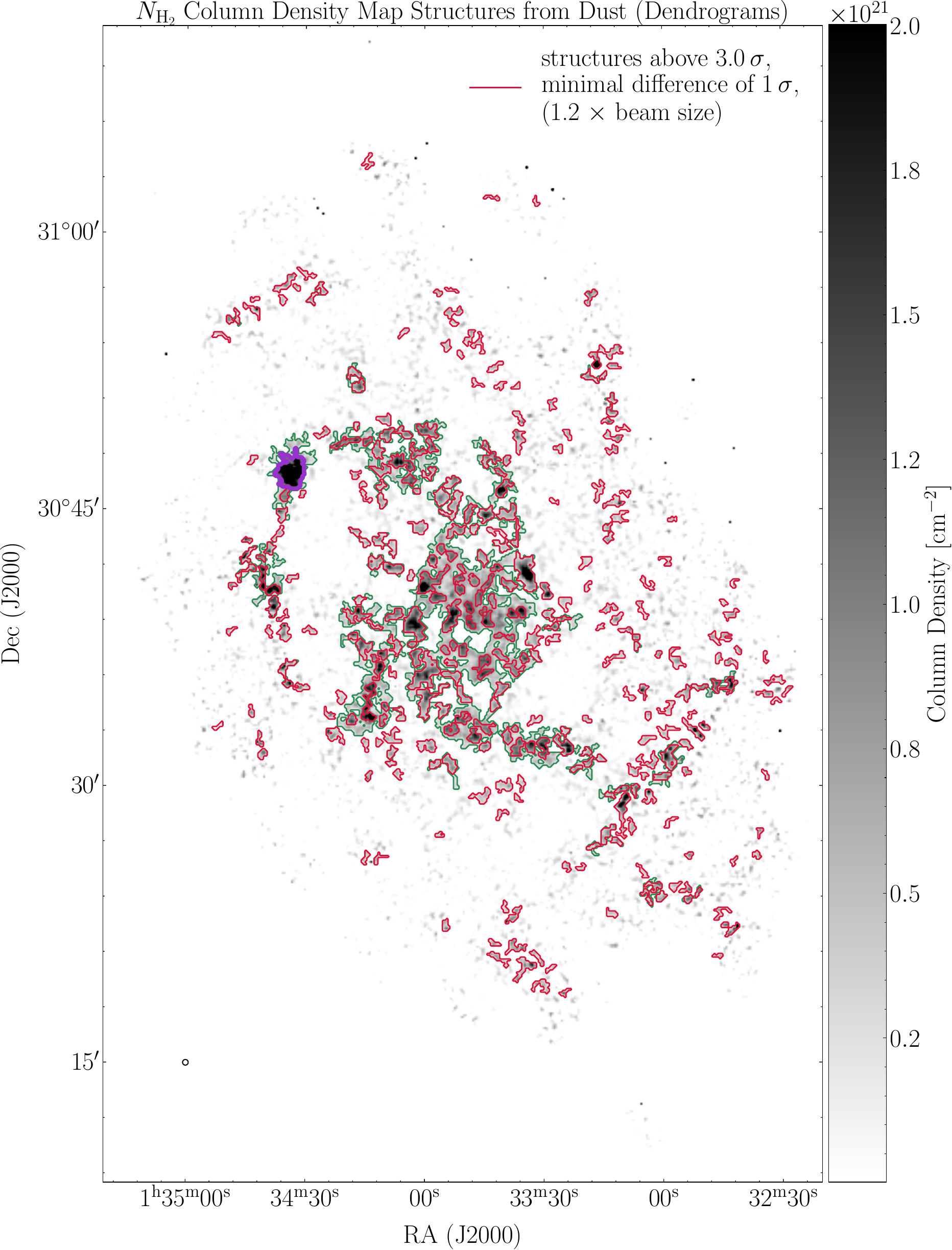}
\includegraphics[width=0.45\linewidth]{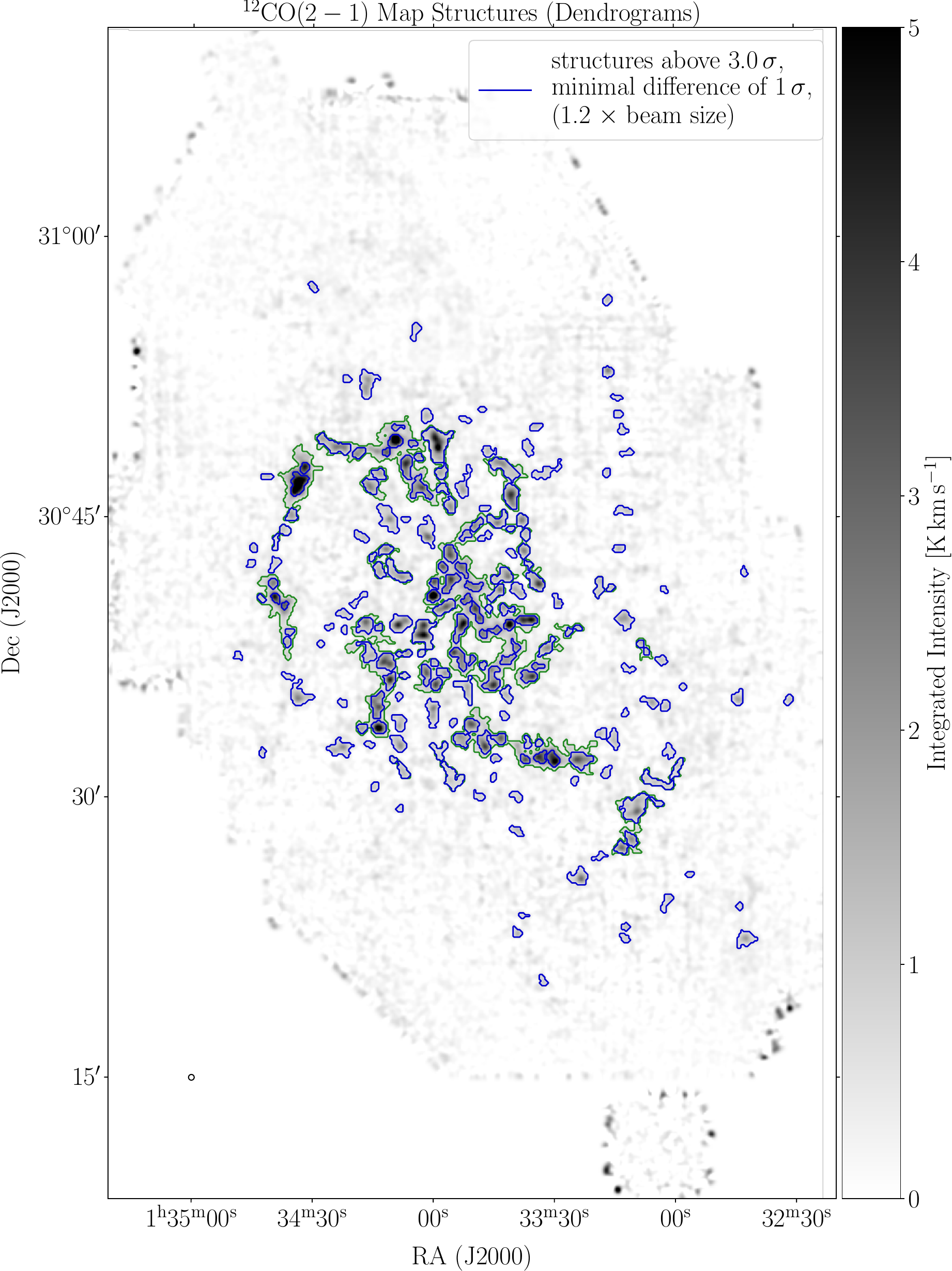}
  \caption{Dendrogram source extraction from the $\Nhtwo$ and $\mathCO$ maps.
  Left: GMC detections outlined by the lowest level branches in the dust-derived $\Nhtwo$ map (green contours) and \numstructdendrodust leaf structures (red contours). NGC604 is marked with a thick pink contour.
  Right: GMC detections identifying \numstructdendroCO leaf structures in the line-integrated $\mathrm{CO(2-1)}$ map (blue contours) and similar to the dust map, the branches (green contours). The small circle in the lower left corner of both figures shows the beam size of $18.2''$. }
  \label{fig:colden_ico_structures_dendro}
\end{figure*}
\begin{figure}[htbp]
 \centering
 \includegraphics[width=0.95\linewidth]{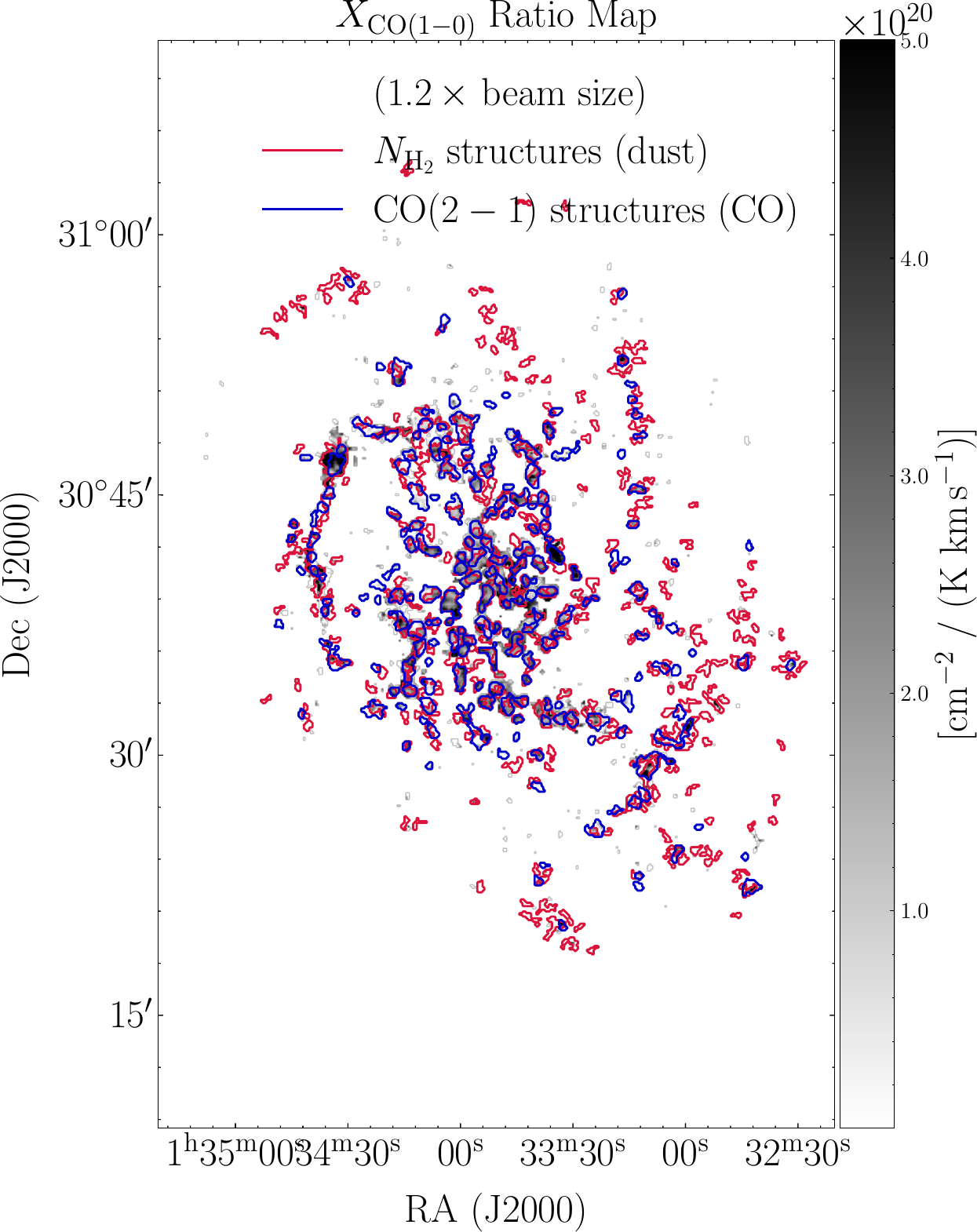}
 \caption{Structures identified using Dendrograms on CO and dust-derived maps superimposed on the \Xco\ map.
 The structures found in the dust-derived $\Nhtwo$ (red) and in the $\mathCO$ map (blue) are mapped onto the \Xco\ factor (ratio) map, which represents the dust-derived H$_2$ column density over CO intensity from Paper I. 
 }
 \label{fig:colden_Xco_coContours}
\end{figure}

Molecular clouds were extracted from the CO spectral data cube of the IRAM 30m telescope at original angular resolution by~\citet{Gratier2012} and~\citet{Corbelli2017} using the CPROPS algorithm~\citep{Rosolowsky2006}.
CPROPS assumes contiguous, bordered emission of clouds by an isosurface in brightness temperature above a threshold, applying moment measurements to derive size, line width and flux from the position-position-velocity data cube. In our study, we employed Dendrograms for extracting GMCs.

\subsection{The Dendrograms method and initial values for M33} \label{subsubsec:dendro}
The Dendrogram algorithm works intrinsically in two and three dimensions. The algorithm searches for the highest value in the map and systematically collects all other data points with lower values as long as three conditions are fulfilled. The first condition is the minimum difference (\command{min\_delta}) in intensity between two identified peaks, which must be satisfied to consider those as two different structures. This retaining level of two structures is set to $1\sigma$ of the rms noise of the maps. We also have tested different values of \command{min\_delta}, ranging from $1\sigma$ to $5\sigma$. 
The second parameter is the minimum level (\command{min\_value}) that a structure must have in order to be considered as a coherent clump/cloud. 
We explored levels from $3\sigma$ to $5\sigma$ and ultimately chose to set the threshold above the $3\sigma$ level of the rms noise, which is $\sim\,$$6.5 \times 10^{19}\,\mathrm{cm^{-2}}$ for the dust-derived $\Nhtwo$ and $\sim\,$$0.35\,\kkms$ for the line-integrated $\mathCO$ map~\citep{Gratier2010,Druard2014,Keilmann2024}. 
Our investigations did not reveal notable variations in the final identification of structures, indicating that the Dendrogram results are not substantially influenced by the selection of these two input parameters when applied to our data.
The last parameter defines the minimum number of pixels to be considered as a structure (\command{min\_npix}), which is related to the width of the structure when we assume circular geometry. This parameter is given by the actual number of pixels, which fit into the full width at half maximum (FWHM) beam width multiplied by $1.2$ and corresponds to $\sim\,$$11$ pixels to ensure that the MCs are well resolved. We also experimented with factors of $1$, $1.5$ and $3$ (see below). Obviously, a factor of $1$ tends to find more smaller structures around the beam size and a factor of $3$ ``blurs'' clouds into fewer but larger structures. We choose a factor of 1.2 as the best compromise to obtain reliable cloud statistics (see also~\citealt{Schneider2004,Kramer1998}). 
The extraction of sources has then been applied to the dust-derived $\Nhtwo$ map and to the line-integrated $\mathCO\mathrm{(2-1)}$ map, for which the final properties of the detected GMCs have been derived using the \Xco\ factor map from~\citetalias{Keilmann2024}.
We investigate the influence of varying Dendrogram parameters in Appendix~\ref{app:dendrogram_parameters}.

Dendrograms distinguish between so called leaves and branches. Branches contain substructures like other branches or leaves, while leaves only consist of themselves. As the distance to M33 is $847\,\mathrm{kpc}$~\citep{Karachentsev2004} and the angular resolution is $18.2''$, the minimum resolved convolved structure that we can identify has a size of $75\,\mathrm{pc}$. This corresponds roughly to the size of GMCs (and small GMAs) in the Milky Way~\citep{Roman_Duval2010,Hughes2013,Nguyen2016,Spilker2022}. We mostly concentrate on leaves in our analysis since they best represent a single GMC/GMA. However, leaves do not capture all the emission. Branches, which contain leaves and comprise larger areas, represent the more diffuse, extended $\htwo$ emission that we define as ``inter-cloud'' medium. An example is the crowded center of M33, where separating individual clouds can become challenging. Branches also include emission just around individual, well-separated GMCs. We refer to this surrounding structure as the ``envelope''. We note that this does not constitute an \HI\ envelope. An example of this is the GMC NGC604 in the northeast spiral arm, where the smaller leave structure is embedded in the larger branch.  
Unless stated otherwise, branches refer to dust, as CO branches show a relationship similar to dust branches as leaves do.

\subsection{Dendrogram extraction of GMCs in the CO and dust-derived $\Nhtwo$ maps} \label{subsect:extraction}

In the following sections, we discuss the morphology of detected GMCs, NGC604 and the differing numbers of detected GMCs in dust and CO.
We focus on leaf structures only because they represent mostly the GMC/GMA population.
Tables~\ref{table:clouds_properties_dust_dendro} and~\ref{table:clouds_properties_CO_dendro} list the main cloud parameters for the first 10 clouds ordered by their surface mass density. See Sect.~\ref{subsubsec:prop} for details on the calculation of the quantities listed in the tables. 
The tables show no one-to-one correlation between dust- and CO-derived GMCs. For example, NGC604 is one single GMC in the dust-derived map, but in CO it splits into two smaller structures. This discrepancy arises because structures identified in both tracers differ in size and mass.

We focus on the distribution statistics in Sect.~\ref{sec:analysis}. A summary of the mean values is given in Table~\ref{table:clouds_mean_properties}. The similarity in this table, especially in masses, is because the structures are well identified as leaves in both tracers (see Sect.~\ref{subsect:extraction}). The CO-derived $\htwo$ column density is lower, evident in the central branch masses, where the CO-derived mass is $\sim\,$$70\%$ of the dust-derived mass. This similarity extends to many parameters, such as radius and densities. Their average values are comparable, varying by less than a factor of $2$.

\subsubsection{Morphological description} \label{subsubsect:morphology}

Figure~\ref{fig:colden_ico_structures_dendro} displays the \numstructdendrodust leaf (red) and 142 branch (green) structures identified in the dust-derived $\Nhtwo$ map (left panel) and the \numstructdendroCO leaf (blue) and 94 branch (green) structures in the $\mathCO\mathrm{(2-1)}$ map (right panel). 
To provide a clearer overview, we only show the lowest level of branch extraction, which may include other branches (and leaves). 
The overlay of CO and dust-derived structures on the \Xco\ map (Fig.~\ref{fig:colden_Xco_coContours}) 
shows a similar morphology for both tracers. However, the dust emission identifies more structures beyond the spiral arms and central region. 
\numstructdendroCO structures in the CO map are more locally concentrated, with fewer structures between spiral arms or in M33's outer regions compared to the dust-derived map.

Furthermore, especially the central region in the dust-derived map shows substantial $\htwo$ column density between the identified leaves. 
The emission distribution contained in branches focuses in the crowded center region of M33, where many GMCs/GMAs potentially overlap along the line of sight, leading to a rather homogeneous plateau of emission. This line of sight effect can be one reason why it is not possible to separate dust emission into smaller leaf structures. 
However,~\citet{Koch2019} employed a Gaussian decomposition on the full spectral CO line cube of the IRAM 30m data and found that only $\sim\,$10\% of the CO spectra show multiple components. This finding does not conclusively rule out the possibility that there are several clouds along the line of sight, but overall this effect is probably less important than for other galaxies. 
It is also plausible that the crowded emission in the center seen in dust constitutes an inter-cloud $\htwo$ medium, similar to what is found in the Milky Way. We note that the flocculent morphology of M33 already points toward an important gas reservoir between the GMCs.
However, another explanation is that the inter-cloud gas is warmer and tends to decrease the CO brightness of the low-J lines, which requires future observations of $\mathCO(3-2)$ or $\mathCO(4-3)$ line emission. In any case, this more widespread gas reservoir in the center contains a significant mass. While dust-derived leaves collectively hold $8.3\times10^7\,\mathrm{M_\odot}$ in total, branches excluding leaves contain $3.1\times10^7\,\mathrm{M_\odot}$ of the H$_2$ gas mass. Especially in the central region of M33, the leaves contain $2.6\times10^7\,\mathrm{M_\odot}$, while the branches in the center comprise $1.5\times10^7\,\mathrm{M_\odot}$.

The CO emission map (Fig.~\ref{fig:colden_ico_structures_dendro}, right) shows less homogeneous material in branches than the dust-derived map and reveals that the CO leaves are typically surrounded by a more extended envelope. This finding supports the one of~\cite{Rosolowsky2008a}, who propose that around 90\% of the diffuse emission to within $100\,$pc of a GMC arises from a population of small, unresolved MCs. However, the CO sensitivity might be too low to detect CO-dark gas or CO might be easily dissociated in the center. Additionally, the H$_2$ emission from dust can be overestimated due to the complex map production process and the subtraction of a VLA \HI\ map, which has its own detection limits. The dust map might still contain \HI, as CO-faint column densities are low ($0.5$ to $1\times10^{21}\,\mathrm{cm^{-2}}$), close to the atomic-to-molecular hydrogen transition level. The CO-identified leaf structures have a total mass of $4.2\times10^7\,\mathrm{M_\odot}$, with branches holding $2.1\times10^7\,\mathrm{M_\odot}$ (50\% of the mass compared to leaves; 37\% in the dust-derived map). In the center, the leaves contain $1.7\times10^7\,\mathrm{M_\odot}$ and branches $1.1\times10^7\,\mathrm{M_\odot}$ (64\% of leaves' mass; 57\% in dust-derived map). 

\subsubsection{The GMA NGC604} \label{subsubsect:ngc604}

The SF region NGC604 stands out with the highest mass and largest area, forming a single structure on the dust-derived $\Nhtwo$ map but several GMCs on the CO map (Fig.~\ref{fig:colden_ico_structures_dendro}), similar to the findings of~\citet{Williams2019}. 
The discrepancy may arise from the greater extent of the GMC in dust compared to CO, as the $3\sigma$ CO signal shows a narrower north-south ridge (see Fig.~\ref{fig:colden_ico_structures_dendro}). Another explanation could be CO-dark gas in the enveloping gas, with dust emission reaching $2\times10^{20}\,$cm$^{-2}$, conducive to CO formation. This and the limited spatial resolution probably explain the divergence of NGC604 from the majority of the GMC population in this study. 

However,~\citet{Relano2013} (and references therein) reported that NGC604 is not a single \HII\ region, but comprised of a few compact knots and filamentary structures joining the different knots. The whole complex has a radius of $280\,$pc and forms the second most luminous \HII\ regions association in the Local Group, surpassed only by 30 Doradus in the LMC. 

Observations of the Atacama Large Millimeter/submillimeter Array (ALMA) in $^{12}\mathrm{CO(2-1)}$ and $^{13}\mathrm{CO(1-0)}$~\citep{Muraoka2020,Phiri2021,Peltonen2023} at an angular resolution of $0.44'' \times 0.27''$ ($1.8\,$pc $\times$ $1.1\,$pc)~\citep{Muraoka2020} and $3.2'' \times 2.4''$ ($13\,$pc $\times$ $10\,$pc)~\citep{Phiri2021} confirm that NGC604 constitutes multiple individual molecular clouds.

\subsubsection{Caveats regarding dust- and CO-derived GMCs} \label{subsubsect:numbers}

Some GMCs are identified only in the CO dataset, while others appear only in the dust-derived dataset. Dust-only detections may indicate CO-dark H$_2$ gas or may be due to smaller $\mathrm{CO(2-1)}$ envelopes given its higher critical density compared to $\mathrm{CO(1-0)}$~\citep{Schinnerer2024}. Regions seen only in CO may reflect underestimated molecular hydrogen column densities, possibly from overestimated $\kappa_0$ values derived from molecular hydrogen regions, casting doubts on the assumption of a constant $\kappa_0$ between atomic and molecular phases.

Furthermore, $\kappa_0$ may be overestimated as it requires regions with CO emission below $2\sigma$, hence assuming no CO emission. This leads to a bias due to generally low CO emission in the disk (see Eq.~16 in~\citetalias{Keilmann2024}). The IRAM CO map might show structures from noise fluctuations. Raising the detection threshold to $5\sigma$ can address this potential issue, but this approach also leads to similar detections when consistently increasing the threshold for dust-derived data. The uncertainty in H$_2$ detection and the prevalent use of the $3\sigma$ threshold for CO and dust data make it challenging to conclusively ascertain the origin of these structures.

Increased noise in the dust-derived $\Nhtwo$ map may cause structures experience a quasi ``beam-diluted'' effect and to blend into the background due to too low $\Nhtwo$ (not fulfilling $\command{min\_value}$) or failing the minimum size condition ($\command{min\_npix}$) to be identified. This aligns with the observation of low column densities in both dust-derived and $\mathCO$-derived $\Nhtwo$ maps. Figure~\ref{fig:colden_Xco_coContours} supports this, showing that structures detected only in the $\mathCO$-derived $\Nhtwo$ data have the lowest \Xco\ factor. The presence of CO-dark gas and a non-changing $\kappa_0$ in both the atomic and molecular phases may explain the greater number of GMCs in the dust-derived data.

\begin{table*}[htbp]
\caption{GMC properties derived from the Dendrogram leaves in dust.}
\label{table:clouds_properties_dust_dendro} 
\centering
\begin{tabular}{ccccccccccccc}
\hline\hline
\textbf{x} & \textbf{y} & \textbf{$\boldsymbol{A}$} & \textbf{$\boldsymbol{R}$} & \textbf{$\boldsymbol{M}$} & \textbf{$\boldsymbol{n}$} & \textbf{$\boldsymbol{\Sigma}$} & \textbf{$\boldsymbol{T_\mathrm{d}}$} & \textbf{AR}\\
{\tiny$[\mathrm{''}]$} & {\tiny$[\mathrm{''}]$} & {\tiny$[10^{4}\,\mathrm{pc^2}]$} & {\tiny$[\mathrm{pc}]$} & {\tiny$[10^{5}\,\mathrm{M_\odot}]$} & {\tiny$[\mathrm{cm^{-3}}]$} & {\tiny$[\mathrm{M_\odot\,pc^{-2}}]$} & {\tiny$[\mathrm{K}]$} & \\
\hline
$305$  &        $-308$  &       $0.23$    &     $27$  & $2.4$    &      $42$  &  $104$  &  $20.4$ &  $4.69$ \\
$-70$  &        $-17$  &        $0.23$    &     $27$  & $1.6$    &      $28$  &  $70$  &  $20.7$ &  $2.16$ \\
$554$ & $432$  &        $12.49$    &    $199$  &        $78.6$    &     $3$  &  $63$  &  $22.0$ &  $1.20$  \\
$284$  &        $-171$    &     $0.23$    &     $27$  & $1.3$    &      $23$  &  $56$  &  $21.0$ &  $1.44$ \\
$304$  &        $-357$  &       $1.69$    &     $73$  & $8.9$    &      $8$  &  $53$  &  $21.0$ &  $1.50$ \\
$-856$  &       $1040$   &      $0.23$    &     $27$  & $1.2$    &      $21$  &  $53$  &  $19.9$ &  $1.71$ \\
$614$  &        $50$ &  $1.44$    &     $68$  & $7.3$    &      $8$  &  $51$  &  $19.4$ &  $1.80$ \\
$652$  &        $156$ & $0.29$    &     $30$  & $1.5$    &      $19$  &  $51$  &  $22.0$ &  $2.21$ \\
$-121$  &       $-97$  &        $0.29$  &       $30$ &  $1.5$ & $18$   &  $50$  &  $21.7$ &  $2.30$ \\
$-194$  &       $-20$ & $0.96$    &     $55$  & $4.6$    &      $9$  &  $48$  &  $22.0$ &  $1.52$ \\
\hline
\end{tabular}
\tablefoot{The table is ordered according to surface mass density and gives the properties of the first ten clouds. 
The offsets, $x$ and $y$, are calculated regarding the center position of M33 of RA(2000)=$01^{\rm h}33^{\rm m}50.62^{\rm s}$ and Dec.(2000)=$30^\circ39'46.45''$. $R$ is the radius, $M$ the mass, $n$ the density, $\Sigma$ the surface mass density, $T_{\rm d}$ the dust temperature and AR the aspect ratio of each GMC. See Sect.~\ref{subsubsec:prop} for details on the calculation of the listed quantities. The full table is provided in electronic form at the CDS. }
\end{table*}
%
\begin{table*}[htbp]
\caption{GMC properties derived from the Dendrogram leaves in CO.}
\label{table:clouds_properties_CO_dendro}
\centering
\begin{tabular}{cccccccccccccc}
\hline\hline
\textbf{x} & \textbf{y} & \textbf{$\boldsymbol{A}$} & \textbf{$\boldsymbol{R}$} & \textbf{$\boldsymbol{X_\mathrm{CO}}$} & \textbf{$\boldsymbol{L_\mathrm{CO}}$} 
& \textbf{$\boldsymbol{M}$} & \textbf{$\boldsymbol{n}$} & \textbf{$\boldsymbol{\Sigma}$} & \textbf{$\boldsymbol{T_\mathrm{d}}$}
& \textbf{AR} \\
{\tiny$[\mathrm{''}]$} & {\tiny$[\mathrm{''}]$} &  {\tiny$[10^{4}\,\mathrm{pc^2}]$} & {\tiny$[\mathrm{pc}]$} & 
{\tiny$[\frac{10^{20}}{\mathrm{cm^2\,K\,km\,s^{-1}}}]$} 
& {\tiny$[10^{4}\,\mathrm{K\,km\,s^{-1}\,pc^2}]$}  
& {\tiny$[10^{5}\,\mathrm{M_\odot}]$} & {\tiny$[\mathrm{cm^{-3}}]$} & {\tiny$[\mathrm{M_\odot\,pc^{-2}}]$} & {\tiny$[\mathrm{K}]$} & \\
\hline
$92$   & $406$   &  $4.47$   &   $119$  & $0.2$  & $10.3$ &  $47.8$ &    $10$   & $107$  & $22.0$ &     $1.30$  \\
$-18$  & $7$     &  $0.35$   &  $33$   & $2.5$  & $4.4$  &  $2.1$  &    $20$  &      $61$  & $22.0$ &         $3.01$  \\
$95$ & $457$   &        $0.96$   &      $55$   & $0.7$  & $3.6$  &  $5.5$  &      $11$  & $57$  & $21.0$ &         $1.35$   \\
$330$  & $-122$   &     $1.81$   &      $76$   & $5.5$  & $5.4$  &  $10.1$  &      $8$  &  $56$  & $21.4$ &         $1.65$   \\
$455$ & $-39$   &       $3.51$   &      $106$  & $0.6$  & $7.2$  &  $18.0$ &       $5$   & $51$  & $19.6$ &         $1.78$   \\
$483$  & $-93$  &       $0.35$   &      $33$   & $0.5$  & $1.2$  &  $1.8$  &      $17$   &        $51$  & $19.6$ &         $1.02$   \\
$301$   & $-290$   &    $0.78$   &      $50$   & $2.0$  & $5.5$  &  $3.8$  &      $11$  & $49$  & $21.3$ &         $2.24$   \\
$112$  & $-223$  &      $0.71$   &      $48$   & $2.2$  & $5.4$  &  $3.4$  &      $11$  & $48$  & $21.7$ &         $1.34$   \\
$-17$  & $289$  &       $0.53$   &      $41$   & $1.3$  & $3.4$  &  $2.5$  &      $12$  & $46$  & $21.6$ &         $2.36$   \\
$399$  & $758$  &       $1.38$   &      $66$   & $0.8$  & $5.3$  &  $5.7$  &      $7$   & $41$  & $21.8$ &         $1.25$   \\
\hline
\end{tabular}
\tablefoot{The table is ordered after surface mass density and gives the properties of the first ten clouds. 
The offsets, $x$ and $y$, are calculated with respect to the center position of M33 of RA(2000)=$01^{\rm h}33^{\rm m}50.62^{\rm s}$ and Dec(2000)=$30^\circ39'46.45''$. See Sect.~\ref{subsubsec:prop} for details on the calculation of the listed quantities. The full table is provided in electronic form at the CDS.}
\end{table*}
%

\begin{table}[htbp]
\caption{Average cloud properties derived with Dendrograms.}
\label{table:clouds_mean_properties}
\centering
\begin{tabular}{lcccc}
\hline\hline
& \textbf{Dust-Derived} & \textbf{CO}  \\
\hline
\textbf{Clouds} & \numstructdendrodust &                                               \numstructdendroCO  \\
\textbf{$\Bar{R}$ $[\mathrm{pc}]$} & $58\pm13$ &                                   $68\pm21$  \\
\textbf{$\Bar{n}$ $[\mathrm{cm^{-3}}]$} & $5.2\pm1.5$ &                             $3.0\pm1.1$  \\
\textbf{$\Bar{M}$ $[\mathrm{M_\odot}]$} & $(2.8\pm0.9)\times10^{5}$ &            $(2.9\pm0.9)\times10^{5}$  \\
\textbf{$\Bar{\Sigma}$ $[\mathrm{M_\odot\,pc^{-2}}]$} & $22\pm5$ &            $16\pm6$  \\
\hline
\end{tabular}
\tablefoot{
$\Bar{R}$, $\Bar{n}$, $\Bar{M,}$ and $\Bar{\Sigma}$ are the average values for the area, radius, beam-averaged column density, and number density as well as mass and surface mass density, determined from the leaves extraction from the dust-derived and CO maps, respectively.}
\end{table}

\section{Determination of physical cloud properties} \label{subsubsec:prop}

For each identified structure, we compute several parameters such as the area $A$ in $\mathrm{pc^2}$, the equivalent radius $R$ in $\mathrm{pc}$, the column density $N$ in $\mathrm{cm^{-2}}$, the (beam-averaged) number density $n$ in $\mathrm{cm^{-3}}$ and the mass $M$ in $\mathrm{M_\odot}$ along with pressure $P/k_\mathrm{B}$ in $\mathrm{cm^{-3}\,K}$. The shape of the identified structures is described by the aspect ratio (AR), that is, the ratio between major and minor axes of the GMC. The following section outlines the methodologies employed to compute these quantities.

To calculate $A$, the area of each pixel of the identified GMC is summed and scaled by the squared distance, $D^2$, to M33.  
This pixel size is denoted $ A_\mathrm{pixel} = \mathrm{d}\theta_\mathrm{ra}\cdot\mathrm{d}\theta_\mathrm{dec}\cdot D^2$, where $\mathrm{d}\theta_\mathrm{ra}$ and $\mathrm{d}\theta_\mathrm{dec}$ represent the angular size of a pixel in radians.
Dendrograms provides information on the location of the identified structure, which serves as a mask for the original dataset. This allows for the calculation of the number of pixels associated with a structure.

The radius $R$ is determined as the equivalent radius of a circle with the area $A$ of the Dendrogram structure, $A = \pi\, R^2$. 
The radius of each structure is de-convolved by $R'_{i} = \sqrt{R_{\mathrm{GMC},i}^2-R_\mathrm{beam}^2}$, where $R_{\mathrm{GMC},i}$ represents the radius of the $i$-th structure and $R_\mathrm{beam}$ corresponds to the beam size.

To calculate the column density of $\htwo$ of a structure using $\mathrm{CO(2-1)}$ data, scaled to $\mathrm{CO(1-0)}$ using the line ratio of 0.8~\citep{Druard2014}, we consider all pixels from the \Xco\ factor map that belong to a detected structure in the line-integrated CO intensity map. We then multiply the corresponding \Xco\ values with the line-integrated intensities of this map on a pixel-by-pixel basis.
This approach provides a more precise estimate compared to using a constant \Xco\ factor for the entire galaxy, and allows us to uncover intriguing variations in the distribution of GMCs within M33, which is further explored and discussed in Sect.~\ref{sec:analysis}.
Additionally, to obtain an average \Xco\ value for each structure, we divided the dust-derived $\mathrm{H_2}$ column density by the corresponding CO line-integrated intensity on a pixel-by-pixel basis. These \Xco\ values for each structure are presented in Table~\ref{table:clouds_properties_CO_dendro} (as discussed in Sect.~\ref{sec:analysis}).

To determine the masses of GMCs using the molecular hydrogen column densities obtained from Dendrograms (both from CO and dust), the pixel size $A_\mathrm{pixel} = \mathrm{d}\theta_\mathrm{ra}\cdot\mathrm{d}\theta_\mathrm{dec}\cdot D^2$ is multiplied by $N(\mathrm{H_2})_{j}$. 
Here, $j$ represents the index of a pixel within an identified structure. This is finally multiplied by the hydrogen mass, $m_\mathrm{H}$, and the mean molecular weight, $\mu$,
\begin{equation}
    M_\mathrm{GMC} =  A_\mathrm{pixel} \cdot \sum_i N(\mathrm{H_2})_{j} \cdot m_\mathrm{H} \cdot \mu~,
    \label{eq:GMC_mass}
\end{equation}
with $\mu=2.8$~\citep{Kauffmann2008} to account for Helium ($\mathrm{He}$) and metals.

To calculate the average number density $n$ of a GMC consisting of $\mathrm{H_2}$, we assume a spherical configuration and use the mass and the equivalent radius obtained above. The average density $n$ is then determined as
\begin{equation}
    \frac{n(\mathrm{\htwo + He})}{\mathrm{cm^{-3}}} = 14.6 \, \frac{M}{\mathrm{M_\odot}} \cdot 
    \left( \frac{4\pi}{3}  \frac{R'^3}{\mathrm{pc^{3}}} \right)^{-1}~,
    \label{eq:number_density}
\end{equation}
where $M$ represents the mass of the cloud in solar masses and $R'$ denotes the de-convolved equivalent radius of the cloud in parsecs.\footnote{The prefactor $14.6$ is derived by multiplying the solar mass, dividing by $\mu$ and the hydrogen mass and converting $\mathrm{pc^3}$ to $\mathrm{cm^3}$. Depending on constants and rounding, the prefactor can vary;~\citet{Roman_Duval2010} determined $15.1$ using rounded values.}
Since our spatial resolution is $75\,$pc, the density can only reflect a beam-averaged density derived by dividing the (beam-averaged) column density by the (beam) size. The detected GMC will be composed of smaller substructures with much higher local densities. We note that, due to the critical density, the density of the clouds should be in the order of $10^3\,\mathrm{cm^{-3}}$ for the low-J CO transitions to be sufficiently excited.

\begin{figure*}[htbp]
  \centering
  \includegraphics[width=0.44\linewidth]{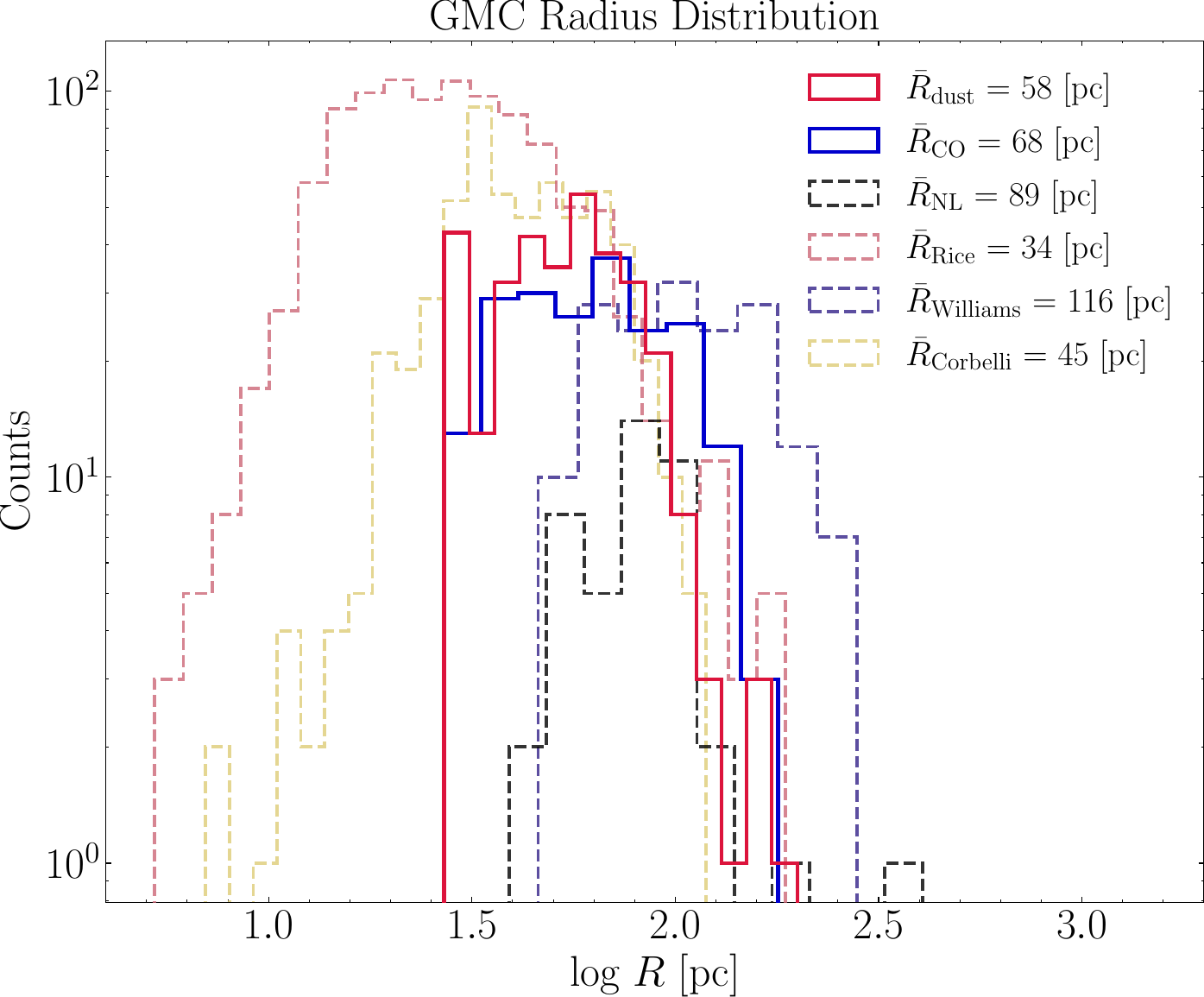}
  \includegraphics[width=0.44\linewidth]{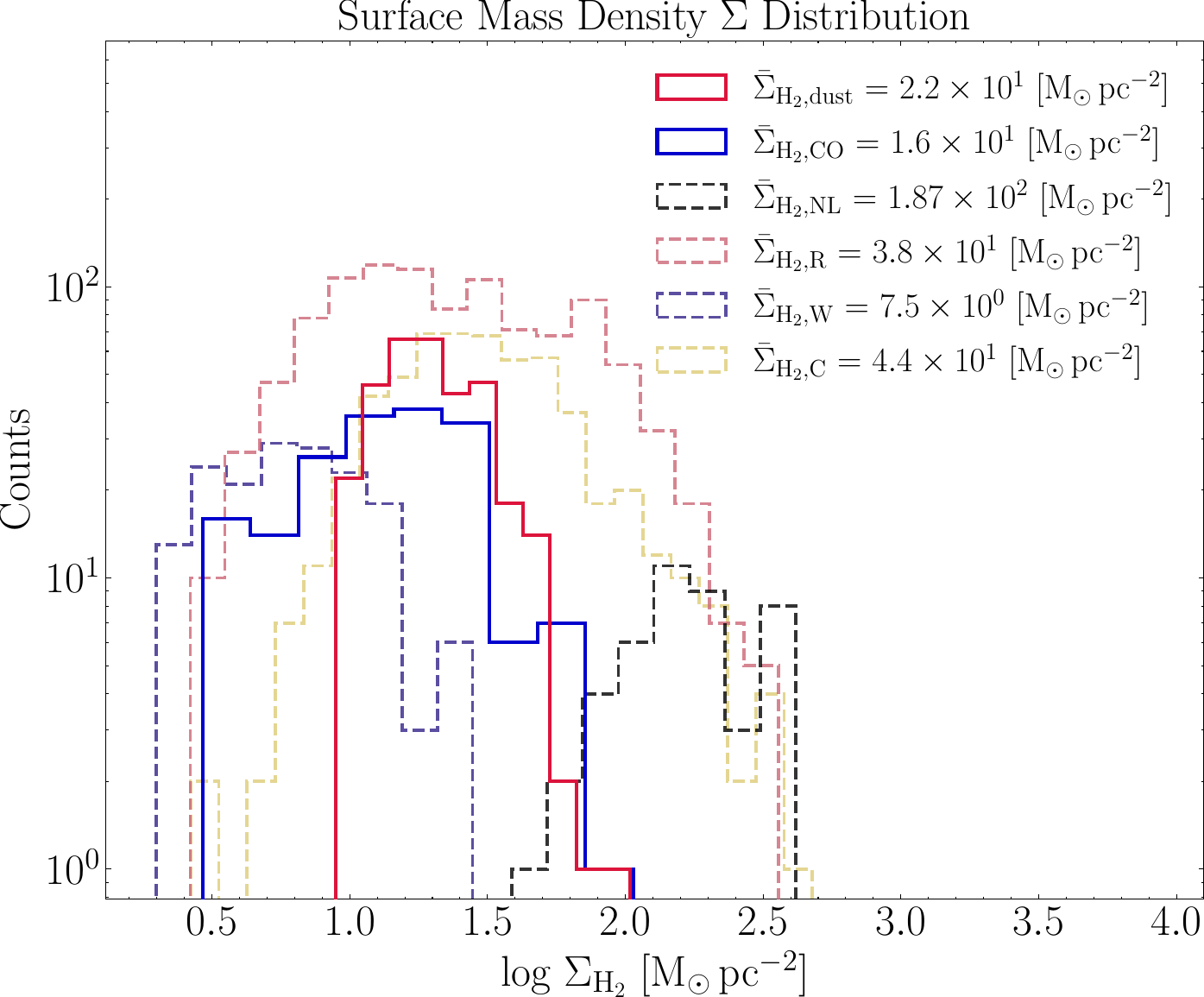}

  \vspace{0.1cm}
  \includegraphics[width=0.44\linewidth]{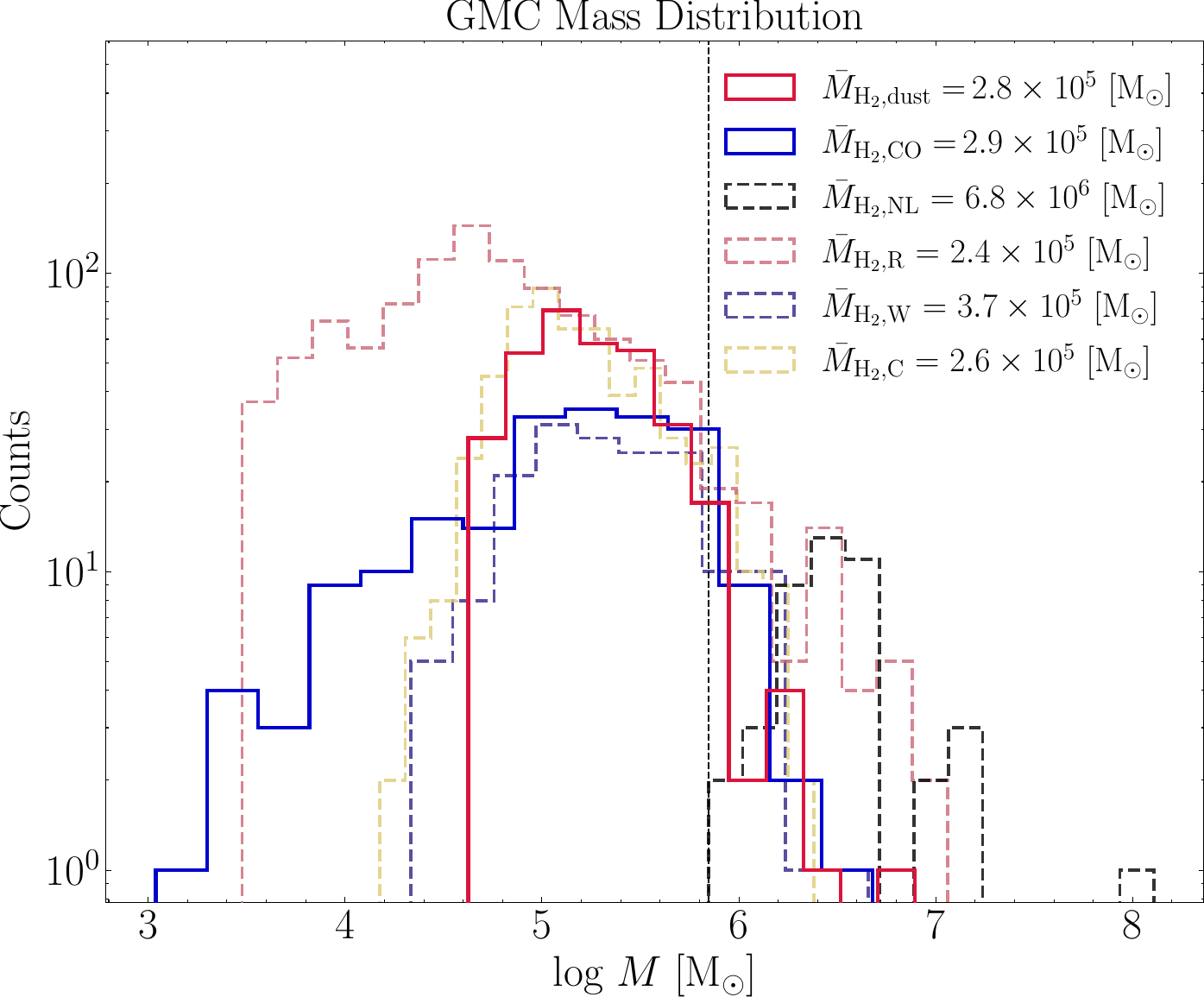}
  \includegraphics[width=0.44\linewidth]{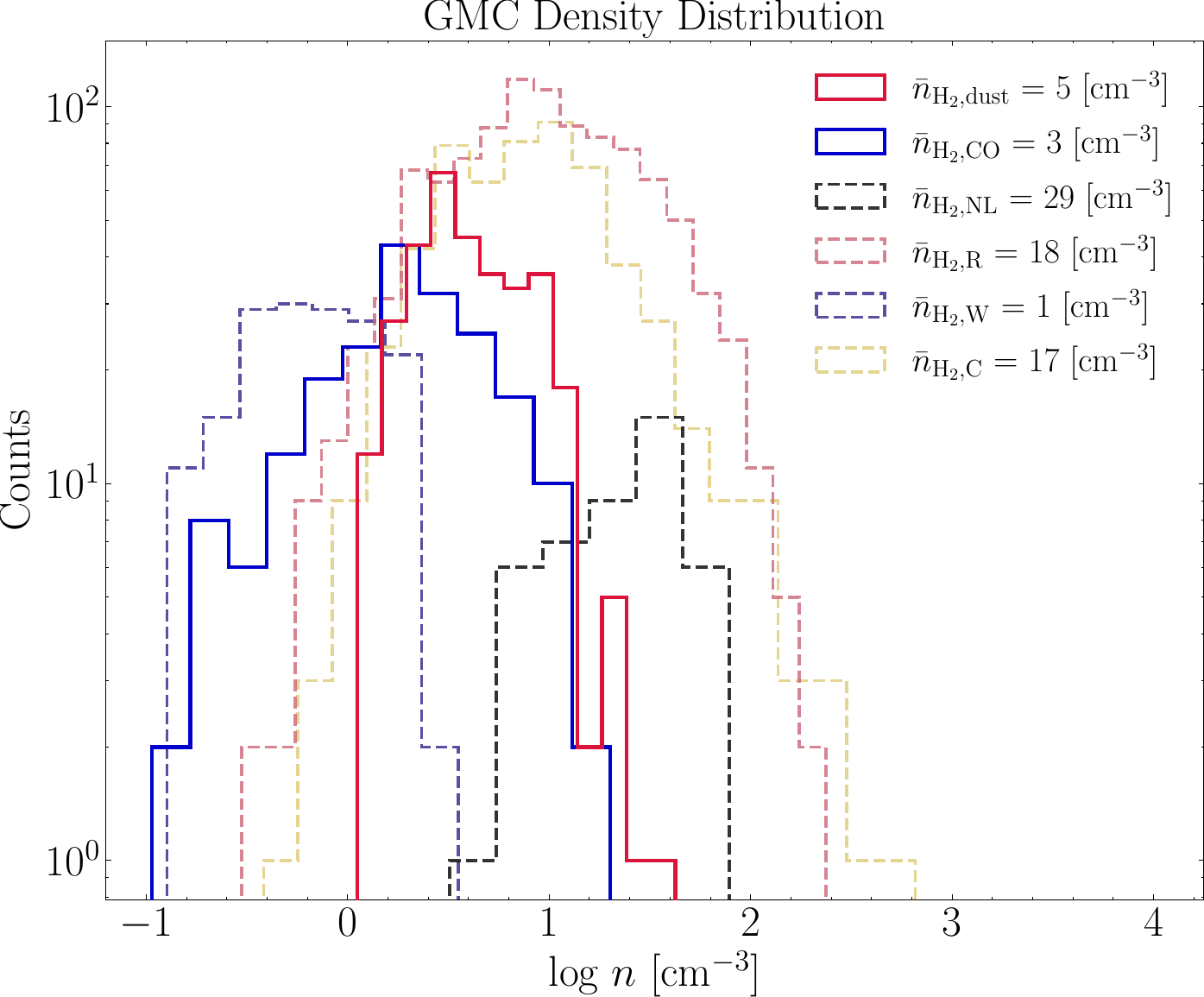}
  \caption{Distributions of GMC properties from our study and the literature.
  The panels show histograms of radius (top left), surface mass density (top right), mass (bottom left) and beam-averaged number density (bottom right) derived from H$_2$ data from dust (solid red) and CO (solid blue) from this study. The dashed purple and golden lines display the distributions obtained for M33 from dust~\citep{Williams2019} and CO~\citep{Corbelli2017}, respectively. The dashed black and light red lines give the distributions for the Milky Way studies of~\citetalias{Nguyen2016} and~\citet{Rice2016}, respectively.  
  The vertical black dashed line in the mass distribution signifies the minimum mass limit of the selected structures in the Milky Way by~\citetalias{Nguyen2016}.} 
  \label{fig:H2_mass_density_area_radius_distribution_dendro}
\end{figure*}

The surface mass density $\Sigma$ of the GMCs is determined via 
\begin{equation}
    \frac{\Sigma_\mathrm{GMC}}{\mathrm{M_\odot\,pc^{-2}}} = 
    \left( \frac{M}{\mathrm{M_\odot}} \right) \left( \frac{A}{\mathrm{pc^{2}}} \right)^{-1}~.
    \label{eq:smd}
\end{equation}

The elongation of detected GMCs is quantified by their column density-weighted aspect ratio of semi-major to semi-minor axis.


Finally, the CO luminosity, $L_\mathrm{CO}$, is calculated as 
\begin{align}
    \frac{L_\mathrm{CO}}{\mathrm{K\,km\,s^{-1}\,pc^2}} &= 
    \frac{D^2}{\mathrm{pc^2}}
    \left( \frac{\pi}{180\cdot3600} \right)^2  \sum_i I_\mathrm{int.} \, \mathrm{d}\theta_\mathrm{ra} \mathrm{d}\theta_\mathrm{dec}~,
    \label{eq:Lco}
\end{align}
where $D$ represents the distance to M33 in parsecs, $\mathrm{d}\theta_\mathrm{ra}$ and $\mathrm{d}\theta_\mathrm{dec}$ are the spatial dimensions of a pixel given in radian and the integrated intensity $I_\mathrm{int.}$ in units of $\mathrm{K\,km\,s^{-1}}$. Hence, we sum over all pixels of a GMC, multiply each pixel by the size of a pixel, scale it by the distance and convert it to units of radians. The resulting $L_\mathrm{CO}$ therefore describes the integrated emission inside a GMC summed over its entire area.

It is formally possible to calculate the gas pressure within the GMCs, $P$, using the equation 
\begin{equation}
    \frac{P/k_\mathrm{B}}{\mathrm{cm^{-3}\,K}} = \frac{n \,\mu}{\mathrm{cm^{-3}}} \frac{T}{\mathrm{K}}~,
    \label{eq:gas}
\end{equation}
where $k_\mathrm{B}$ is the Boltzmann constant, $n$ the number density of Eq.~\ref{eq:number_density} 
and $T$ the ``mass-weighted'' dust temperature. We note, however, that the latter is only the cold component of the dust temperature (Fig.~9 (left) of~\citealt{Tabatabaei2014}), which has also been used to produce the column density maps in~\citetalias{Keilmann2024}. 
The dust temperature is around $20\,$K, which corresponds to a similar gas temperature only when gas and dust are thermally well coupled by collisions~\citep{Goldsmith2001,Goicoechea2016}, which is the case in cool, dense regions. In less dense regions, gas and dust temperatures can differ significantly due to the inefficiency of collisional energy transfer. We thus expect a difference between the inner regions within a GMC, with typical temperatures of around $10-20\,$K, and the inter-cloud gas, which can be significantly higher, corresponding to gas temperatures of $>\,$$100\,$K. 
In addition, there is possible unresolved substructure in the beam and the density is rather low because of the beam-averaging. 
Thus, it is not surprising that our pressure results in lower values compared to~\citet{Hughes2013},~\citet{Sun2020b} and~\citet{Sun2020}. The latter two use velocity-resolved CO data from which they obtain higher pressures. 
The pressure values we derive are therefore only valid for the cool, molecular GMCs and we do not go into great detail in our interpretation.

\section{Dendrogram analysis and comparison with the Milky Way} \label{sec:analysis}

In this section, we discuss the distributions of the key physical cloud properties from the Dendrogram leaves extraction of the dust-derived $\Nhtwo$ and CO maps individually (Sects.~\ref{subsec:sizes} and \ref{subsec:masses_density}).  
We compare our results with the Milky Way GMC statistics from~\citet{Rice2016} and~\citetalias{Nguyen2016} that rely on the Columbia (CfA) $^{12}\mathCO(1-0)$ survey~\citep{Dame1986,Dame2001}, which provides the most comprehensive Milky Way GMC catalog. \citetalias{Nguyen2016} derived the cloud properties by eye inspection of line-integrated CO maps and focuses on large ($R\gtrapprox50\,$pc) and massive ($M\gtrapprox10^6\,\mathrm{M_\odot}$) MCCs. Those MCCs with a SFR larger than $1\,\mathrm{M_\odot\,yr^{-1}\,kpc^{-2}}$ are called ``mini-starbursts'', an example is the W43 region. However, what we find in M33 are more MCCs without a high SFR~\citep{Corbelli2017}, though the SFR densities of MCCs are comparable with the
SFR of super giant \HII\ regions in M33~\citep{Miura2014}. \citet{Rice2016} performed a Dendrogram analysis on the velocity-resolved CO data and also included smaller ($<\,$$50\,$pc) and lighter MCs, down to a limit of a few $10^4\,\mathrm{M_\odot}$, which are beyond our resolutions.   
Since there are other CO surveys of the Milky Way with extensive datasets, we also partly compare our findings with those. Nevertheless, these studies mainly detect smaller molecular clouds, posing a challenge in making meaningful comparisons with the GMCs we can resolve. For a comprehensive overview of the current CO surveys of regions in the Milky Way, see~\citet{Park2023}. The most relevant studies utilize data from the Galactic Ring Survey (GRS); see~\citet{Simon2001} and~\citet{Roman_Duval2010} and cloud compilations presented in~\citet[e.g.,][]{Kramer1998,Schneider2004,Su2019}.

We also compare our results with~\citet{Dobbs2019}, who studied molecular clouds in a simulation of a M33-type galaxy and from the same IRAM CO data of M33 we use. Their models, based on Smooth Particle Hydrodynamic (SPH) codes SPHNG~\citep{Bate1995} and GASOLINE2~\citep{Wadsley2017}, are detailed in~\citet{Dobbs2018}. 
They used Friends-of-Friends (FoF) and CPROPS algorithms to determine cloud properties of the simulations.

For completeness and to compare with other studies, Appendix~\ref{sec:CO_lumi} shows and discusses the $^{12}\mathrm{CO(1-0)}$ luminosity of M33.

\subsection{GMC radii}
\label{subsec:sizes} 

The calculated mean of the beam-deconvolved cloud equivalent radius reveal a similar overall distribution and mean values of around $58\pm13\,\mathrm{pc}$ and $68\pm21\,\mathrm{pc}$ for dust- and CO-derived GMCs, respectively. 
The largest structure observed from dust data (NGC604) exhibits the most notable difference, featuring a radius of approximately $200\,\mathrm{pc}$.
For completeness, we report that the branches have a mean radius of $354\pm152\,\mathrm{pc}$.

For comparison,~\citet{Gratier2012} and~\citet{Corbelli2017} found mean radii of $42\pm13\mathrm{\,pc}$ and $45\pm12\,$pc, respectively, from the IRAM $\mathCO$ map using CPROPS. These sizes are smaller compared to our findings, primarily due to the higher spatial resolution of $12''$ of the unsmoothed IRAM CO map they used. 
\citet{Williams2019} report a median GMC size of $105\,$pc for their identified GMCs in M33, while we derived a mean value for this catalog of $116\pm29\,$pc. They identified with Dendrogram the clouds in the SPIRE $250\,\mum$ map at $18''$ resolution and then performed an SED fit on the averaged flux values of 160, 250, 450 and 850$\,\mum$ within one identified structure and determined the cloud mass with a fitted DGR and \Xco\ factor. 
They find an \Xco\ value of $\sim\,$$6\times10^{20}\,\cmKkms$ by fixing the dust absorption coefficient $\kappa_0$ and emissivity index $\beta$ and bin the GMCs at 500$\,$pc. A DGR and \Xco\ factor are radially determined via scatter analysis fitting both parameters simultaneously, resulting in possibly degenerate values since different combinations can lead to the same result (their Eq.~6). They subtracted an \HI\ map without short-spacing from~\citet{Gratier2012}. A source extraction was also performed on the higher resolution (13$''$) 450$\,\mum$ map, reporting a similar size distribution of the clouds.   

\citet{Rice2016} has the most complete GMC catalog of the Milky Way, with mean radii of $34\pm6\,$pc. Since the subset of~\citetalias{Nguyen2016} only concentrates on large and massive clouds, the mean value is higher around $90\pm20\,\mathrm{pc}$ (see Table~\ref{table:clouds_mean_properties} and Fig.~\ref{fig:H2_mass_density_area_radius_distribution_dendro}). 
Despite this, the trend of the distribution closely mirrors the patterns observed in the M33 data derived from dust and CO data for larger GMCs. There appears to be a size limit of around $150\,$pc for the largest GMCs/GMAs, in the Milky Way as well as in M33, though both galaxies are different in terms of size, mass and age. Interestingly,~\citet{Dobbs2019} find a similar threshold in their simulations of M33 and their cloud extractions of the IRAM CO data set (their Fig.~4). The three distributions exhibit a comparable decline in both the shape and the number of structures as they increase in size. 
We further do not clearly detect Giant Molecular Filaments (GMFs), which can reach lengths of up to $200\,$pc in the Milky Way~\citep{Wang2020}. However, some of our GMCs have an elongated geometry and aspect ratios larger than 3 so that they formally fit to the definition of GMFs. We come back to this point in Sect.~\ref{subsubsec:pressure_elongation_temp}. 

A potential mechanism that explains the growth of GMCs in alignment with the results can be attributed to supernovae. \citet{Kobayashi2017} and~\citet{Kobayashi2018} show that \HI\ gas is an important mass reservoir for growing GMCs and they show that supernovae can accumulate the \HI\ gas to molecular clouds. In this case, the GMC growth is assumed to depend on the maximum potential sizes of supernovae remnants. Hence, most GMCs are predicted to show sizes of $\lesssim\,$$100\,$pc, with a few exceptions of up to 150 to $200\,$pc. Another potential explanation is that GMC growth depends on the galactic gas disk scale height, $h_z$. When a GMC is smaller than $h_z$, it can grow in all three dimensions. Once it reaches the size of $h_z$, its ability to grow in the vertical direction will drop. Only the two remaining directions allow the GMCs to expand, but this slows their growth, giving time for stellar feedback or other mechanisms to destroy and regulate cloud sizes. The gas scale height of the galactic disk in the Milky Way ranges from 300 to $400\,$pc~\citep{Carroll2007}. M33 shows a comparable scale height of $320\pm80\,$pc~\citep{Berkhuijsen2013}. Therefore, this rationale could account for the analogous shape and upper size limit of the largest GMCs in both galaxies. 
We note, however, that~\citet{Koch2019} determined a CO/\HI\ line width ratio of around 0.7 and suggest that M33 has a marginal thick molecular disk and not a thin disk dominated by GMCs and a thicker diffuse molecular disk as seen for the Milky Way and other more massive spirals.

However, we caution that the GMCs identified in M33 potentially have line of sight effects due to limited resolution, the inclination and the increased thickness of the central region, which can blend distinct GMCs into a larger structure that is not one coherent GMA. In addition, as mentioned in~\citet{Rice2016}, the mass obtained for some GMCs can be inaccurate by up to an order of magnitude due to challenges of reliably determining the correct distances.

\subsection{GMC masses and densities}
\label{subsec:masses_density} 

Figure~\ref{fig:H2_mass_density_area_radius_distribution_dendro} (bottom row) shows the mass and average density distributions of H$_2$ derived from dust and CO. The black dashed line represents the minimum mass selection used in~\citetalias{Nguyen2016}. 
The average number density for the binned data set (Fig.~\ref{fig:H2_mass_density_area_radius_distribution_dendro} bottom right) and the individual clouds (Tables~\ref{table:clouds_properties_dust_dendro} and \ref{table:clouds_properties_CO_dendro}) are low, typically below $30\,\mathrm{cm^{-3}}$ for both tracers. The mean of the average densities are similar, with values of $n=5.2\pm1.5\,\mathrm{cm^{-3}}$ for dust-derived and $n=3.0\pm1.1\,\mathrm{cm^{-3}}$ for CO-derived GMCs. 
Our maps have a spatial resolution of $75\,$pc, and therefore the identified structures are likely composed of smaller substructures with higher local densities. 
The densities of GMCs in the Milky Way ($29.1\pm8.0\,\mathrm{cm^{-3}}$) have been calculated using the same methodology, based on the data presented in Table~1 of~\citetalias{Nguyen2016}. The branches have a low average density of $1.1\pm0.4\,\mathrm{cm^{-3}}$, which is reasonable given that they span larger areas than the leaves and incorporate a significant amount of inter-cloud and envelope material, both of which are expected to have lower densities. 

The mass distributions in M33 (Fig.~\ref{fig:H2_mass_density_area_radius_distribution_dendro}, bottom left) show no significant differences between CO and dust for our study. The maximum GMC mass from CO is $\approx5\times10^{6}\,\mathrm{M_\odot}$, whereas for dust it is NGC604 with $\approx8\times10^{6}\,\mathrm{M_\odot}$. We note that there are only a few GMCs in M33 above $10^6\,\mathrm{M_\odot}$ in both tracers. The mean values derived from the dust data are very similar, with $M=(2.8\pm0.9)\times10^{5}\,\mathrm{M_\odot}$ compared to $M=(2.9\pm0.9)\times10^{5}\,\mathrm{M_\odot}$ for CO.
The branches have a mean mass of $M=(1.3\pm0.2)\times10^{7}\,\mathrm{M_\odot}$.

~\citetalias{Nguyen2016} selected only Milky Way GMCs/GMAs with masses of larger than around $10^6\,\mathrm{M_\odot}$, and thus it is not surprising that the distribution contains only GMCs in this mass range (GMCs with lower masses are not absent but were not included in the survey). Notably, M33 lacks a significant high-mass GMC population. The procedure for mass determination is the same for our study and that for~\citetalias{Nguyen2016}: for CO, the line-integrated intensity was used to derive the CO column density and then finally the mass using an \Xco\ factor; and for the dust, the $\Nhtwo$ column density was derived from an SED fit. 
Interestingly, the lack of significant high-mass GMCs in M33 also becomes evident by comparing with the comprehensive~\citet{Rice2016} Milky Way catalog, which arises from a velocity-based identification of GMCs from CO data, which in addition shows a mean mass similar to our results of $(2.4\pm1.0)\times10^5\,\mathrm{M_\odot}$.
The difference in mass thus stems from the lower overall CO luminosity and hydrogen column density in M33. 

The H$_2$ gas mass in the center of M33 (see Fig.~\ref{fig:gal_envi} for an outline of the center) is $\sim\,$$25\%$ of the total dust-derived H$_2$ mass and amounts to $4.3\times10^7\,\mathrm{M_\odot}$.
This is an order of magnitude lower than the central molecular zone (CMZ) of the Milky Way ($\sim\,$$1.3\times10^8\,\mathrm{M_\odot}$). 
We note that the overall mass of M33 is one order of magnitude lower than that of the Milky Way.
The total H$_2$ mass of the Milky Way is suggested to be 1.4 times higher than the values found in earlier studies~\citep{Sun2021}. Applied to the results reported in~\citet{Garcia2014}, this leads to a total H$_2$ mass of $4.2\times10^8\,\mathrm{M_\odot}$. Consequently, the proportion of the CMZ of the Milky Way to this mass is $\sim\,$$30\%$.

The Milky Way also shows higher number densities, with a mean density of about $30\pm11\,\mathrm{cm^{-3}}$ and $18\pm6\,\mathrm{cm^{-3}}$ for the dataset presented in~\citetalias{Nguyen2016} and~\citet{Rice2016}, respectively. Mean values from dust and CO data are roughly five times smaller than those in the Milky Way datasets (and not one order of magnitude). 
According to the mass-size relation discussed in Sect.~\ref{subsec:mass_size_relations}, the density decreases with size. This might explain why mean densities do not show the same trend like GMC masses, central region mass or total H$_2$ mass of M33, all of which are consistently an order of magnitude lower compared to those of the Milky Way.

\citet{Williams2019}, using dust data, found in M33 GMC masses shifted to higher values averaging to ranges from $(3.7\pm1.4)\times10^{5}\,\mathrm{M_\odot}$ and low mean number densities of $1\pm0.4\,\mathrm{cm^{-3}}$, while the average cloud mass in~\citet{Gratier2012} from CO data is $(2.4\pm0.9)\times10^{5}\,\mathrm{M_\odot}$ with a mean density of $30\pm7\,\mathrm{cm^{-3}}$. \citet{Corbelli2017} find similar results using the same data at the same angular resolution with $(2.6\pm1.1)\times10^{5}\,\mathrm{M_\odot}$ and $17\pm5\,\mathrm{cm^{-3}}$. The masses match our findings, but the higher number densities are due to detecting smaller structures. This may result from the $12''$ resolution of the unsmoothed CO data and a different cloud extraction method (CPROPS).

The effects of limited resolution of our data do not cause non-detections of GMCs with similar masses and densities in M33 compared to the Milky Way.
Using a larger beam would inaccurately merge smaller structures into fewer larger structures, consequently inflating the overall mass. The dissimilarity in mass between M33 and the Milky Way, with M33 having only around $10$\% of the mass of the Milky Way, probably originates from variations in the sizes and evolutionary stages of the galaxies. The diameter of M33 is approximately two-thirds the size of the Milky Way.
\citetalias{Nguyen2016} used a dataset with an angular resolution of $8'.8$, translating into a spatial resolution of $\sim\,$$60\,$pc for the most distant GMCs in the Milky Way. For these distant GMCs, our spatial resolution is similar.

By considering sweeping the \HI\ medium by supernovae as we discussed in Section~\ref{subsec:sizes}, the typical maximum mass is limited by the gas scale height so that $n_\mathrm{ISM} \cdot h_z^3$, where $n_\mathrm{ISM}$ is the volume density of the ISM and $h_z$ is the galactic gas disk scale height. The gas disk scale heights of both galaxies are similar, as discussed above. Thus, the remaining factor influencing the mass growth may be attributed to the density of galaxies. Given that the Milky Way has a higher $\htwo$ density and total mass (from which a greater column density and ultimately a higher number density can be expected), we anticipate that the Milky Way will show higher densities. Therefore, we propose this mechanism as a possible driver.
Meanwhile,~\citet{Kobayashi2017} and~\citet{Kobayashi2018} performed a semi-analytic theory to investigate the impact of cloud-cloud collisions. They show that, even in the Milky Way galaxy, cloud-cloud collisions have a minor impact on GMC growth and are only effective to clouds more massive than $10^6\,\mathrm{M_\odot}$. We therefore suspect that cloud-cloud collisions are mostly ineffective for M33.

\subsection{Mass--size relations}
\label{subsec:mass_size_relations}

\begin{figure}[htbp]
  \centering
  \includegraphics[width=0.9\linewidth]{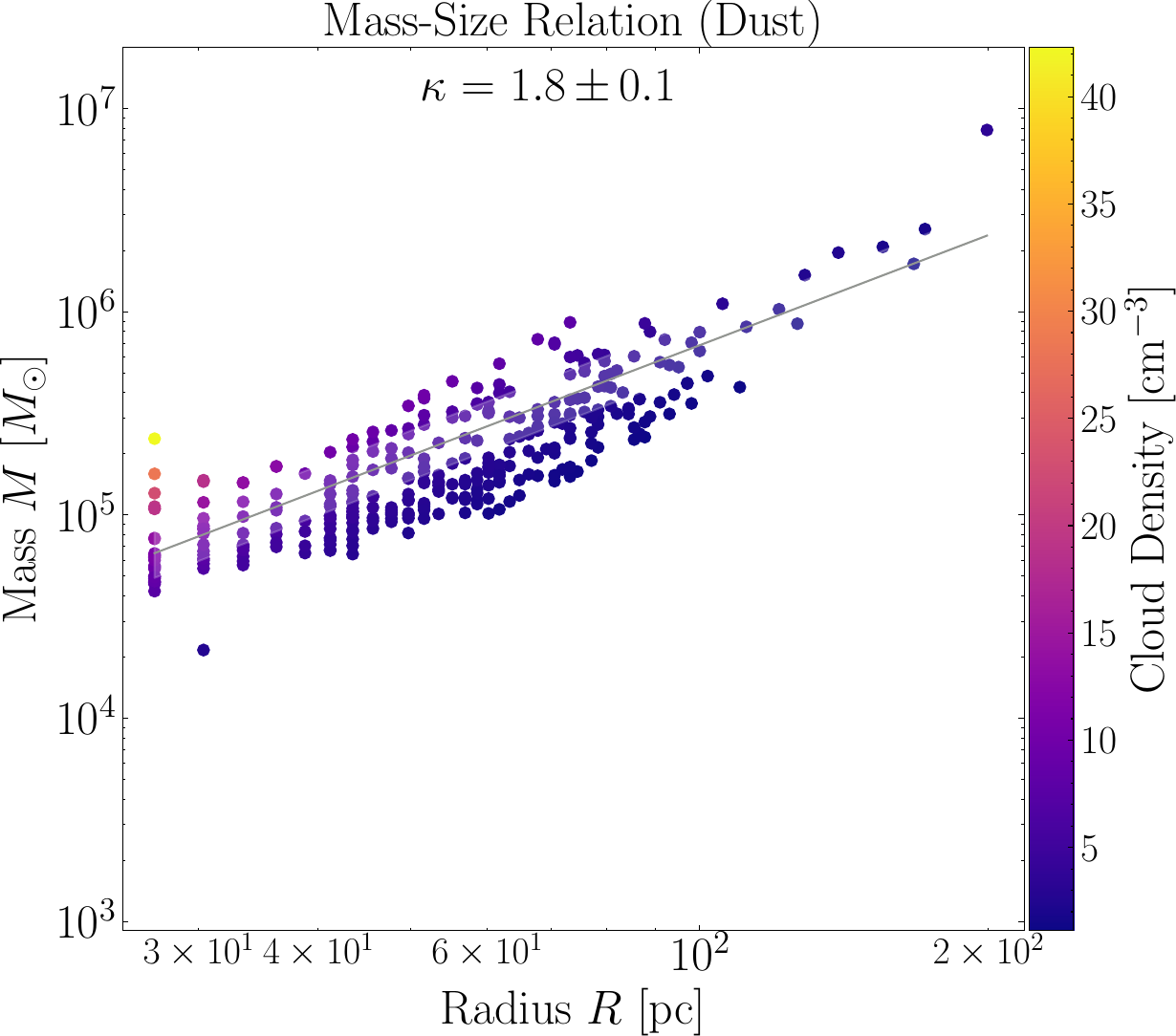}
  \includegraphics[width=0.9\linewidth]{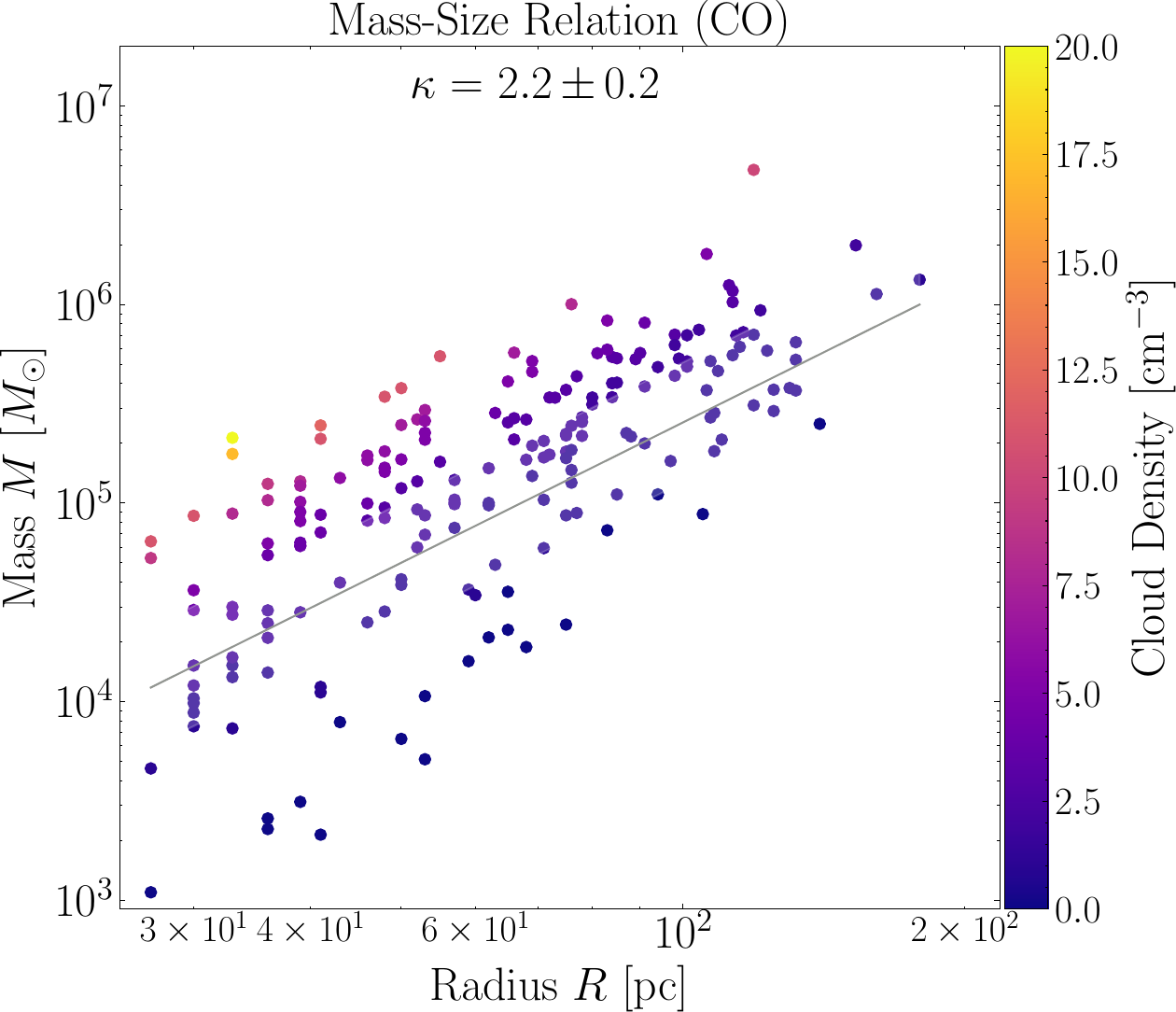}
  \caption{Mass--size relation of the identified GMCs.
  {\sl Top}: Mass--size Larson relation for GMCs derived from dust. 
  {\sl Bottom}: Mass--size Larson relation for GMCs derived from CO data. The various colors indicate the average density of individual clouds. The panel displays the slope $\kappa$ and its corresponding error.}
    \label{fig:mass_radius_relation_dendro}
\end{figure}
Figure~\ref{fig:mass_radius_relation_dendro} illustrates Larson's~\citep{Larson1981} relationship between mass and size for data derived from dust and CO. For dust-derived GMCs the slope is $1.8 \pm 0.1$, while it is $2.2 \pm 0.2$ for CO-derived GMCs. In the simulations of M33,~\citet{Dobbs2019} also found a clear mass-size relation in the observations and simulations. However, they did not quantify the slope of this relation so that we extracted the data points from their Fig.~4 and fitted a linear function with a slope of $1.4 \pm 0.1$. This estimation carries a significant level of uncertainty, attributed to manual data extraction and the overlapping of numerous data points, preventing clear individual identification. As previously noted in~\citet{Dobbs2019}, the larger GMCs appear too extended, also compared to the GMCs we identified, leading to a less steep slope. 

A slope of about $1.6$~\citep{Lombardi2010,Kauffmann2010} in the mass-size relation suggests the presence of substructures within individual clouds, while a slope of around $2.4$ was identified for GMCs in the GRS Galactic plane survey~\citep{Roman_Duval2010}. \citetalias{Nguyen2016} determined a slope of $1.9$ for GMCs and a slope of $2.2$ for MCCs.
Given that the third Larson relation indicates a power-law connection between mass and size, represented as $M\propto R^\kappa$, with a typical power-law exponent usually around $2$, it suggests similar gas surface mass densities for all GMCs. Furthermore, in line with assumptions for a spherical object, the mass can be linked to the size by $M\sim n/R^3 \sim N/R^2$, leading to $M\sim R^2$. This finding aligns well with observations that incorporate a column density threshold (see~\citet{Schneider2004} and references therein).

\subsection{GMC surface mass densities}
\label{subsec:surface_mass_density}

Comparing cloud masses and sizes across studies can be unreliable due to differing GMC definitions and boundary settings. Resolution limits may also cause undetected clouds or beam smearing.
The concept of cloud surface densities inherently considers the cloud size per definition, $\Sigma_\mathrm{GMC} = M/A$, thereby mitigating the impact of varying resolutions across studies. However, complete resolution uniformity is not achieved for instance when the galaxy is not perfectly face-on, as some large clouds may still merge into one single larger cloud when the beam size is large, resulting in a higher surface mass density. Conversely, smaller clouds, if sufficiently spaced from others, may get smeared within the beam, causing dilution and a decrease in surface mass density. This can be mitigated by excluding too small structures, which we do by only accepting structures 1.2 times the beam size.
Nonetheless, comparing surface mass densities can facilitate a less biased evaluation of clouds occupying similar spatial areas.

In Fig.~\ref{fig:H2_mass_density_area_radius_distribution_dendro}, we compare the GMC surface mass densities. For Milky Way GMCs~\citepalias{Nguyen2016}, the mean value of $187\pm51\,\mathrm{M_\odot\,pc^2}$ is approximately one order of magnitude higher than our dust-derived ($22\pm5\,\mathrm{M_\odot\,pc^2}$) and CO-derived values ($16\pm6\,\mathrm{M_\odot\,pc^2}$) in M33. Whereas compared to the more complete cloud catalog obtained by~\citet{Rice2016}, the mean value is $38\pm14\,\mathrm{M_\odot\,pc^2}$, approaching similar high surface densities at the higher end of the spectrum as the clouds presented in~\citetalias{Nguyen2016}. 
Branches show consistent mean surface mass densities of $19\pm5\,\mathrm{M_\odot\,pc^2}$.

For comparison,~\citet{Hughes2013} report a gas surface mass density for M33 of $46\pm20\,\mathrm{M_\odot\,pc^2}$ using $\mathrm{CO(1-0)}$ data published by~\citet{Rosolowsky2007}. This value is roughly a factor of $2$ higher than our results. 
\citet{Corbelli2017} similarly find $44\pm15\,\mathrm{M_\odot\,pc^2}$. 
Although they identify smaller GMCs with the CO data at $12''$ angular resolution, they still find similar masses, resulting in higher surface mass densities. The fact that they find surface mass densities about twice as high as our data are likely attributed to their application of a \Xco\ value twice that of the Galactic standard value. However, this has been disputed in~\citetalias{Keilmann2024}, which finds an average value nearly identical to the Galactic one.
\citet{Gratier2012} find similar values for these properties for the same reasons.
The data of~\citet{Williams2019} exhibit the lowest mean surface mass densities of all with $7.5\pm2.5\,\mathrm{M_\odot\,pc^2}$ which is probably due to the large sizes of the GMCs.
\citet{Roman_Duval2010} find for their Milky Way data a median surface mass density of $144\pm20\,\mathrm{M_\odot\,pc^2}$. 
Although the mass-size relations indicate a comparable surface mass density, there is an observed dissimilarity in the distribution shapes, with mean values varying by a factor of approximately one order of magnitude between M33 and the Milky Way. It should be noted that this finding aligns with the GMC masses we find in M33 being approximately an order of magnitude lower compared to the Milky Way GMCs and with the total masses of the two galaxies found by other studies mentioned above. We note that in the simulations of~\citep{Dobbs2008} the GMCs are more massive in galaxies with stronger spiral shocks or higher surface densities. 

Increased SF activity and higher pressures correlate with increased molecular gas surface mass densities~\citep{Heyer2004,Lehnert2015,Wang2017,Krumholz2018}.
The difference between our results and the subset in~\citetalias{Nguyen2016} is most likely due to manual selection of GMCs, involving a threshold applied to their masses. However, considering the~\citet{Rice2016} Milky Way catalog reveals a similar range of especially high surface mass densities between both galaxies. 
Given the smaller mean sizes and higher masses of this catalog compared to our results, both mass and size lead to increased surface mass densities by a factor of $\sim\,$2.
However,~\citet{Corbelli2017} find a distribution similar to the GMCs in the~\citet{Rice2016} catalog. 
We attribute this to the higher spatial resolution of the observations by~\citet{Corbelli2017}, which yield smaller GMC sizes relative to ours, although they still report mean masses comparable to ours probably due to the use of an \Xco\ value twice the Galactic standard value.

\subsection{Power-law mass spectra}
\label{subsec:power_law_mass_spectra}

\begin{figure}[htbp]
  \centering
  \includegraphics[width=0.9\linewidth]{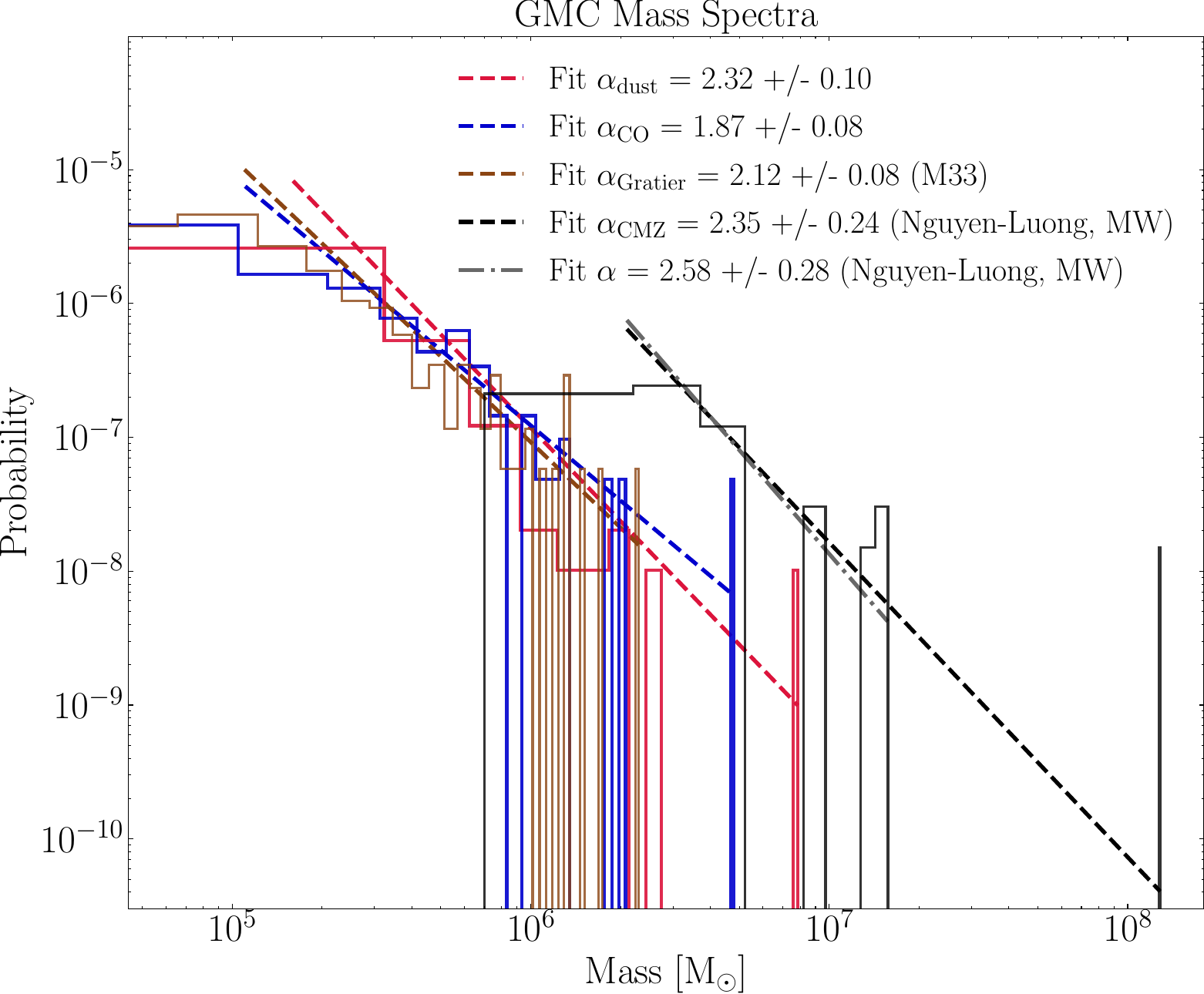}%
  \caption{Power-law mass spectra of detected GMCs from dust-derived data (red) and CO (blue) as well as from~\citetalias{Nguyen2016} (Milky Way, black). In black, the fit is performed with all listed structures in~\citetalias{Nguyen2016}, while the light gray dotted-dashed line shows the fit, where the CMZ of the Milky Way is excluded.
  }%
  \label{fig:all_powerlaws}
\end{figure}
We aim to determine the mass spectrum of GMCs in M33 identified via dust and CO, which may relate to cluster and star mass functions~\citep[and references therein]{kennicutt2012}.
Differences in the mass spectrum across regions might reflect variations in the processes that govern the formation, evolution and destruction of clouds~\citep{Rosolowsky2005,Colombo2014a}.
The mass spectrum typically conforms to a power-law probability distribution. To determine the power-law exponent $\alpha$ and its standard error ($\sigma/\sqrt{N}$) we first linearize the function
\begin{equation}
    p(x) \propto x^{-\alpha} 
,\end{equation}
and then employ a least-squares approach to fit a linear slope to the data.

Figure~\ref{fig:all_powerlaws} shows the distributions with an index determined to be $\alpha=2.32\pm0.10$ for the dust-derived and $\alpha=1.87\pm0.08$ for the CO-derived mass spectrum. The steeper slope of the dust-derived data indicates a larger number of less massive structures. This result is somewhat unexpected, because the low angular resolution of our data would have moreover suggested that many smaller molecular clouds along the line of sight would be artificially grouped into larger complexes, resulting in a flatter slope. \citet{Dobbs2019} reported $\alpha=1.59$, using the CPROPS algorithm to identify clouds in M33 from the IRAM CO data. For the data of~\citet{Corbelli2017}, using the same data and extraction method, a curved pattern with an index of 1.6 has been identified.
For the results reported in~\citet{Gratier2012}, using IRAM CO data as well as CPROPS, we determine a single power-law of $\alpha=2.12\pm0.08$. We note that in the simulations of~\citet{Dobbs2019}, the spread in $\alpha$ is large, between 1.66 and 2.27 (with an uncertainty of 0.2) and depends on the simulation (SPHNG or GASOLINE2) and the identification algorithm (FoF or CPROPS). 
\citet{Williams2019} find a higher slope of 2.83. Their result suggests a poorer ability of M33 to form massive clouds.
\citet{Rosolowsky2005} report a similarly steep mass spectrum slope of $2.9\pm0.4$, which may be biased by only sampling the high-mass end of the mass spectrum where the slope tends to be steeper.

For the power-law index of Milky Way clouds, including the CMZ, we determine $\alpha=2.35\pm0.24$ using the data presented in~\citetalias{Nguyen2016}. Excluding the CMZ results in an exponent of $\alpha=2.58\pm0.28$. We note that this comparison relies solely on the 44 structures manually selected by~\citetalias{Nguyen2016} for structures more massive than $0.7\times10^6\,\mathrm{M_\odot}$, which could introduce a potential bias and an undetected systematic error. For the results reported in~\citet{Roman_Duval2010}, we derive $\alpha=1.61\pm0.03$ (not shown in the figure) for the Milky Way CO data from the Galactic Ring Survey. 
These findings are in alignment with~\citet{Rice2016}, who found a slope of $1.6\pm0.1$ for their entire catalog. For the outer Galaxy, they reported a higher slope of $2.2\pm0.1$, whereas for the inner Galaxy, the slope remained at $1.6\pm0.1$. Similarly,~\citet{Fujita2023} found generally higher slopes, yet they show a consistent pattern with an index of $\alpha=2.30\pm0.11$ derived from $^{12}$CO data for distances below $8.15\,\mathrm{kpc}$ and $\alpha=2.51\pm0.14$ for distances less than $16.3\,\mathrm{kpc}$.
The power-law indices found in several other studies of the Milky Way, all using CO data, typically range between 1.6 and 2~\citep{Kramer1998,Simon2001,Schneider2004,Roman_Duval2010}.

The efficiency of cloud formation has been associated with various processes. As discussed in~\citet{Williams2019}, the influence on the GMC population of the spiral density wave amplitude~\citep[e.g.,][]{Shu1972} can be excluded to explain the tentatively higher slopes in M33 due to modeling efforts, which indicate that the spiral arms of M33 are likely due to gravitational instabilities~\citep{Dobbs2018}. The interstellar gas pressure might also be influential~\citep{Elmegreen1996,Blitz2006}. 
\citet{Kasparova2008} report increased interstellar pressure compared to the Milky Way, potentially leading to the formation of more massive clouds. Thus, interstellar pressure may not be the primary cause of a potential inefficient cloud formation. This contrasts with findings by~\citet{Blitz2006} and~\citet{Sun2018}, indicating M33 lies within a lower pressure regime. This scenario aligns with the upper cloud mass limit being influenced by interstellar pressure.
Another factor could be the role of metallicity in the transformation of \HI-to-$\mathrm{H_2}$~\citep{Krumholz2008,Kobayashi2023}. If M33 indeed has subsolar metallicity, this conversion would be less efficient, resulting in similarly inefficient cloud formation. However, the determined metallicity of M33 shows a very high dispersion~\citep{Willner2002,Crockett2006,Rosolowsky2008a,Magrini2010}. Furthermore, it is proposed that merging \HI\ clouds could form $\htwo$~\citep[e.g.,][]{Heitsch2005}, suggesting that larger \HI\ velocity dispersions could lead to more massive clouds. In M33, however, the 
average \HI\ velocity dispersion is around $\mathrm{13\,km\,s^{-1}}$ with minimal radial variation~\citep{Corbelli2018}. 
This is consistent with the velocity dispersion of 11 nearby galaxies of $\sim\,$$10\,\mathrm{km\,s^{-1}}$ presented in~\citet{Tamburro2009}. Typical velocity dispersions measured for the Milky Way are in the same range~\citep{Malhotra1995,Marasco2017}.
Another potential mechanism remains within supernovae.
The power-law index may be considered to represent the balance between GMC mass-growth and destruction by massive stars~\citep{Kobayashi2017,Kobayashi2018}. The supernova frequency per unit volume varies across the galactic disk and the expansion of supernovae remnants compresses the ISM initiating the transition of \HI-to-$\mathrm{H_2}$~\citep{Kobayashi2020,Kobayashi2022}.
In this case, the power-law slopes of the GMC mass functions are determined by the balance between the transition rate from \HI-to-$\mathrm{H_2}$ and the destruction rate by stellar feedback from massive stars, mainly radiative feedback~\citealp{Kobayashi2017}.
Additional mechanisms like shear may also set the maximum mass and lifetimes~\citep{Jeffreson2018}, especially in a region where the shear rate is high and the orbital speed is fast (e.g., the outer regions of the CMZ in case of the Milky Way galaxy).
We cannot determine which of these mechanisms primarily drive the potentially inefficient cloud formation in M33, as suggested by some of the findings discussed above.

In summary, the power-law index $\alpha$ shows a large spread for both M33 (1.6 to 2.9) and the Milky Way (1.6 to 2.5), due to differences in datasets and methods. Despite errors, there is no significant variance between M33 and the Milky Way, except for a slight tendency for higher values in M33. Both exhibit self-similarity from molecular clouds ($\sim\,$$50\,$pc) to larger GMAs, suggesting similar physical mechanisms for massive GMCs in both galaxies and a limit in sizes and masses despite their high difference in mass.
Given the values in the existing literature, it is difficult to determine whether the cloud mass distribution in M33 is significantly different from that in other large spirals within our Local Group. 

\begin{figure*}[htbp]
  \centering
  \includegraphics[width=0.8\linewidth]{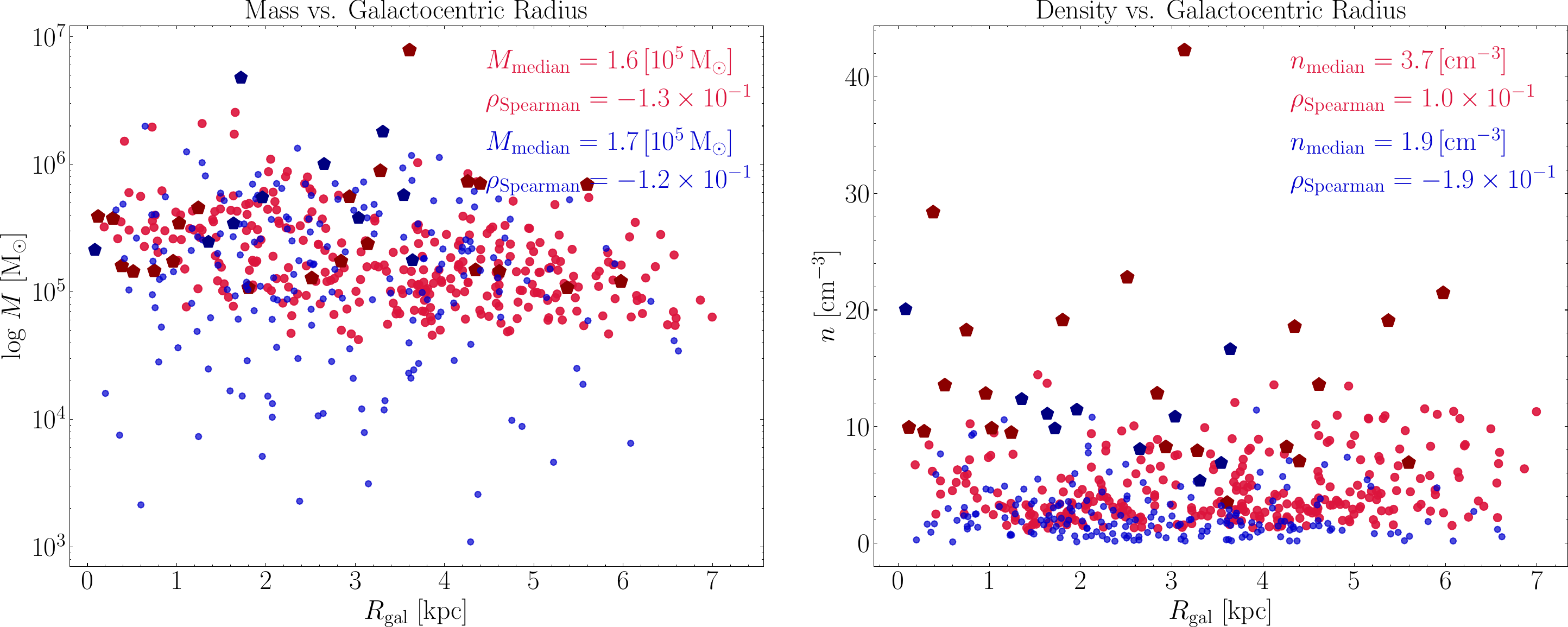}
  \includegraphics[width=0.8\linewidth]{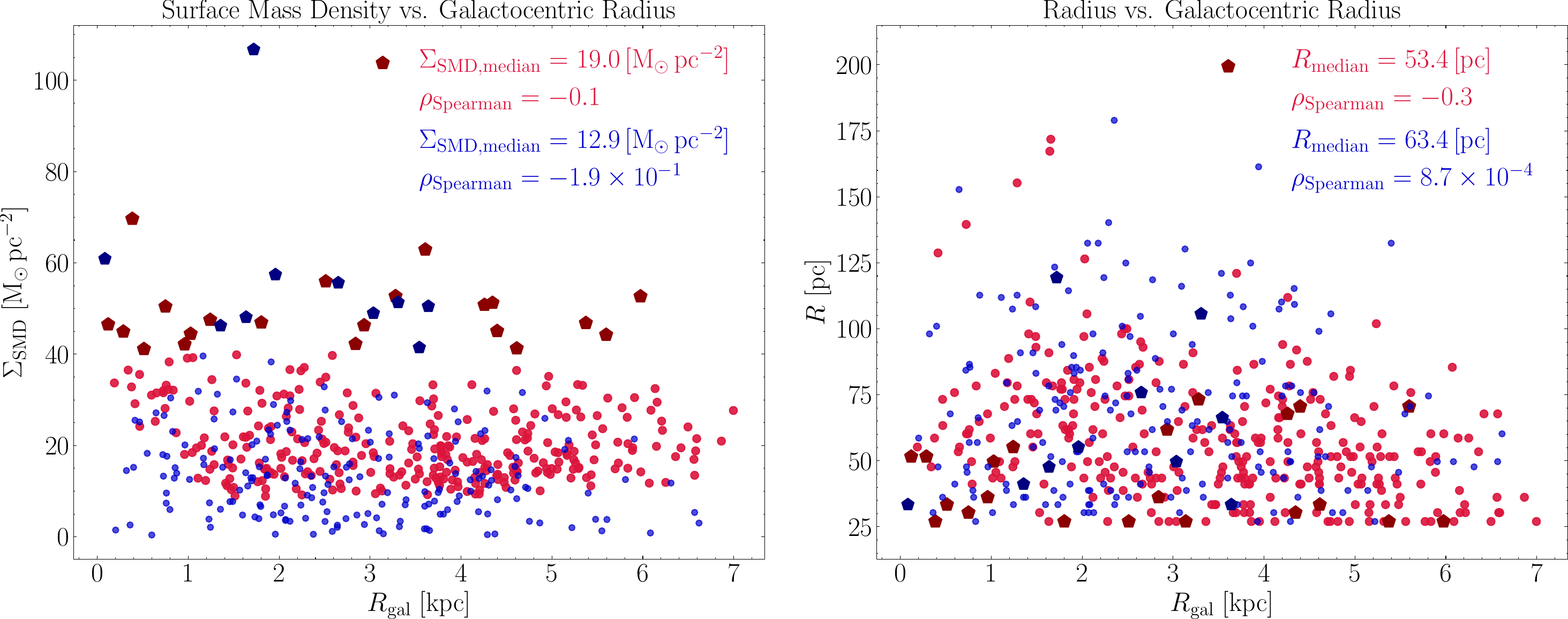}
  \caption{Mass, average density, surface mass density, and radius of the identified GMCs along the galactocentric radius.
  The dust-derived GMCs are shown in red, whereas the $\mathCO$-derived GMCs are represented in blue.
  The bigger dark red and blue data points mark the GMCs that have a surface mass density exceeding $40\,\mathrm{M_\odot\,pc^2}$.
  }
  \label{fig:mass_density_SMD_radius_vs_Rgal}
\end{figure*}

\begin{figure*}[htbp]
  \centering
  \hspace{0.14cm}
  \includegraphics[width=0.38\linewidth]{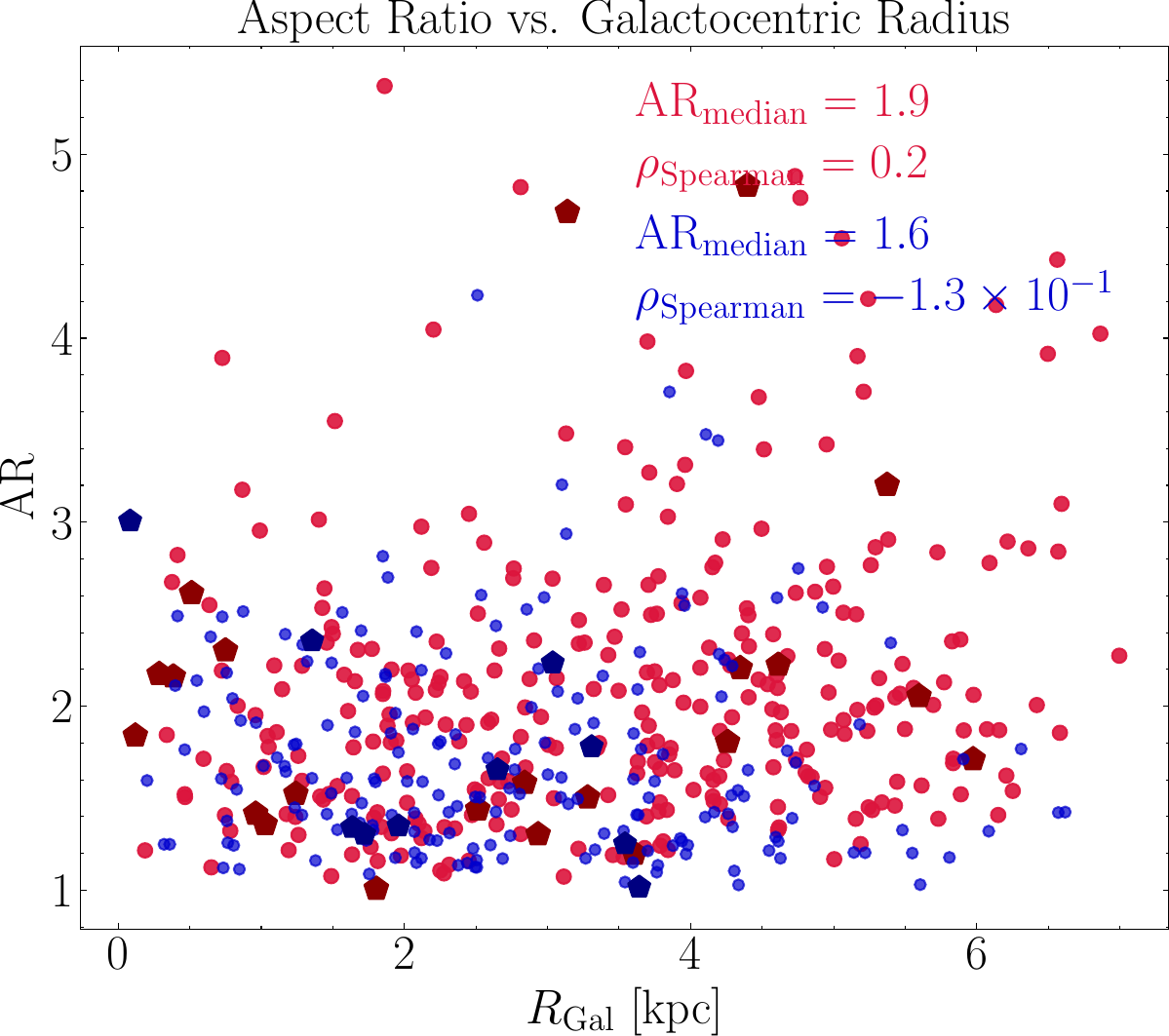} 
  \hspace{0.2772cm}
  \includegraphics[width=0.3855\linewidth]{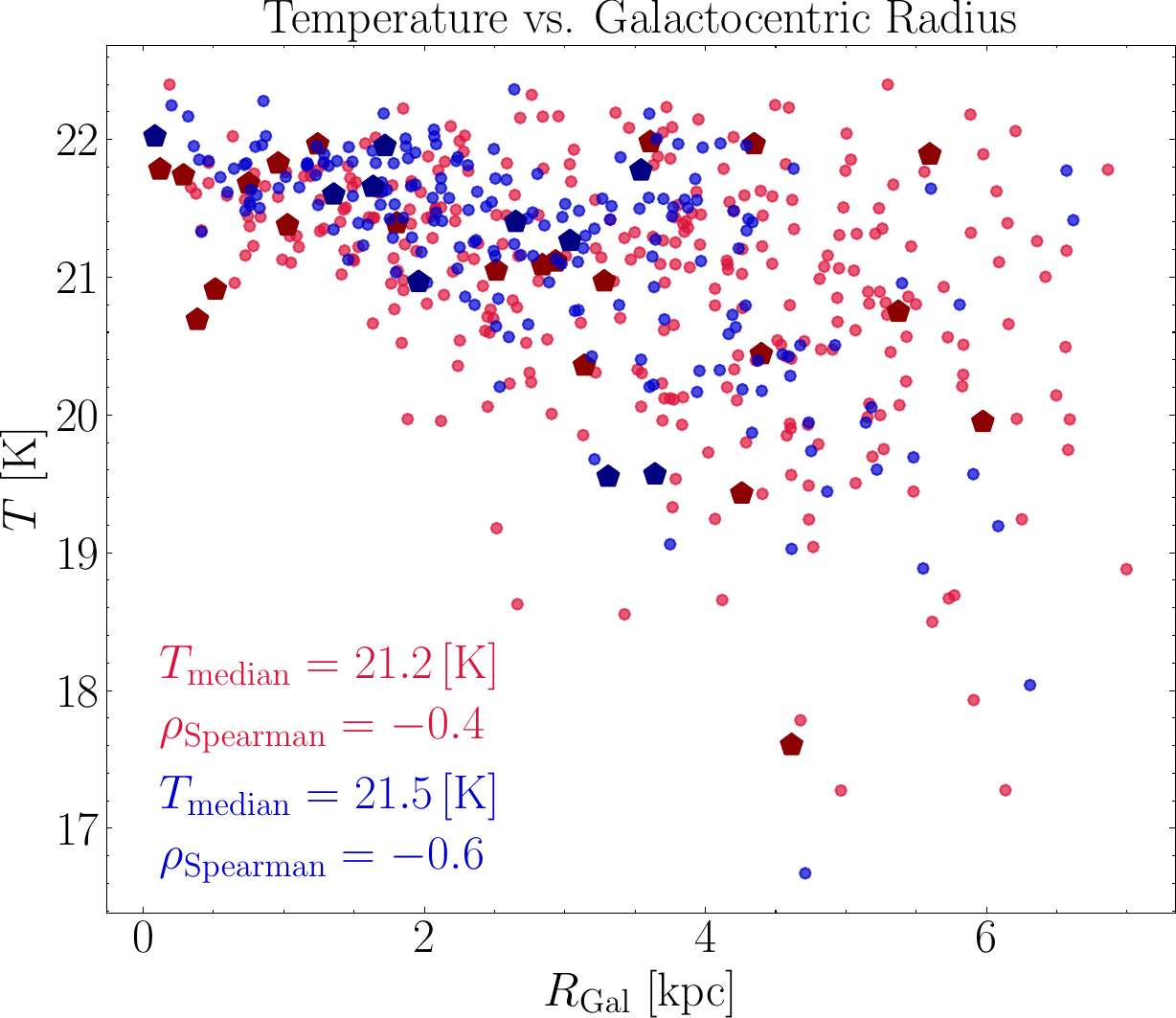}%
  \caption{
  Aspect ratio and temperature from dust and CO-derived data along the galactocentric radius.
  The bigger dark red and blue data points mark the GMCs that have a surface mass density exceeding $40\,\mathrm{M_\odot\,pc^2}$.}
  \label{fig:pressure_AR_temp_vs_Rgal}
\end{figure*}

\section{Trends with galactocentric radius and galactic environment in M33}
\label{sec:comparison_Rgal}

Molecular clouds do not possess a perfectly spherical shape. Instead, their morphology is often influenced by complex processes such as merging or turbulent flows~\citep[e.g.,][]{Vazquez-Semadeni1995,Heitsch2006,Clark2019,Schneider2023} 
or cloud-cloud collisions~\citep{Casoli1982,Fukui2021}, leading to irregular shapes characterized by clumps and filaments. Variations in cloud properties under different environmental conditions within a galaxy offer valuable insight into the factors shaping cloud formation and evolution \citep[e.g.,][]{Sun2020}. 

The molecular gas, for example, forms huge associations as a result of the gravitational attraction of the spiral arm. As the gas exits the spiral arms and experiences significant shear forces, it breaks apart and reverts to smaller elongated structures~\citep{LaVigne2006}. Numerous observational and computational studies emphasize the presence of filamentary structures in the areas between the arms~\citep{Ragan2014,DuarteCabral2016,DuarteCabral2017} and the presence of high-mass structures within the spiral arms~\citep{Dobbs2011,Miyamoto2014}. Apart from structure variations, metallicity gradients within a galaxy can also lead to variations in the physical properties of the molecular cloud. 
We thus examine in the following sections the physical properties of the GMCs in M33 as a function of the galactocentric radius and the galactic environment of M33.

\subsection{Trends with galactocentric radius}    \label{subsec:comparison_Rgal}

Figures~\ref{fig:mass_density_SMD_radius_vs_Rgal} and~\ref{fig:pressure_AR_temp_vs_Rgal} 
display the mass, average density, surface mass density, radius as well as 
aspect ratio and dust temperature as a function of the galactocentric radius. The relationship between GMC properties and galactocentric radius has also been examined by~\citet{Gratier2012},~\citet{Corbelli2017} and~\citet{Braine2018}.
A comprehensive discussion of specific properties can be found in Appendix~\ref{app:trends_Rgal}.

In summary, the parameters show only a weak (for high-$\Sigma$ GMCs) or non-existing (for low-$\Sigma$ GMCs) trend with the distance from the galaxy's center, raising the question whether SF is influenced by the galactocentric radius. Only GMCs with the highest surface mass densities (above $40\,\mathrm{M_\odot\,pc^2}$) show a tendency to have higher values for 
density and $\Sigma$ 
in the center of M33. This finding is similar to what is observed in the Milky Way. In both galaxies, self-gravity and cloud-cloud collisions become more important for these high-$\Sigma$ GMCs in the respective CMZ. 

In the following section, we discuss the more significant trends we observe for different regions (center, spiral arms, outskirts) in M33, as a radial dependence on the galactocentric radius does not entirely unveil systematic differences in the galactic environments.

\subsection{Trends with galactic environment}  \label{subsec:comparison_envi}

It is not yet clear whether SF is more efficient in particular regions of galaxies and to which extent the SFR and SFE are linked to the physical properties of the GMC population. Observations and simulations indicate that GMCs are concentrated in spiral arms, often with regular spacing, which can be explained when GMCs are formed by gravitational instabilities~\citep{Elmegreen1990,Kim2002}. On the other hand, GMCs can also form by agglomeration of smaller clouds or merging of flows (see references above). A higher SFR can then be just a by-product of the higher material reservoir in the spiral arms. 
While some studies~\citep{Koda2009,Pettitt2020,Colombo2022} report variations between their spiral arms and inter-arm populations, others~\citep{DuarteCabral2016,Querejeta2021} find no discernible differences in the overall properties of the cloud population. 

In this section, we systematically investigate if there are variations in the physical properties of the GMC population in certain regions of M33. For that, we use our dust-derived column density map and split the galaxy by eye-view into a central region, the two main spiral arms and the outskirts (Fig.~\ref{fig:gal_envi}). 
The two main spiral arms are approximated to extend to a galactocentric radius of roughly $4\,$kpc, whereas the central area of M33 can roughly be described as an equivalent circle with a galactocentric radius of around $1.3\,$kpc. The outskirts are considered to be the remaining area of M33's disk.\footnote{Since the inter-arms are faint and challenging to distinguish from the surrounding diffuse gas, we refer to this area as the ``outskirts'' or outer region. We emphasize that this mask is not meant to be considered as a precise delimitation.}
To determine the spiral arm structure more quantitatively, we additionally employ a similar approach as in~\citet{Querejeta2021} and model the spiral arms with a log-spiral function and perform a fit to this model. Details of this procedure and the results are given in Appendix \ref{app_c:env} and in Fig.~\ref{fig:gal_envi}.
The visually estimated borders of the spiral arms already capture the fitted log-spirals very well. We therefore continue to use the masks presented in Fig.~\ref{fig:gal_envi} to study the spiral arms and outskirts.

\subsubsection{Column density complementary cumulative distributions}

\begin{figure}[htbp]
  \centering
  \includegraphics[width=0.9\linewidth]{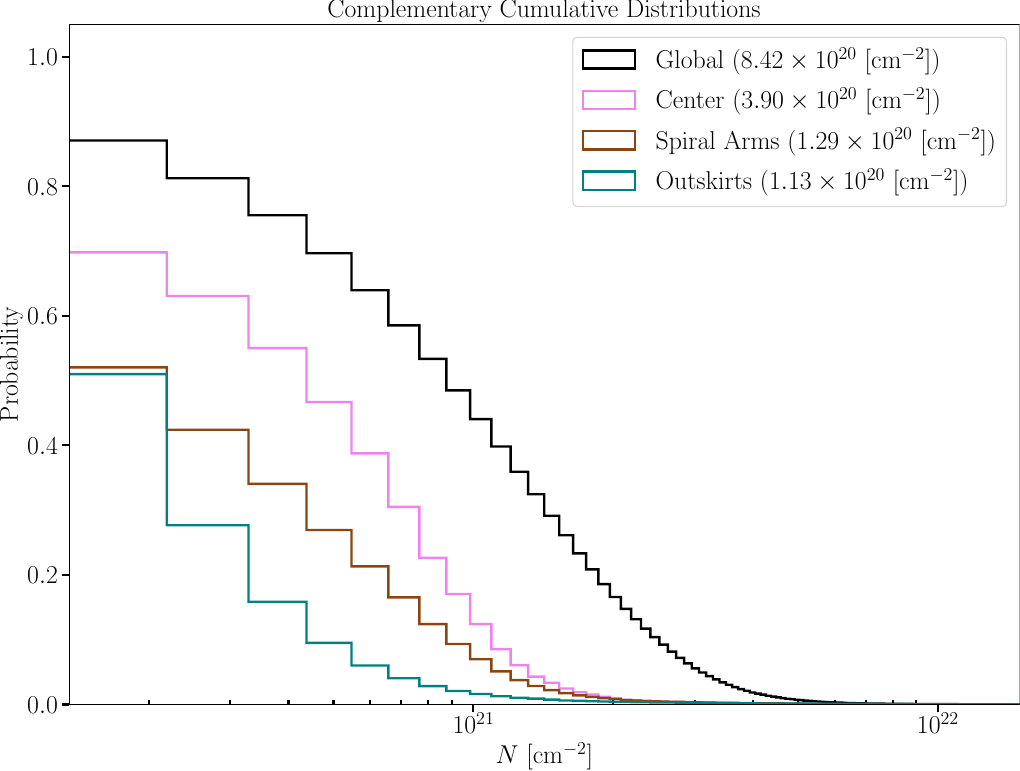}
  \caption{Complementary cumulative $\Nhtwo$ distributions of the entire galaxy (based on dust-derived data) and the three galactic environments. We note that these distributions are solely based on pixels and are not connected to GMCs.}%
  \label{fig:ccdf_envi}
\end{figure}

\citet{Querejeta2021} reported increased gas surface densities closer to the central regions of galaxies by analyzing the $\mathrm{CO(2-1)}$ data obtained from the PHANGS-ALMA survey~\citep{Leroy2021}. 
\begin{figure*}[htbp]
  \centering
  \includegraphics[width=0.29\linewidth]{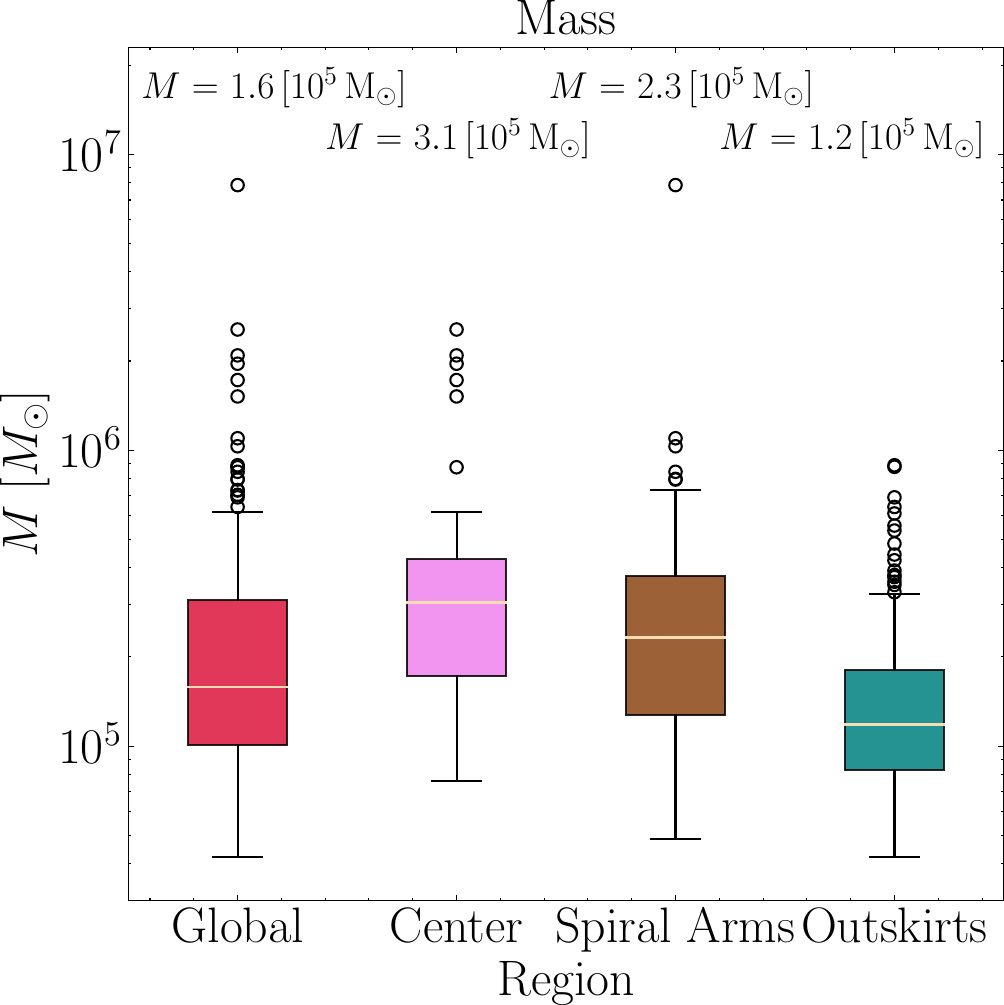}
  \includegraphics[width=0.29\linewidth]{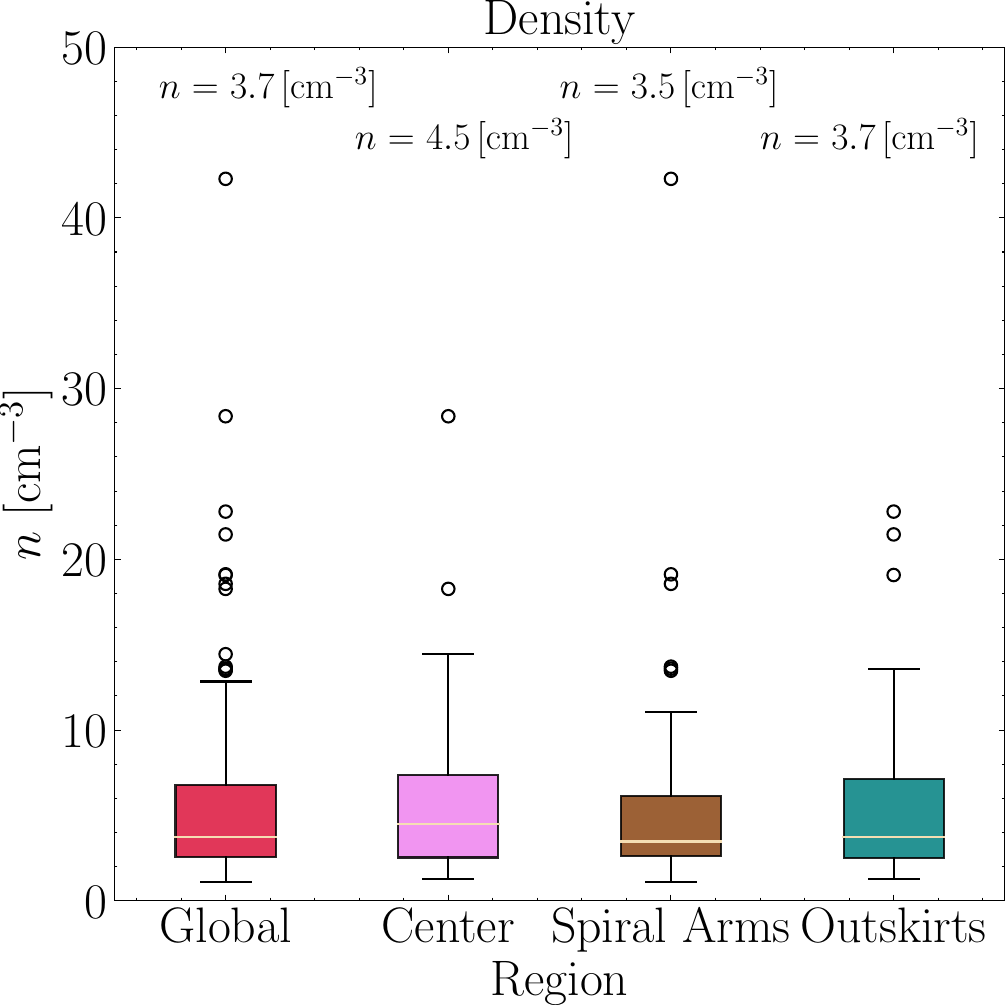}
  \includegraphics[width=0.29\linewidth]{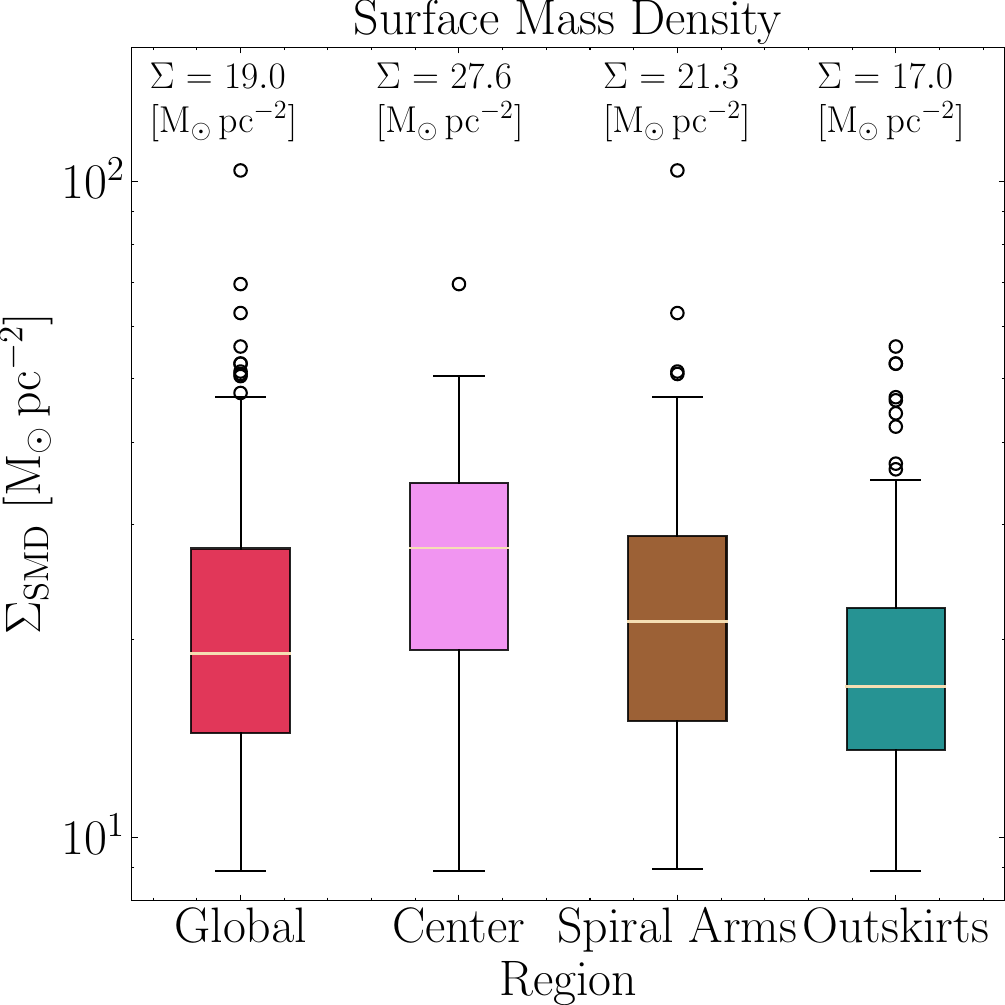}
  \includegraphics[width=0.29\linewidth]{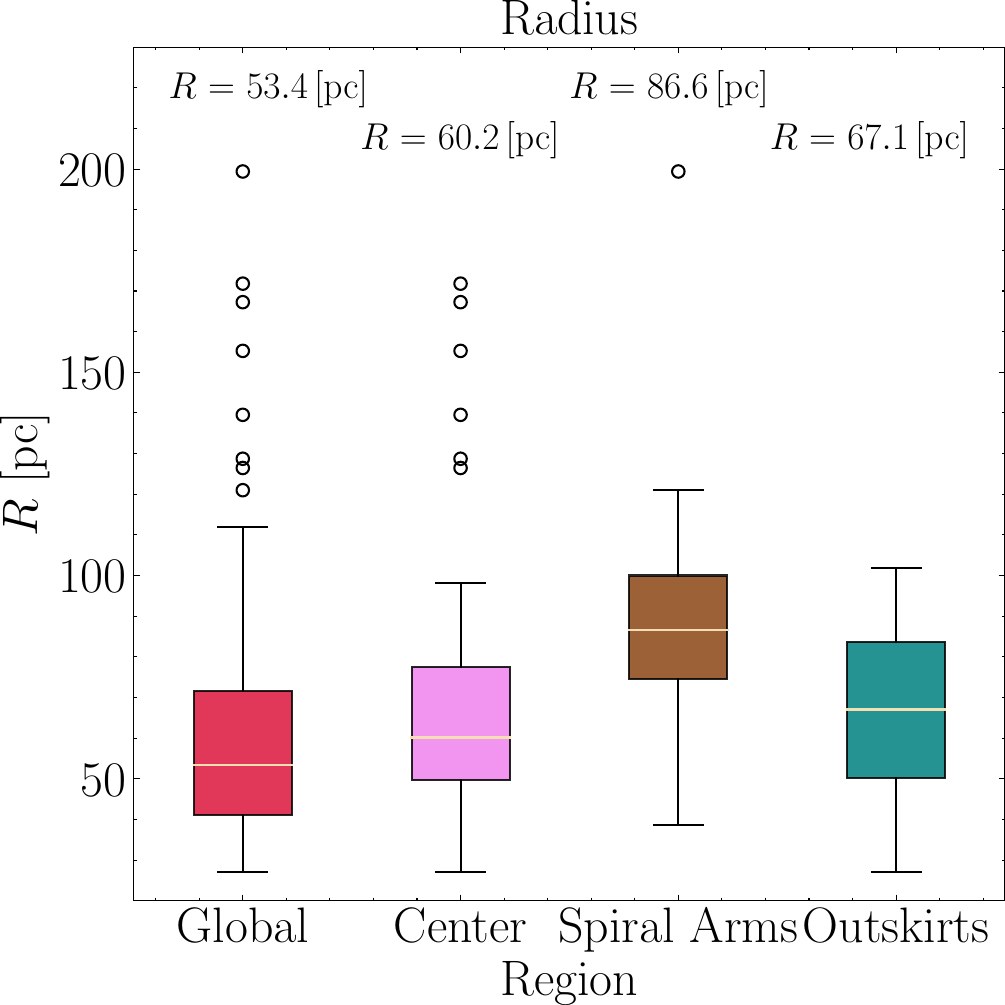}
  \includegraphics[width=0.29\linewidth]{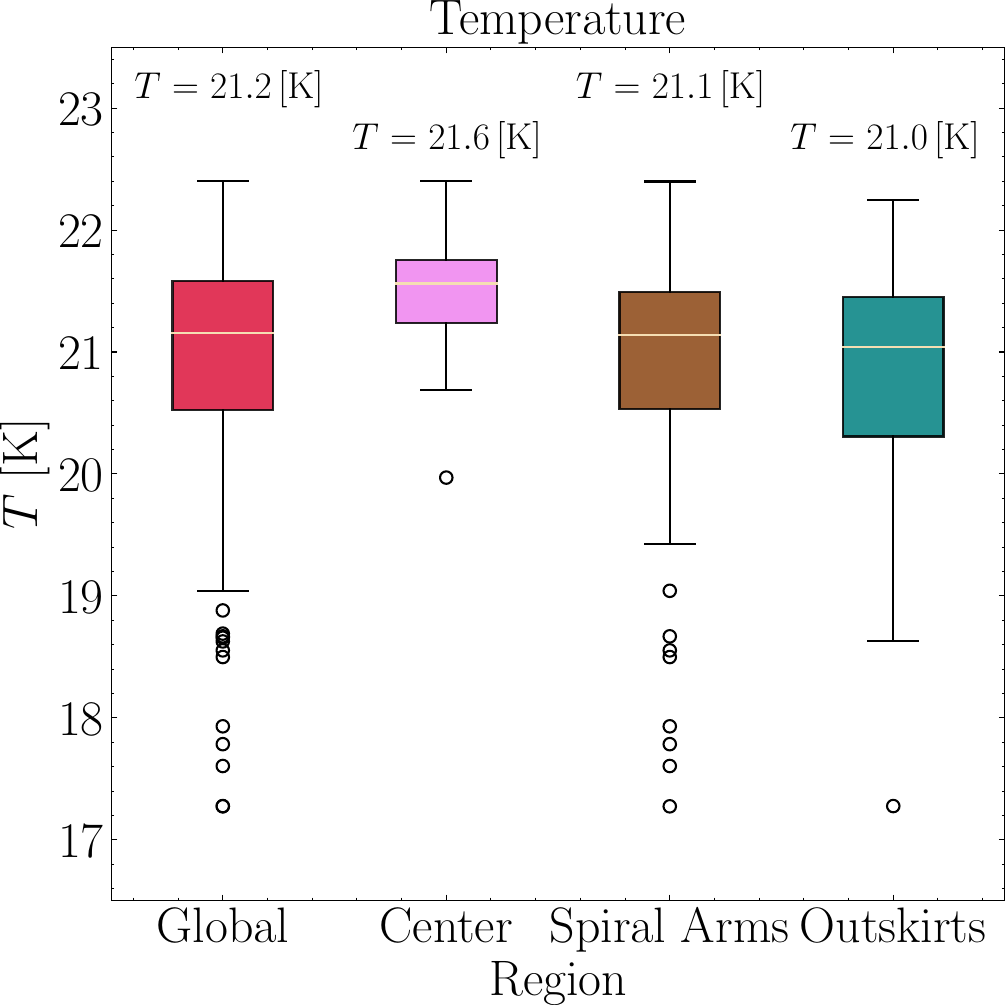}
  \includegraphics[width=0.29\linewidth]{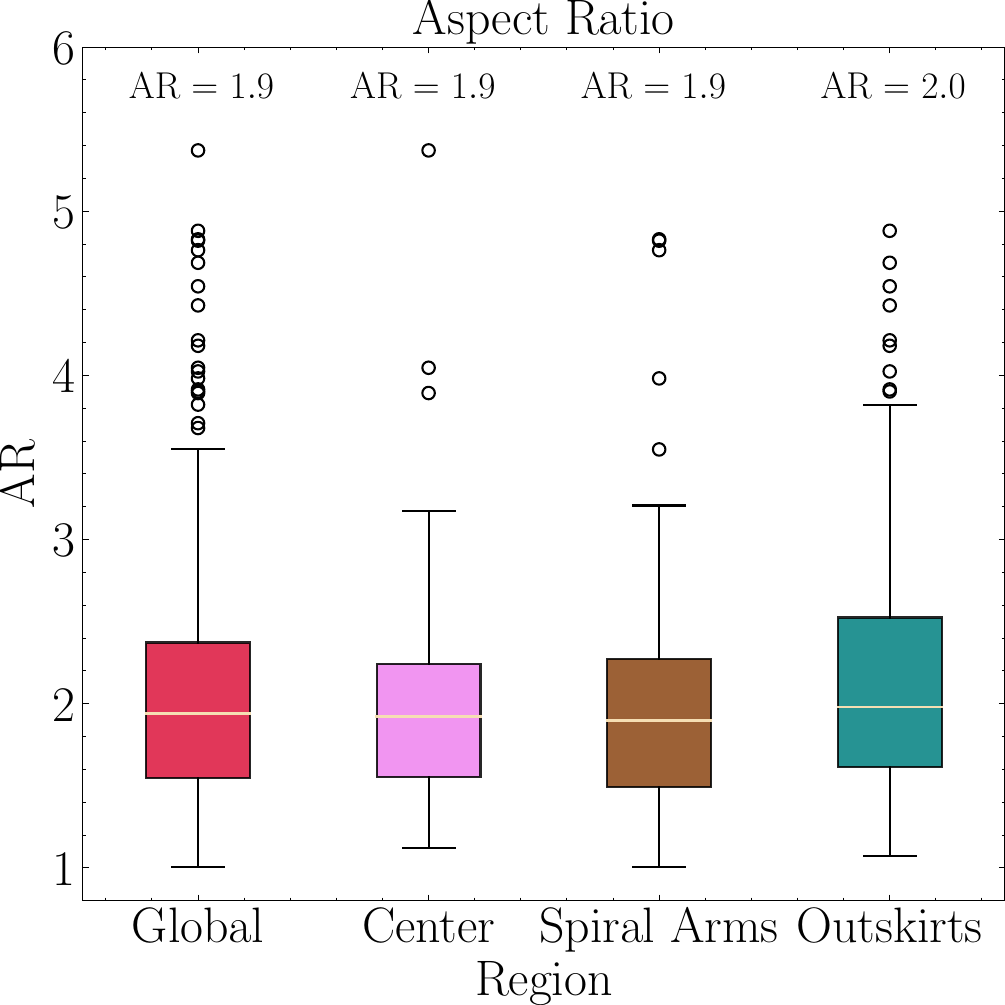}
  \caption{
  Box plots of the determined dust-derived parameters categorized based on galactic environments. 
  The lower and upper whiskers of the box plot represent the lowest and maximum values of the dataset, respectively. The colored box shows the distribution's interquartile spread, or the range from the 25th to the 75th percentile; the median is indicated by the solid beige line inside the box. The distributions' outliers are shown as circles.}%
  \label{fig:mass_density_smd_rad_pressure_AR_vs_regions}
\end{figure*}
We confirm this finding for M33 using our dust-derived high-resolution $\Nhtwo$ map (Fig.~\ref{fig:ccdf_envi}), which shows the complementary cumulative distributions of the entire disk of M33 and the three defined environments.\footnote{
We focus hereafter on dust, since the results of dust and CO are very similar and we want to avoid to overcharge the paper.} The complementary cumulative distribution function provides the likelihood that an observation from a sample exceeds a certain value on the x-axis. It becomes evident that the central region exhibits column densities throughout the spectrum higher than those of the spiral arms and outer regions. The spiral arms and the outskirts display comparable levels of low $\Nhtwo$ below approximately $2\times10^{20}\,\mathrm{cm^{-2}}$. Beyond this threshold, the spiral arms diverge, maintaining higher column density values. This finding aligns with the results reported in~\citet{Leroy2021}. The median value for the central region (provided in the panel for all distributions) is roughly three to 3.5 times higher than for the other two regions. Furthermore, the central region shows the steepest slope among all distributions, while the spiral arms and outer regions demonstrate a shallower slope towards higher column densities.

\subsubsection{GMC properties in different environments}

The distributions of GMC properties (mass, average density, surface mass density, radius and aspect ratio) are shown in Fig.~\ref{fig:mass_density_smd_rad_pressure_AR_vs_regions} as a function of galactic environment. 
The global distribution of the entire disk of M33 is shown in red on the left for comparison. GMCs located in the center are represented in violet, those in the two main spiral arms are in brown and those in the outskirts, excluding the center and the two main spiral arms, are depicted in turquoise. The median is displayed as a straight line within the boxes in beige.

Most of the properties show a weak variation for the median values in different environments. Only the central region of M33 exhibits larger masses and surface mass densities of the GMCs compared to the regions in the remaining disk (see also Sect.~\ref{subsec:comparison_Rgal}, where we have already observed this trend.)  
Overall, the GMCs in the center are denser, those in the spiral arms are larger, while those in the outskirts are more elongated. Generally, the GMC populations in the spiral arm and outer regions do not exhibit large variations in their properties.

\subsubsection{GMC masses in different environments}
\label{subsubsec:mass_vs_region}

The masses of the GMCs are noticeably higher in the central region of M33. 
The median and minimum values 
indicate significantly higher masses compared to the other two regions. Apart from the exceptional case of NGC604 in the spiral arm, the highest mass values are comparable to those in the central area, while the GMCs with the lowest masses have even lower values. The outer regions exhibit GMCs with similarly low masses as those in the spiral arms but lack GMCs with such high masses.

One hypothesis is that spiral arms, which contain a larger amount of material, increase the occurrence of cloud-cloud collisions, thereby supporting the creation of high-mass entities~\citep{Dobbs2008}. This would result in a tendency for the most massive clouds to be situated in spiral arms. However, the spiral arms exhibit lower densities. This is also true for the surface mass density compared to that in the central region. If larger GMCs gather more mass and thus support SF, then this should yield higher surface mass densities. Since the GMCs in spiral arms are merely larger without possessing higher column densities, this results in lower masses and surface mass densities, which correlate with SF, suggesting that SF should be lower. As discussed above, the impact of cloud-cloud collisions in the Milky Way have been investigated by~\citet{Kobayashi2017} and~\citet{Kobayashi2018}, for which an effective impact has only been found for GMCs more massive than $10^6\,\mathrm{M_\odot}$.
Furthermore, while the most massive GMC (NGC604) is located in the northern spiral arm, the other GMCs in these environments do not support this picture. Both the median and the 75th percentile values are lower than those of the center. Additionally, most outliers, except for NGC604, have less mass compared to those in the center. This discrepancy may be due to the limited resolution of $75\,$pc, whereas~\citet{Dobbs2008} simulate molecular clouds with higher resolution.
\citet{Corbelli2019} suggested that the formation of more massive clouds in the center may occur due to the rapid rotation of the disk relative to the spiral arm pattern, allowing the clouds to grow further as they traverse the arms. 

\subsubsection{GMC densities, surface mass densities, and radii in different environments}

For GMC densities, the environments show minimal variation. The median values are similar across different regions. The main distinction is observed in the outliers at the outskirts, where the densities do not peak as high as those in the GMCs in other areas.

The surface mass densities demonstrate a pattern similar to that of the masses. The central region contains GMCs with the highest masses, whereas the median and the values at the lower ends of the spectrum decrease in the spiral arms and decrease even more in the outer regions. This finding aligns with the radial trends that have been discussed in Sect.~\ref{subsec:comparison_Rgal} and confirms~\citet{Querejeta2021}, reporting increased gas surface densities closer to the central regions of galaxies by analyzing the CO(2-1) data obtained from the PHANGS-ALMA survey~\citep{Leroy2021}.

On the other hand, the GMC radii differ most significantly in the spiral arms, with NGC604 as the outlier. The center and outskirts have smaller GMC radii. The median radius in the spiral arms is $\sim\,$$90\,$pc, while in the outskirts it is $\sim\,$$63\,$pc.

\subsubsection{GMC elongations/aspect ratios and temperatures in different environments}
\label{subsubsec:pressure_elongation_temp}

The center has GMCs with the least elongation at the higher end of the spectrum, while the spiral arm and outskirts have slightly stronger elongated GMCs. 
It is not clear to which extent the GMCs with an AR larger than 3 represent GMFs that were found in the Milky Way~\citep{Ragan2014,Goodman2014,Zucker2015,Wang2020} and in external galaxies~\citep{Hughes2013,Leroy2016}. These studies typically define GMFs as long filamentary structures with lengths exceeding $50\,$pc and masses above $10^5\,\mathrm{M_\odot}$ and suggest that they trace the denser spine region of the spiral arms and the mid-plane of the gravitational potential in the galaxy. 
We note that while some Galactic GMFs exhibit widths down to $\sim\,$$1\,$pc, a scale which remains indistinguishable from our current resolution, other GMFs possess notably larger widths~\citep{Zucker2018}.
In particular,~\citet{Wang2020} presented dust and dense gas tracers of one filament in the Milky Way with an AR of about 3 and a length of $68\,$pc which would fit formally to some of the GMCs we detect. 
Figure \ref{fig:mass_density_smd_rad_pressure_AR_vs_regions} shows that there is only a very weak environmental dependency of the AR. However, the most elongated GMCs are found within the spiral arms and the outskirts and this could indicate (as discussed in Sect.~\ref{subsubsec:AR_vs_Rgal}) a stretching effect due to shear forces on the massive GMCs (or molecular clouds) as they transition from the spiral arms to the inter-arm regions~\citep{Koda2009}. It could also be the result of disruption caused by feedback from stars~\citep{Meidt2015,Chevance2020,Bonne2023}.
\citet{DuarteCabral2016} discovered in a computational simulation of GMCs within a two-armed spiral galaxy that, while the average characteristics of the inter-arm and spiral arm GMCs are comparable in terms of their ARs, the extremely elongated GMCs in their dataset are predominantly associated with the inter-arm regions.
Given that the outskirts exhibit highly elongated GMCs it is possible that stellar feedback contributes to the disruption of GMCs in both environments. 

Hence, our proposition is that the shear forces in the center do not account for disrupting the GMCs, since the GMCs in the center exhibit the lowest median elongation. Additionally, the least elongated and most massive GMCs are located in the center (excluding NGC604), indicating that the center is conducive to the formation of high-mass GMCs. 
As detailed in Section~\ref{subsubsec:temp_vs_Rgal}, we argue that the strong galactic potential subjects GMCs to an isotropic pressure, which accounts for the observed distribution of GMC elongation. 
It is unlikely that stellar feedback plays a significant role in elongating GMCs in the center, as the feedback would be uniformly distributed throughout the center, resulting in disrupted GMCs across the region and dynamically altering GMC boundaries. After typical GMC lifetimes, any cloud detection algorithm would identify new segments of an original GMC as a new GMC, incorporating parts of previously disrupted GMCs. With this iterative process and the isotropic galactic potential, extreme elongation tendencies are expected to diminish, resulting in the non-increasing elongation of GMCs, unless specific conditions, such as the presence of a bar, exist in the central area.

In summary, it is observed that while the overall dynamics on a large scale influences cloud properties, there is no clear indication that SFE is notably enhanced in any specific environment.

\subsection{Power-law mass spectra with galactic environment}
\label{subsec:powerlaw_envi}

\begin{figure}[htbp]
  \centering
  \includegraphics[width=0.9\linewidth]{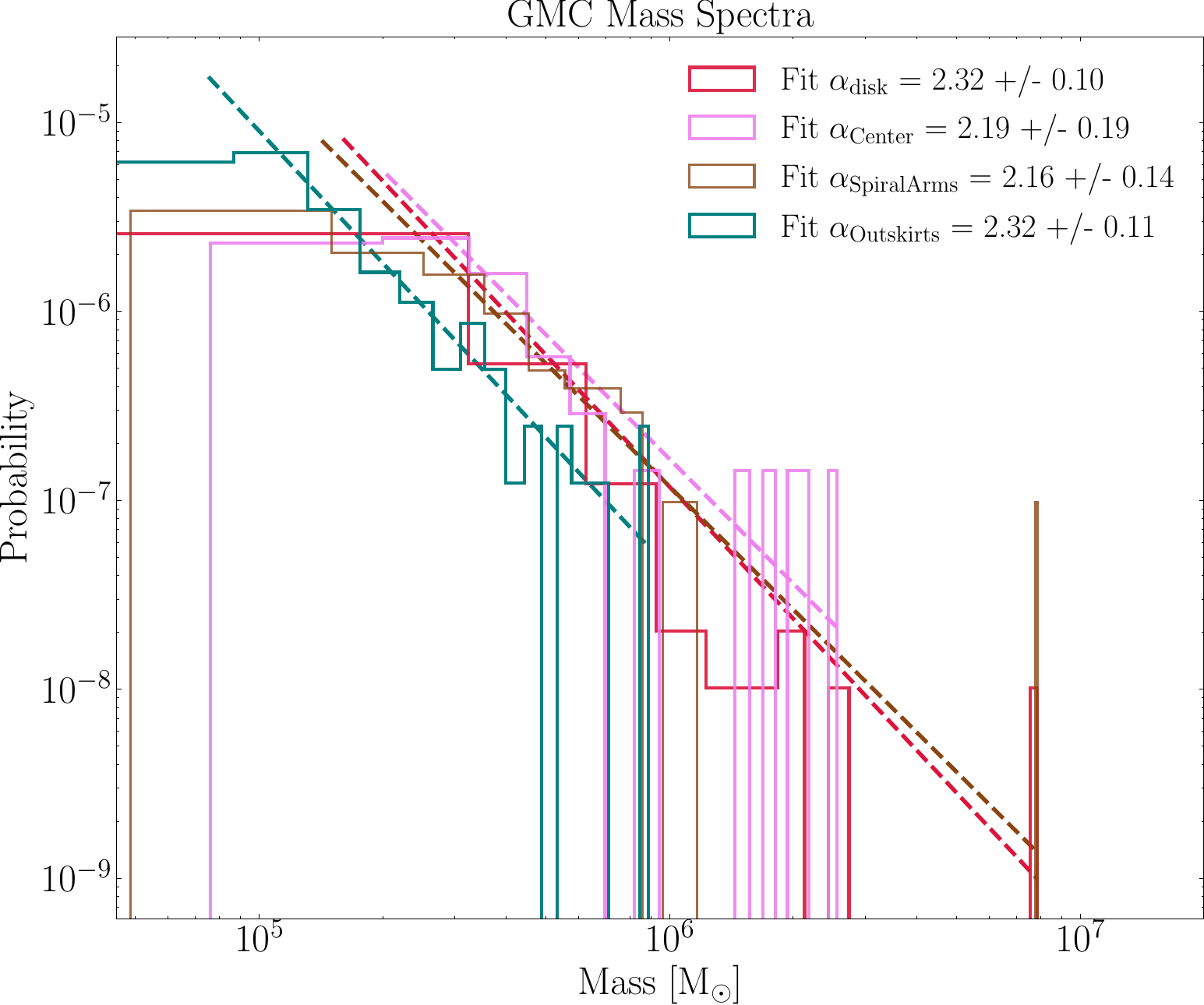}%
  \caption{Power-law mass spectra of GMCs in the dust-derived map for the three galactic environments.
  The power-law mass spectra is split into the three galactic environments center (pink), spiral arms (brown) and outskirts (turquoise). For comparison, the fit for the whole disk is shown again in red. 
  }%
  \label{fig:powerlaws_envi}
\end{figure}

As pointed out in Sect.~\ref{subsubsec:mass_vs_region}, there are noticeable differences in the masses of the GMCs between the center and the remaining disk of M33. 
We therefore also investigate the distribution of GMC masses within each large-scale environment by conducting power-law mass spectra. 

Figure~\ref{fig:powerlaws_envi} illustrates the mass spectra of the GMCs in the different environments of M33. 
The central area and the spiral arms of M33 exhibit the highest abundance of high-mass clouds, with $\alpha_\mathrm{Center} = 2.19 \pm 0.19$ and $\alpha_\mathrm{SpiralArms} = 2.16 \pm 0.14$ having relatively shallow slopes. In contrast, there is a significant reduction in high-mass GMCs toward the outer regions, where GMCs have mainly lower to moderate masses with a steeper slope of $\alpha_\mathrm{SpiralArms} = 2.32 \pm 0.11$. 
\citet{Bigiel2010} observed smaller GMCs at larger galactocentric radii of M33, suggesting a steeper slope in the outskirts of M33, which is supported by the slope we determine. This higher slope suggests that high-mass objects in the outskirts may face challenges in their formation or are rapidly destroyed after formation. 
This aligns with the findings presented in~\ref{subsubsec:mass_vs_region} and with the results of~\citet{Rosolowsky2021} in their examination of GMCs across spiral galaxies within the PHANGS dataset.

The spectra shown in Fig.~\ref{fig:powerlaws_envi} indicate that GMCs have higher masses in areas with lower galactocentric radii, closer to the center.
In the outskirts, the most massive GMC reaches a mass of $\sim\,$$1\times10^6\,\mathrm{M_\odot}$, the lowest of the three environments. This is consistent with having the steepest slope of all three. In the central region, GMCs that are only about three times more massive are found, while in the spiral arm, the most massive GMC (NGC604) has a mass of $\sim\,$$8\times10^6\,\mathrm{M_\odot}$. This observation indicates that cloud growth may be prevented or that large GMCs are being disrupted in the central area, at least to reach such high masses as observed in NGC604. This phenomenon could be attributed to complex dynamics and shear forces or to the enhanced interstellar radiation field in the central region. 
In contrast, the mass distribution of GMCs in the spiral arms, excluding NGC604, consists of less massive clouds than in the center. NGC604 leads to a flattening of the slope in the spiral arms. Despite the predominance of low-mass objects in the spiral arms, the conditions in this region appear to be conducive to the growth of larger clouds, maybe due to the absence of a strong interstellar radiation field and/or shear forces disrupting the clouds. 
However, since the center hosts the GMCs with the highest masses, with the exception of NGC604, this conclusion remains uncertain.

\citet{Dobbs2019} found a decrease in the power-law index after incorporating SF into their simulations. As clouds become dense, the index drops to values between $\alpha \approx 1.8$ and $\approx 2$. Considering delayed SF results in an index that agrees better with observations. This suggests, in general, that SF occurs in later stages of GMC formation. These authors also divided the clouds into ``star-forming'' and ``non-star-forming'' clouds (SF clouds and non-SF clouds hereafter). While SF clouds inject energy into the clouds, heating them locally, this leads to a flattened slope of $\alpha=1.8$, whereas non-SF clouds exhibit a slope of $\alpha=2.68$. They identified that the non-SF clouds tend to reside at a larger galactocentric radius, indicating higher SF activity in the center.
Increased surface mass densities are associated with this phenomenon and we also observe higher surface mass densities toward the center, in accordance with this. Additionally, a stronger galactic potential towards the center could account for this finding.
Compared with our results, this suggests a higher SF activity in the center and spiral arms than in the outskirts. 
\citet{Dobbs2019} also provide reasons as to why larger clouds tend to host more SF: these larger clouds are statistically more inclined to have dense areas, thus increasing the likelihood of SF. These clouds probably accumulate more mass as they begin to form stars, suggesting that clouds not undergoing SF may just be in an earlier phase of their lifetime when they have lower masses.

This is consistent with what~\citet{Braine2018} found by analyzing the IRAM 30m CO data of M33. Detected GMCs (from the catalog of~\citealt{Corbelli2017}) have been divided into three radial bins: $R_\mathrm{gal}<2.2\,$kpc, $2.2<R_\mathrm{gal}<3.7\,$kpc and $R_\mathrm{gal}>3.7\,$kpc. The power-law indices for these bins are $\alpha=1.36$, $\alpha=1.68$ and $\alpha=1.87$, respectively, showing an increase with radius. They also subdivided the GMCs into three SF classes -- no obvious SF (A), embedded SF (B) and exposed SF (C) -- based on~\citet{Corbelli2017}. More evolved GMCs accumulate more mass and show shallower slopes. Star-forming GMCs lie closer to the center than non-SF GMCs. However, A-class GMCs consistently show steep slopes regardless of their position, while C-type GMCs also have similar slopes regardless of their position, indicating that SF activity is more important than galactic environment.

As also discussed in Sect.~\ref{subsec:power_law_mass_spectra},~\citet{Fujita2023} found distinct power-law indices of $\alpha=2.30\pm0.11$ and $\alpha=2.51\pm0.14$ in the Milky Way within a galactocentric radius of $<8.15\,\mathrm{kpc}$ and beyond $<16.3\,\mathrm{kpc}$, respectively. Taking the errors into account, this is also consistent with our data split into the environments, which represent distinct regions along the galactocentric radius.

\section{Conclusion and summary}  \label{sec:summary} 

In~\citetalias{Keilmann2024}, we presented a novel technique to use the {\sl Herschel} flux maps and $\mathCO\mathrm{(2-1)}$ data of M33 to produce $\Nhtwo$ maps at $18.2''$ ($\sim\,$$75\,$pc) resolution, resolving GMCs. A complete \Xco\ map was applied to the $\mathCO$ map to compute the $\Nhtwo$ map with values in the range of $1.6-2\times10^{20}\,\cmKkms$. 
This \Xco\ factor is close to the Milky Way value and thus questions the usual approach of applying a single, adopted \Xco\ factor for the whole galaxy and simply using a two-times-higher value for M33 due to its lower metallicity. 

We then employ the Dendrograms algorithm to identify GMCs from these maps, calculate the physical properties, and compare the results between dust and CO and with Milky Way data from the $\mathCO\mathrm{(1-0)}$ Columbia survey presented in~\citet{Nguyen2016}. In addition, an investigation was conducted to explore the potential influences of the galactocentric radius and galactic environment on GMC properties.


\begin{enumerate}
  \item[\tiny{\ding{108}}] 
We find that M33 lacks the more massive ($>10^6\,\mathrm{M_\odot}$) and denser GMCs that are present in the Milky Way.
The mean GMC masses of M33 are about an order of magnitude lower than those of the Milky Way.
A power-law fit to the mass spectrum
gives values of $\alpha=2.32\pm0.10$ for dust and $\alpha=1.87\pm0.08$ for CO. These indices align with those found in other studies of M33, Milky Way values, and simulations, which all show a large spread.

  \item[\tiny{\ding{108}}] 
There appears to be a limit to the sizes of GMCs of around $150\,$pc, as the distributions for the largest GMCs of M33 and Milky Way show similar shapes and a decline above $\sim\,$$100\,$pc. We do not find the equivalent of GMFs in the Milky Way but note that there is an inter-cloud medium at column densities of around $10^{21}\,\mathrm{cm^{-2}}$ that contains a significant mass, in particular in the central region of M33. In the outskirts, the lower-column-density material encloses the GMCs, which is particularly evident in dust. 

\item[\tiny{\ding{108}}] 
The surface mass densities for M33 are $22\pm5\,\mathrm{M_\odot\,pc^2}$ from dust and $16\pm6\,\mathrm{M_\odot\,pc^2}$ for CO, which are about an order of magnitude lower than the same values for the Milky Way. The increased surface mass density may suggest an increase in SFR. Finally, M33  shows similar patterns in some alternative characteristics to those observed in other nearby galaxies in the PHANGS survey.

\item[\tiny{\ding{108}}] We find no or only weak correlations between physical properties and galactocentric radius, but some results indicate a dependence on the larger-scale environment. 

\item[\tiny{\ding{108}}] The central region of M33 displays slightly higher median values for parameters such as mass, average density, surface mass density, 
and dust temperature, but contains the GMCs with the smallest aspect ratios. The center hosts the most massive GMCs (except for NGC604), which also exhibit the highest surface mass densities. However, as the center seems to be the region with the highest influence on star formation, the variations in physical parameters across the environments are predominantly minor in nature.
The spiral arms mainly host the largest GMCs, while they contain most of the extreme outliers across different parameters, such as mass, surface mass density, size, and elongation. On the contrary, the outskirts generally feature the lowest median values, with the exception of average density 
and elongation. However, the majority of the GMCs, despite some outliers, do not seem to be significantly affected by the conditions of the galactic environment. 

\item[\tiny{\ding{108}}] The power-law fits to the mass spectra derived from CO and dust vary with the galactic environment ($\alpha=2.19\pm0.19$ for the center, $\alpha=2.16\pm0.14$ for the spiral arms, and $\alpha=2.32\pm0.11$ for the outskirts). 
These results are consistent with observations in the Milky Way, suggesting similar indices for both the inner and outer disk of our galaxy. However, the slope of the spiral arms decreases due to the high mass of GMC NGC604. The remaining high-mass GMCs in the spiral arms have lower masses than those in the galaxy center. This complicates the identification of the physical mechanisms at work, as high interstellar radiation fields and shear forces are likely to disrupt more massive GMCs, whereas the absence of these mechanisms would enhance these parameters in the spiral arms. 
\end{enumerate}

Overall, we conclude that the center seems to have a slightly greater influence on GMC properties than the other environments, but that mechanisms operating at the cloud scale ---notably stellar feedback--- may have a similar or greater impact on GMCs than large-scale dynamics inherent to galactic environments.

\section{Data availability}
Full Tables~\ref{table:clouds_properties_dust_dendro} and~\ref{table:clouds_properties_CO_dendro} and the updated $\htwo$ column density data are available in electronic form at the CDS via anonymous ftp to cdsarc.u-strasbg.fr (130.79.128.5) or via http://cdsweb.u-strasbg.fr/cgi-bin/qcat?J/A+A/.

\begin{acknowledgements}
E.K. and N.S. acknowledge support from the FEEDBACK-plus project of the BMWI via DLR, Project Number 50OR2217.  \\
E.K. acknowledges support by the BMWI via DLR, project number 50OK2101. \\
S.K. acknowledges support from the Orion-Legacy project that is supported by the BMWI via DLR, project number 50OR2311. \\
We gratefully acknowledge the Collaborative Research Center 1601 (SFB 1601 subprojects A6, B2 and B3) funded by the Deutsche Forschungsgemeinschaft (DFG, German Research Foundation) – 500700252.\\

This research made use of astrodendro, a Python package to compute Dendrograms of Astronomical data (http://www.dendrograms.org/)\\
\end{acknowledgements}

\bibliography{main.bib}{}


\begin{appendix}


\section{IRAM $^{12}\mathrm{CO(2-1)}$ line-integrated intensity map of M33}
\label{app_a:COmap}

\begin{figure}[htbp]
  \centering
  \includegraphics[width=0.95\linewidth]{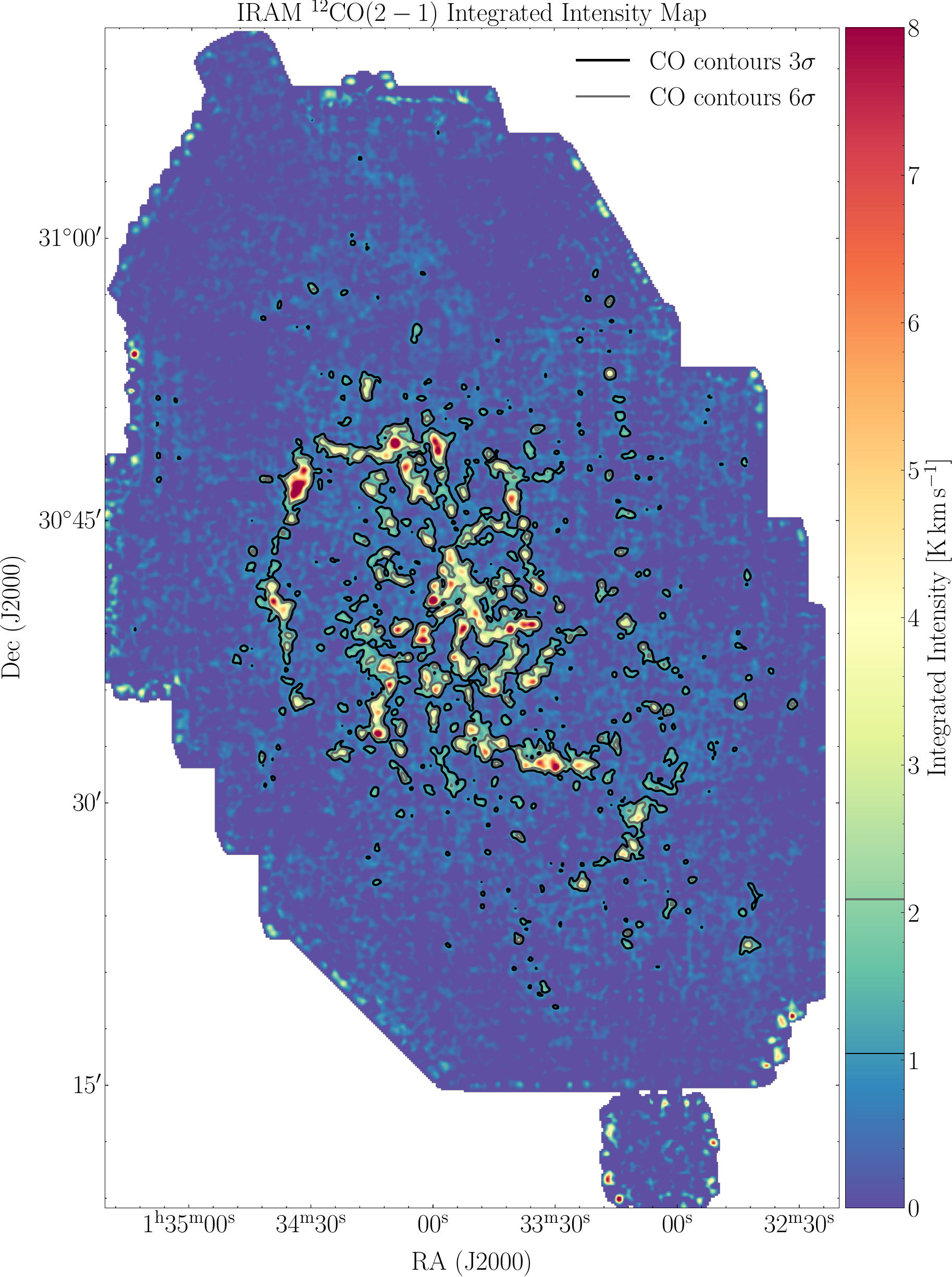}
  \caption
      {$^{12}\mathrm{CO(2-1)}$ line-integrated intensity map of M33~\citep{Druard2014}. The map has been smoothed to the resolution of $18.2''$ and re-gridded to the coordinate grid of the SPIRE 250$\,\mum$ map.}
    \label{fig:co_intint_map}
\end{figure}

Figure~\ref{fig:co_intint_map} shows the $\mathrm{^{12}CO(2-1)}$ map of M33 obtained with the IRAM 30m telescope~\citep{Gratier2010,Druard2014}.

\section{\Xco\ factor map of M33}
\label{app_b:xco} 

\begin{figure}[!htb]
  \centering
  \includegraphics[width=0.95\linewidth]{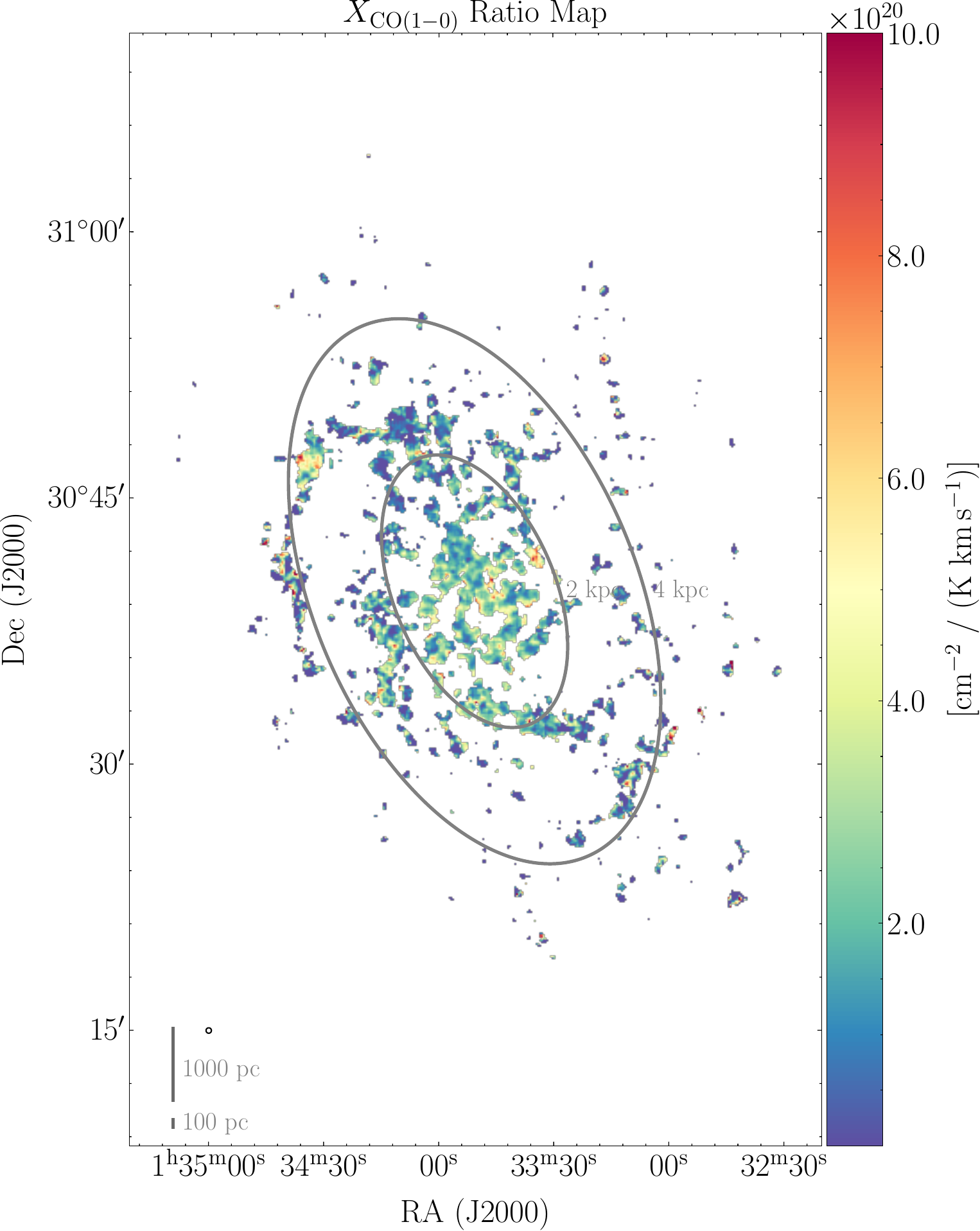}
  \caption{\Xco\ factor (ratio) map of Method I of~\citetalias{Keilmann2024}. The two ellipses represent a circular radius equivalent to $2$ and $4\,$kpc.} 
    \label{fig:ratioMap}
\end{figure}

Figure~\ref{fig:ratioMap} displays the \Xco\ factor map defined as the dust-derived $\Nhtwo$ over $\mathrm{CO}$ line-integrated intensity at each position in M33 at $18.2''$ and scaled with the $\mathrm{CO(\frac{2-1}{1-0})}$ line ratio~\citep{Druard2014} to $\mathrm{CO(1-0)}$ intensity. See~\citetalias{Keilmann2024} for more details.

\section{Influence of Dendrogram parameters on the GMC statistics}
\label{app:dendrogram_parameters}

\begin{figure*}[!htbp]
  \centering
  \includegraphics[width=0.4\linewidth]{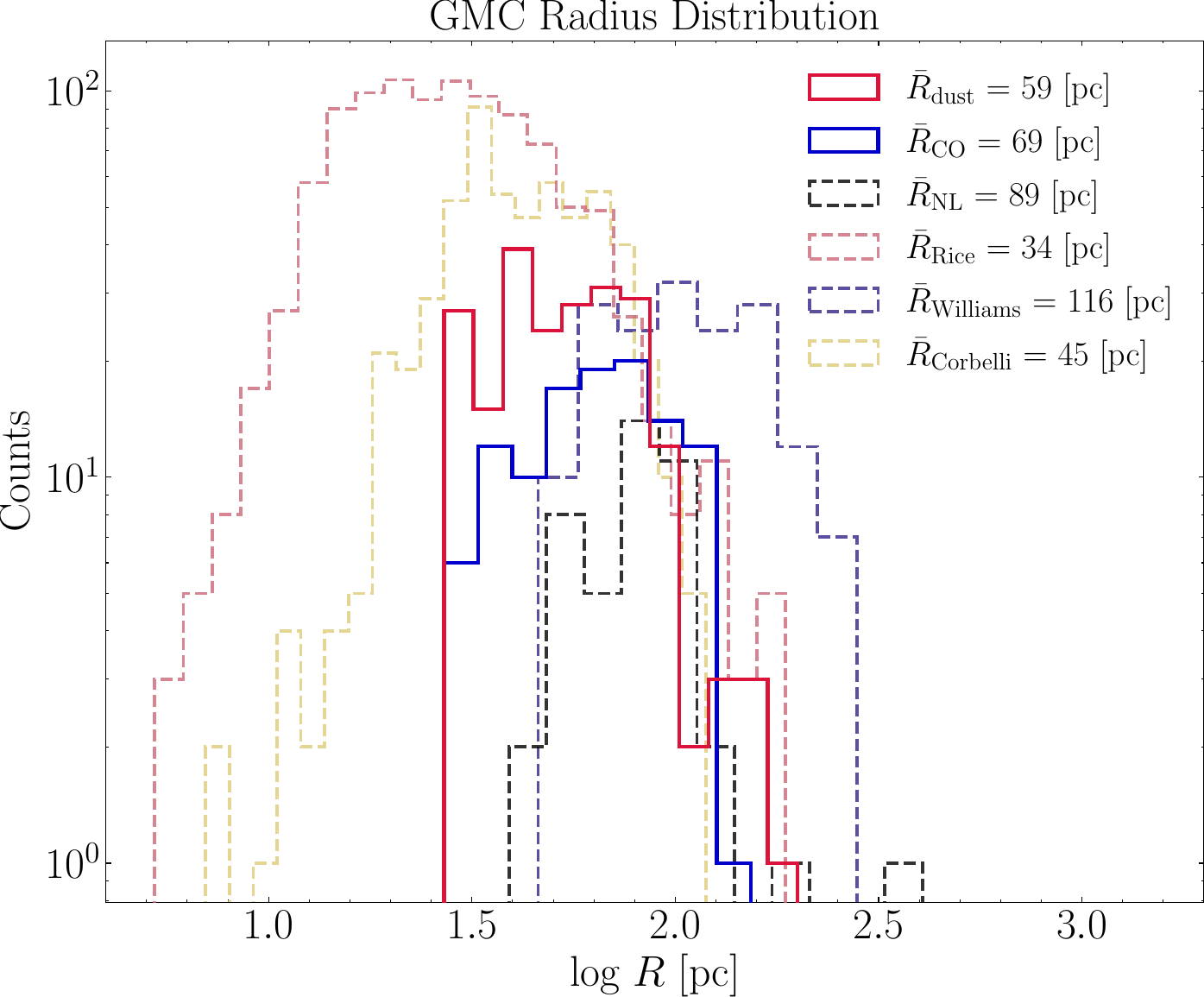}
  \includegraphics[width=0.4\linewidth]{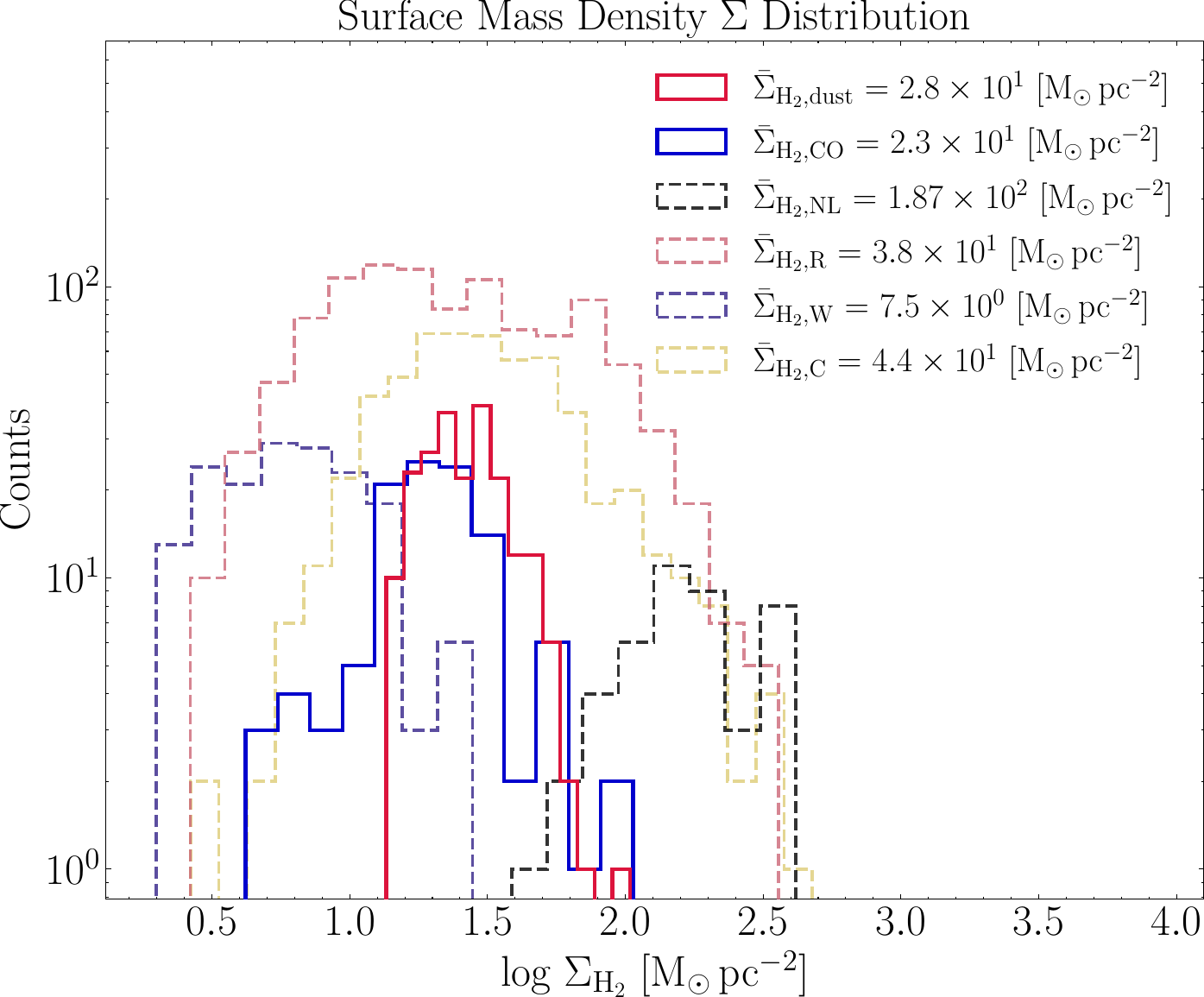}

  \vspace{0.1cm} 
  \includegraphics[width=0.4\linewidth]{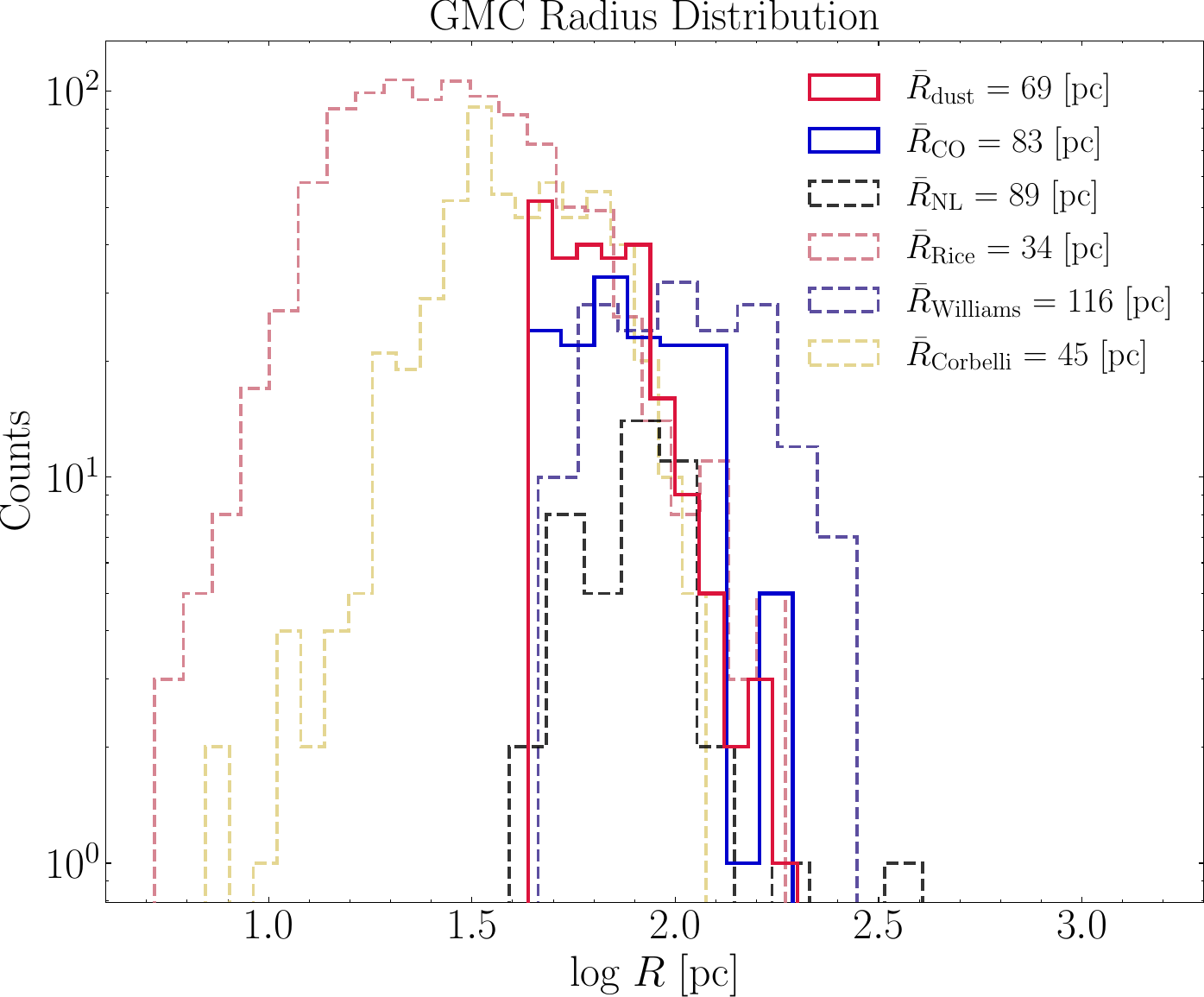}
  \includegraphics[width=0.4\linewidth]{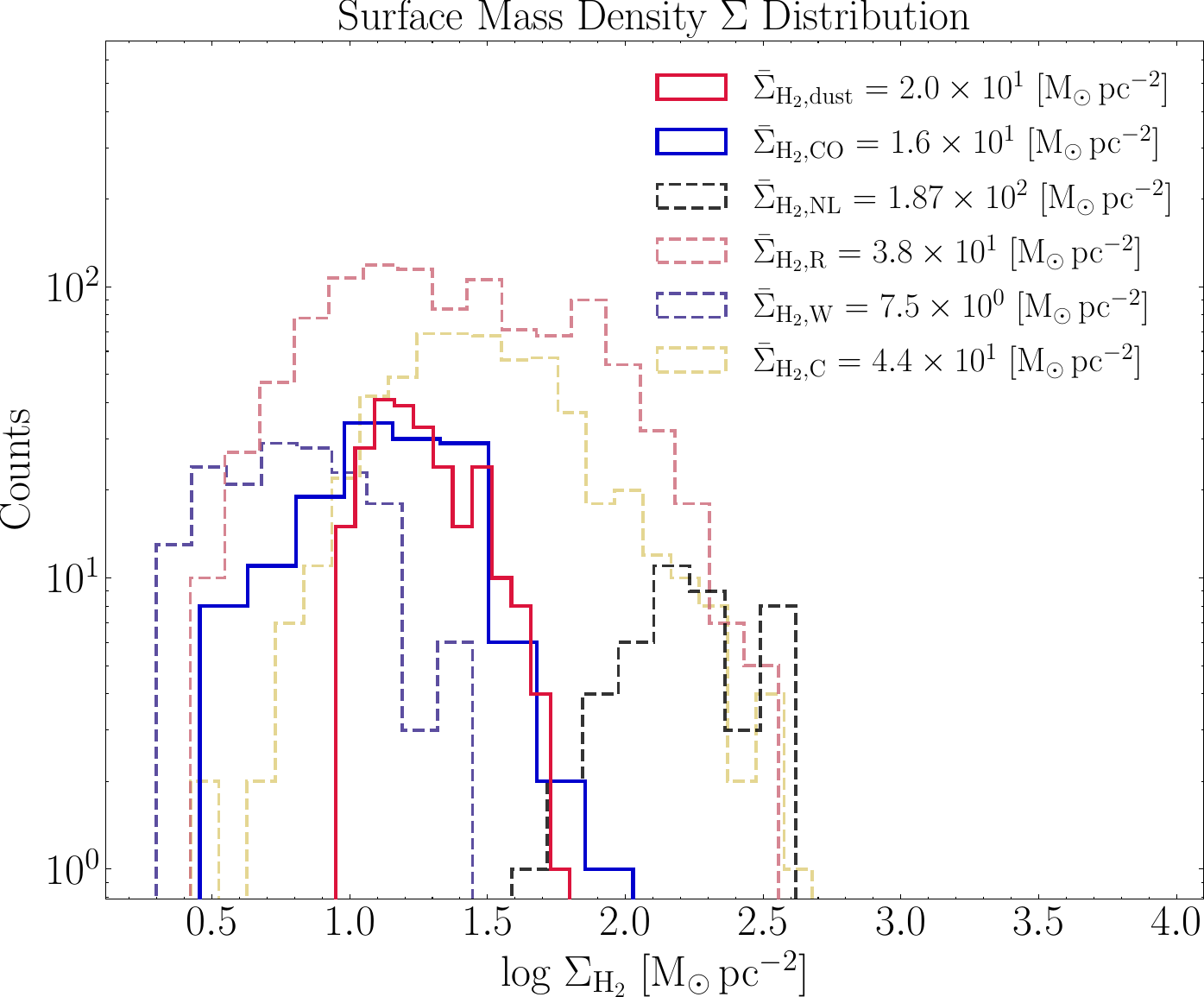}
  \caption{Distributions of GMC properties with varying Dendrogram parameters.
  The upper panels show histograms of radius and surface mass density with a \command{min\_value} of $5\sigma$. The lower panels display the same properties for a beam factor of 1.5.}
  \label{fig:dist_5sigma}
\end{figure*}

We conducted a Dendrograms parameter study by changing the \command{min\_value} and the beam factor for cloud selection. Obviously, increasing these parameters results in the identification of larger and more massive GMCs, whereas changing \command{min\_delta} practically does not alter the results. Nevertheless, the undetected residual emission became more substantial, while the results change non-significantly considering the uncertainties. Consequently, we conclude that the use of $3\sigma$ for \command{min\_value} and a beam factor of 1.2 are the optimal settings for the Dendrogram analysis. Tables~\ref{table:params_dust} and~\ref{table:params_co} list the mean values of the main properties of the dust- and CO-derived GMCs for a \command{min\_value} of $5\sigma$ and a beam factor of 1.5.
We further illustrate the results for a subset of the plots discussed above. Figure~\ref{fig:dist_5sigma} shows the distributions of the radii and surface mass densities for both varied parameters. The size is mainly unchanged, while a \command{min\_value} of $5\sigma$ excludes the low-mass GMCs, leading to an overall shift towards higher values in mass and (surface mass) densities. However, the change is still low.

The effect on the galactocentric radius dependence as an example for the mass and density
is marginally shifted to higher values. Furthermore, there is no notable trend with the galactocentric radius for any of the other properties, similar to the result found with a \command{min\_value} of $3\sigma$ and a beam factor of 1.2.

For the power-law slopes, the increase in the slope for dust-derived data is insignificant. For CO, the slope rises from 1.87 to 2.03. However, considering the uncertainties, the change falls within the margin of error.

This analysis confirms that our selected Dendrogram parameters are robust and produce reliable results.

\begin{table}[htbp]
\caption{Mean properties of dust-derived GMCs.}
\label{table:params_dust} 
\centering
\begin{tabular}{lcccccccccccc}
\hline\hline
 & \textbf{$\boldsymbol{\command{min\_value}=5\sigma}$} & \textbf{$\boldsymbol{\mathrm{beam\,factor} = 1.5}$}  \\
\hline
GMCs                                    &   $214$     &  $242$  \\
$M\,\mathrm{[\times10^{5}\,M_\odot]}$   &       $3.5\pm1.1$     &  $3.6\pm1.1$  \\
$n\,\mathrm{[cm^{-3}]}$                 &       $7\pm3$       &  $3\pm1$    \\
$\Sigma\,\mathrm{[M_\odot\,pc^{-2}]}$   &       $28\pm7$      &  $20\pm5$   \\
$R\,\mathrm{[pc]}$                      &       $59\pm11$      &  $69\pm12$   \\
$\kappa_\mathrm{Mass-Size}$             &       $1.8\pm0.1$     &  $2.0\pm0.1$  \\
AR                                      &   $1.9$           & $1.9$  \\
$\alpha$                                &       $2.38\pm0.13$    &  $2.42\pm0.12$ \\
\hline
\end{tabular}
\tablefoot{The table shows the mean values for the parameters obtained with varying the Dendrogram parameters.}
\end{table}

\begin{table}[htbp]
\caption{Mean properties of CO-derived GMCs.}
\label{table:params_co} 
\centering
\begin{tabular}{lcccccccccccc}
\hline\hline
 & \textbf{$\boldsymbol{\command{min\_value}=5\sigma}$} & \textbf{$\boldsymbol{\mathrm{beam\,factor} = 1.5}$}  \\
\hline
GMCs                                    &   $111$       &  $153$      \\
$M\,\mathrm{[\times10^{5}\,M_\odot]}$   &       $4.0\pm2$     &  $3.8\pm2$   \\
$n\,\mathrm{[cm^{-3}]}$                 &       $4\pm2$       &  $2\pm1$   \\
$\Sigma\,\mathrm{[M_\odot\,pc^{-2}]}$   &       $23\pm9$      &  $16\pm7$   \\
$R\,\mathrm{[pc]}$                      &       $69\pm15$      &  $83\pm17$   \\
$\kappa_\mathrm{Mass-Size}$             &       $2.0\pm0.2$     &  $2.2\pm0.2$   \\
AR                                      &   $1.5$           & $1.7$  \\
$\alpha$                                &       $2.03\pm0.12$    &  $2.13\pm0.12$   \\
$L_\mathrm{CO}$                         &   $1.5\pm0.1$     &  $1.8\pm0.1$  \\
\hline
\end{tabular}
\tablefoot{The table shows the mean values for the parameters obtained with varying the Dendrogram parameters.}
\end{table}
%

\section{$\mathCO$ luminosity}
\label{sec:CO_lumi}

\begin{figure}[htbp]
  \centering  
  \includegraphics[width=0.95\linewidth]{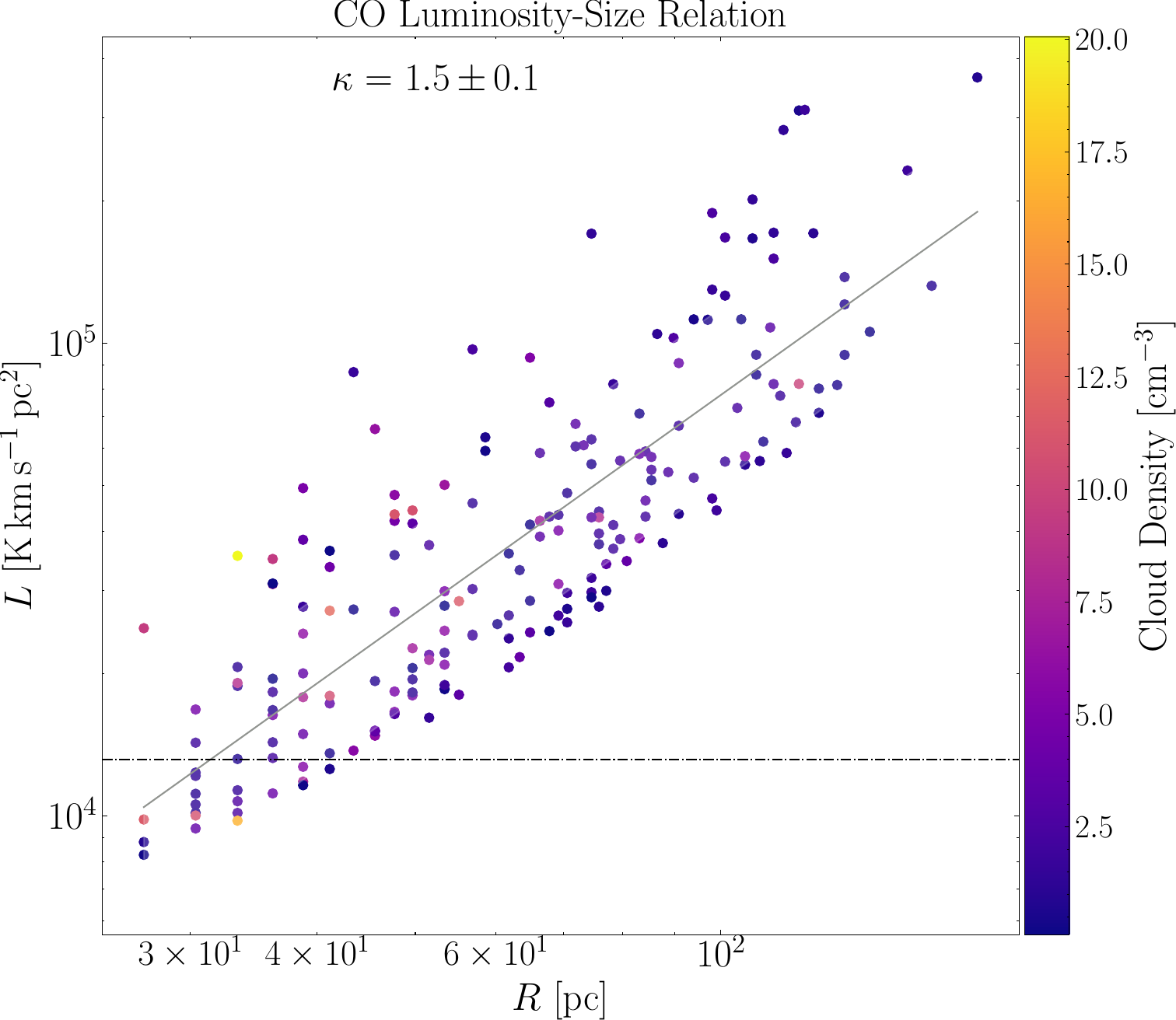}%
  \caption{$\mathCO(1-0)$ luminosity-size relation of the structures identified in the CO data. The horizontal line represents the $2\sigma$ sensitivity limit. 
  }%
  \label{fig:Lco}
\end{figure}

For completeness and to enable comparison to other studies, we display in Fig.~\ref{fig:Lco} the $^{12}\mathrm{CO(1-0)}$ luminosity $L_\mathrm{CO}$ of M33, using the $\mathrm{CO(2-1)/CO(1-0)}$ line ratio of $0.8$ from~\citet{Druard2014}. 
The mean CO luminosity is $(9.2\pm2.0)\times10^4\, \mathrm{K\,km\,s^{-1}\,pc^2}$, which is lower than the value of $3\times10^5\,\mathrm{\kkms\,pc^2}$ for the $\mathrm{CO(2-1)}$ IRAM data at $12''$, which corresponds to $3.75\times10^5\,\mathrm{\kkms\,pc^2}$ for $\mathrm{CO(1-0)}$ applying the same line ratio of 0.8.
\citet{Hughes2013} present M33 data from their observations of $\mathrm{CO(1-0)}$, indicating values around $1\times10^5\,\mathrm{\kkms\,pc^2}$. A slope of 1.5 indicates that the CO emission of smaller GMCs is faster saturated compared to larger GMCs. In other words, more CO emission per area can be accounted for in the outer regions of GMCs.

\section{Trends with galactocentric radius of M33}
\label{app:trends_Rgal}

It is evident across all parameters that the majority of GMCs do not exhibit a significant trend with the galactocentric radius. This observation is supported by Spearman correlation coefficients\footnote{The Spearman correlation coefficient is suitable for all types of monotonic relationships, whether linear or nonlinear, and does not require the data to follow a normal distribution. It ranges from $-1$ to $1$, where $-1$ indicates a strong negative correlation, $0$ no correlation and $1$ a strong positive correlation. Given that the Spearman correlation coefficient is effective for both linear and nonlinear relationships, it does not differentiate between these types of correlations.} ranging from $|8.7\times10^{-4}|$ to $|0.4|$, showing no or only low correlation at best (except for the temperature, which clearly shows a gradient; see~\citealt{Tabatabaei2014,Keilmann2024}). Nonetheless, upon closer inspection of the GMCs at the extreme ends of the spectra, there is a subtle trend of mostly decreasing values with increasing galactocentric radius, especially for the surface mass density and average number density. 
Extreme GMCs, that is, GMCs at the tails of the distributions, enhanced in particular large-scale galactic environments, may suggest the presence of physical mechanisms directly enabling the development of particular cloud types in specific galactic regions, potentially influencing SF. We consider GMCs with surface mass densities above $40\,\mathrm{M_\odot\,pc^2}$ to account for an analysis of whether and how extreme clouds may depend on the galactocentric radius. The corresponding GMCs are depicted as thicker and darker pentagons in the following figures.

\subsection{GMC masses with galactocentric radius}

Specifically, for dust-traced masses, GMCs located within approximately $2\,$kpc demonstrate an increase toward the center for those with the lowest masses. 
This pattern is absent in $\mathCO$-traced GMCs. This could be due to strong interstellar radiation fields, which photo-dissociate $\mathCO$~\citep{Offner2014}, leading to less $\mathCO$ emission in the center. 
The remaining GMCs beyond $2\,$kpc do not show a dependence on the galactocentric radius for both tracers. The Spearman correlation coefficients show practically no correlation with galactocentric radius over the whole data range. For the most massive GMCs observed with both tracers, their highest masses decrease with galactocentric radius, supporting~\citet{Corbelli2017}, who found a similar decrease beyond $4.5\,$kpc.
The GMCs with the most extreme surface mass density values are not those with the highest masses; they appear to be arbitrarily distributed in terms of mass; see Fig.~\ref{fig:mass_density_SMD_radius_vs_Rgal} (top left).
The data point around $4\,$kpc is NGC604.
In the case of branches, the median mass is $2\times10^5\,\mathrm{M_\odot}$. Additionally, there are structures with increased masses within low galactocentric radii below $1\,$kpc. The remaining structures show a similar trend to the GMCs (leaves) with a slightly higher correlation coefficient of $-0.5$, which is still only a moderate correlation.

According to their simulation,~\citet{Dobbs2019} report that the masses of GMCs are influenced by the distance from the center of M33. Their findings exhibit a resemblance to our results, particularly in the case of the most massive GMCs, showing a correlation with the galactocentric radius. Nevertheless,~\citet{Dobbs2019} did not provide a quantitative assessment of this dependency, making direct comparisons difficult. Due to the similarity in the plots showing that the majority of the clouds do not seem to depend strongly on the galactocentric radius, it is possible that the actual dependency they state is similar in magnitude to what we quantify.

The lack of correlation between cloud mass and $R_\mathrm{gal}$ likely results from a balance between the more compact and therefore more luminous clouds at the galaxy center and the diffuse and hence more extended sources in the outskirts.
\citet{Williams2019} identified a higher, yet still weak correlation between mass and galactocentric radius with a Kendall rank correlation coefficient\footnote{The Kendall rank correlation coefficient measures non-parametrically 
how well a monotonic function describes the relationship between two variables without assuming their probability distributions. It indicates the similarity in rank orderings of data when sorted by each quantity. High Kendall correlation means similar ranks between variables (correlation of 1), while low means dissimilar ranks (correlation of $-1$). 
} of 0.12.

\subsection{GMC average densities with galactocentric radius}

While, as in the case of masses, a subtle trend of the least dense GMCs is also noticeable in dust-traced GMCs in terms of average density, it is not observed in those traced by $\mathCO$. In contrast, GMCs with the highest densities tend to be more concentrated toward the inner disk of M33, as indicated by dust-derived GMCs and to a lesser extent by the $\mathCO$-derived GMCs. Almost all of these most dense GMCs are also those which have the highest surface mass densities. This is not surprising, as the number and the surface mass density are closely related. However, the overall correlation is absent, as in the case of the masses. The Spearman correlation coefficients practically do not quantify the correlation.
Structures identified as branches show a median density of $1.1\,\mathrm{cm^{-3}}$ and exhibit a slightly higher correlation coefficient of $0.3$, which is still only a weak correlation.

\subsection{GMC surface mass densities with galactocentric radius}

Since the masses are divided by the area of a GMC, the surface mass density is somewhat less dependent on the resolution, making it comparably easier to compare with other studies.

The surface mass density (Fig.~\ref{fig:mass_density_SMD_radius_vs_Rgal}, bottom left) exhibits a similar pattern as for the mass among dust-traced GMCs considering the GMCs with the lowest surface mass densities found within $2\,$kpc. For both tracers, GMCs located at the higher end of the spectrum demonstrate a slight tendency to exhibit a higher surface mass density as the galactocentric radius decreases. 

The only strong connections to the other parameters are the radius (see Sect.~\ref{subsec:radii_vs_Rgal} and bottom right of Fig.~\ref{fig:mass_density_SMD_radius_vs_Rgal}) \text{and}
the averaged number density (upper right of Fig.~\ref{fig:mass_density_SMD_radius_vs_Rgal}). 
The number density is related to the surface mass density in a natural way, which does not reveal new surprising insights. 

The overall correlation for all data points in CO is practically absent, while for dust-derived GMCs a low correlation of $-0.2$ is determined. Thus, also the surface mass density seems to have no strong dependence on the galactocentric radius.

In the case of branches, the dependency is similar with a median value of $18\,\mathrm{M_\odot\,pc^{-2}}$ and a doubled but still only lower moderate correlation coefficient of $-0.4$.

\subsection{GMC radii with galactocentric radius}
\label{subsec:radii_vs_Rgal}

Considering the radius, the data points suggest that there is no clear pattern with the galactocentric radius, which is further supported by the nearly nonexistent or very weak correlation of $-0.3$. However, there are also a small number of GMCs with the largest radii close to the center, which are not those with the highest surface mass densities. Instead, the GMCs with the highest surface mass density are preferably the smallest ones.

The branches show a median radii of $185\,$pc with very large structures within a galactocentric radius below $1\,$kpc. This mainly increases the correlation coefficient to a moderate value of $-0.5$.

\subsection{GMC elongations with galactocentric radius}
\label{subsubsec:AR_vs_Rgal}

In terms of AR (Fig.~\ref{fig:pressure_AR_temp_vs_Rgal}, left), dust-derived GMCs that are most elongated tend to be situated farther away from the galaxy's center, typically beyond approximately $3\,$kpc, whereas the majority of GMCs do not show a clear trend, showing a weak correlation coefficient of $0.2$. 
CO-derived GMCs exhibit a similar distribution with little to no correlation, with the most elongated GMCs to a low extent found in the mid-range between $\sim\,$$2\,$kpc and $\sim\,$$5\,$kpc of the galaxy. However, the most extreme GMCs do not follow a consistent pattern.
For both tracers, there seems to be no distinct pattern regarding the GMCs with the highest surface mass densities.
The overall elongation of CO-derived GMCs is less compared to dust-derived GMCs with a median of $1.6$, which aligns with the contours of the CO-derived structures in Fig.~\ref{fig:colden_ico_structures_dendro}.

An almost unchanged elongation is found for the branches with a median value of 2.1 and a reduced correlation coefficient of 0.1.

\subsection{GMC temperatures with galactocentric radius}
\label{subsubsec:temp_vs_Rgal}

The data points for the temperature show a dependency with galactocentric radius to some extent (Fig.~\ref{fig:pressure_AR_temp_vs_Rgal}, right). However, some data points beyond roughly $2.5\,$kpc deviate distinctly from the remaining data points. This leads to the most significant correlation with the galactocentric radius among all the parameters analyzed, which is $-0.4$ and $-0.6$ for dust- and CO-derived GMCs, respectively. This is a weak to moderate correlation. Interestingly,~\citet{Williams2019} find only a weak correlation of $-0.26$.
In the case of branches, a similar trend is found with a slightly higher median temperature of $21.5\,$K and a correlation coefficient of $-0.7$. This is the strongest correlation found in this study. 
A natural explanation for the decreasing dust temperature is the overall decrease in intensity of the interstellar radiation field with increasing galactocentric radius~\citep{Rice1990}.

A compelling relationship emerges when higher pressures appear to result in less elongated GMCs (indicated by higher densities and temperatures in the central region). Such a phenomenon could possibly be attributed to the pervasive pressure within the central region~\citep{Sun2020b,Sun2020} as a result of the stronger galactic potential, which acts uniformly, resulting in more isotropically shaped GMCs. In contrast, GMCs located in the outer regions appear to be influenced by pressure originating predominantly from a specific direction, causing forces that are not uniformly distributed across all GMCs. Those at mid-range distances (presumably located in the spiral arms) could be elongated when exiting the spiral arm and therefore its gravitational potential. It is also possible that stellar feedback causes those GMCs to elongate.
Nevertheless, this phenomenon appears to be relevant solely to the GMCs lying at the tail of the spectra, specifically those with the greatest 
elongation.

\section{Galactic environments of M33}
\label{app_c:env}

\begin{figure}[htbp]
  \centering
  \includegraphics[width=0.95\linewidth]{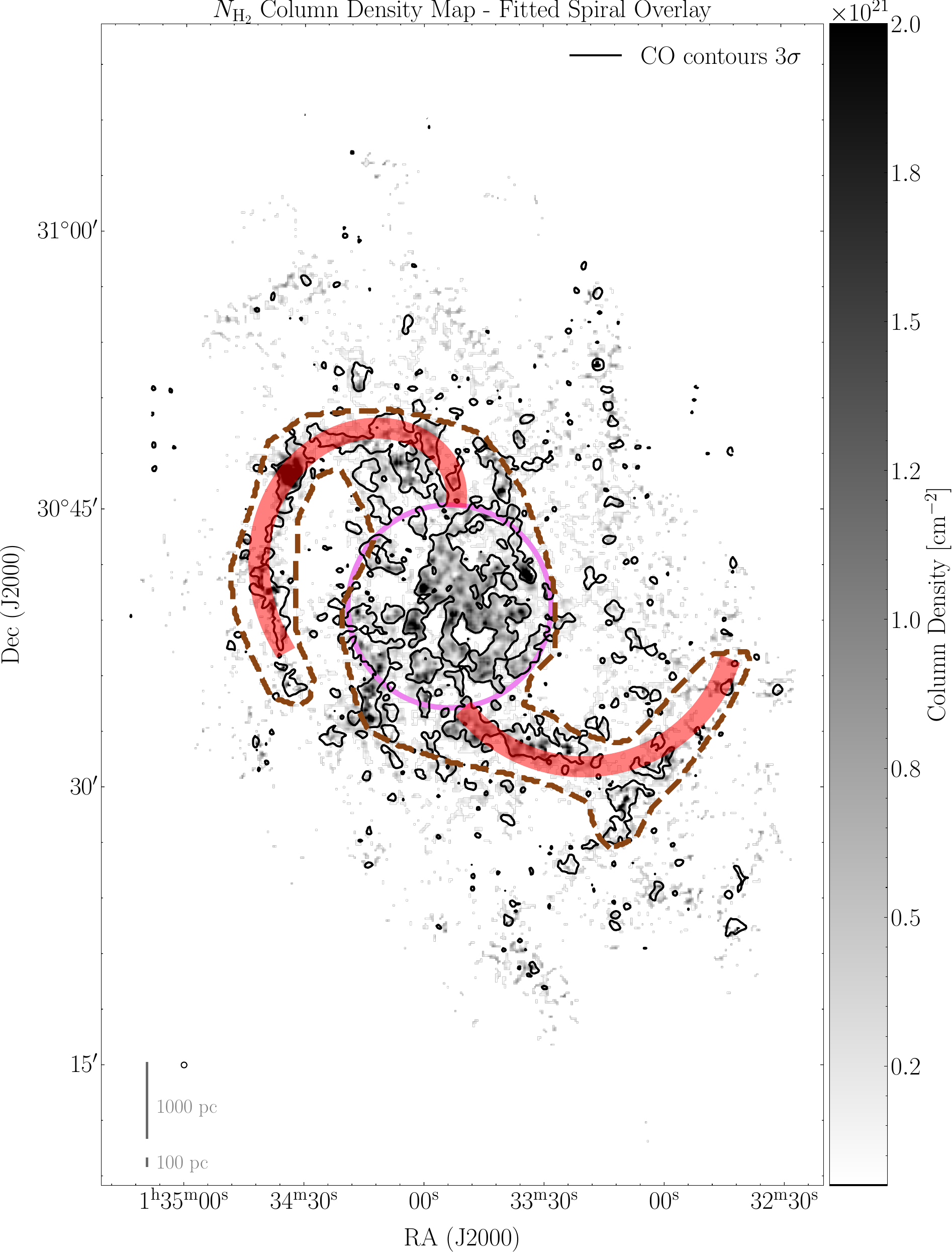}%
  \caption{The dust-derived $\Nhtwo$ map shows the boundaries of three galactic environments: the center (dashed pink lines), spiral arms (dashed brown lines) and outskirts (remaining area). Center coordinates are RA(2000)$\,$=$\,$1$^h$33$^m$50$^s$, Dec(2000)$\,$=$\,$30$^\circ$39$'$37$''$~\citep{simbadM33}. The result of the log-spiral fit is shown in red.
  }%
  \label{fig:gal_envi}
\end{figure}

We constructed an unsharp-masked image of the $N_\htwo$ map and identified the densest points along the spiral arms on our $N_\htwo$ map. This is different to~\citet{Querejeta2021}, who use stellar densities and a morphological decomposition based on Spitzer $3.6\,\mum$. However, since we intend to define the spiral arms for GMCs, which are located in H$_2$ gas, we use our $N_\htwo$ map for this decomposition. The coordinates of the densest points were deprojected to the plane of the galaxy (using an inclination of 56$^\circ$ and a position angle of 23$^\circ$). The fit was then performed in logarithmic polar coordinates. The log-spiral fit was projected back to the plane of the sky and is shown in Fig.~\ref{fig:gal_envi}. It matches the areas of the spiral arms very well by eye-inspection. From this result, the asymmetry of both spiral arms becomes obvious. The northern spiral arm is wound stronger with a higher pitch angle starting from the center compared to the southern spiral arm. Overall, we distinguish three galactic environments: center, spiral arms and outskirts onto the dust-derived $N_\htwo$ map. 
We have chosen a circle of $1.3\,$kpc for the center to distinguish from the spiral arms and outskirts. This radius was selected since it encompasses the maximum column density distribution of the central region, while still maintaining reasonable borders for the fitted spiral arms.

\end{appendix}

\end{document}